\documentclass[aps,pra,amsmath,superscriptaddress,amssymb,preprint,floatfix]{revtex4-1}     

\usepackage{graphicx}
\usepackage{color}
\usepackage{multirow}
\usepackage{setspace}

\newcommand\nmax{n_{\rm max}}
\newcommand\cu{{\cal B}}
\newcommand\cc{{\cal C}}
\newcommand\ccb{{\cal A}}

\newcommand\ca{{\cal A}}
\newcommand\cb{{\cal B}}

\begin{document}
\title{Phase transitions in fluctuations and their role in two-step nucleation} 

\author{Daniella James}
\affiliation{Department of Physics, St. Francis Xavier University, Antigonish, NS, B2G 2W5, Canada}

\author{Seamus Beairsto}
\affiliation{Department of Physics, St. Francis Xavier University, Antigonish, NS, B2G 2W5, Canada}

\author{Carmen Hartt}
\affiliation{Department of Physics, St. Francis Xavier University, Antigonish, NS, B2G 2W5, Canada}

\author{Oleksandr~Zavalov}
\affiliation{Department of Physics, St. Francis Xavier University, Antigonish, NS, B2G 2W5, Canada}

\author{Ivan Saika-Voivod}
\affiliation{Department of Physics and Physical Oceanography, Memorial University of Newfoundland, \\St. John's, Newfoundland A1B 3X7, Canada}

\author{Richard K. Bowles}
\affiliation{Department of Chemistry, University of Saskatchewan, Saskatoon, SK, 57N 5C9, Canada}

\author{Peter H. Poole}
\affiliation{Department of Physics, St. Francis Xavier University, Antigonish, NS, B2G 2W5, Canada}

\begin{abstract}
We consider the thermodynamic behavior of local fluctuations occurring in a stable or metastable bulk phase.  For a system with three or more phases, a simple analysis based on classical nucleation theory predicts that small fluctuations always resemble the phase having the lowest surface tension with the surrounding bulk phase, regardless of the relative chemical potentials of the phases.  We also identify the conditions at which
a fluctuation may convert to a different phase as its size increases, referred to here as a  ``fluctuation phase transition" (FPT).
We demonstrate these phenonena in simulations of a two dimensional lattice model by evaluating the free energy surface that describes the thermodynamic properties of a fluctuation as a function of its size and phase composition.  We show that a FPT can occur in the fluctuations of either a stable or metastable bulk phase and that the transition is first-order.  We also find that the FPT is bracketed by well-defined spinodals, which place limits on the size of fluctuations of distinct phases.  Furthermore, when the FPT occurs in a metastable bulk phase, we show that the superposition of the FPT on the nucleation process results in two-step nucleation (TSN).  We identify distinct regimes of TSN based on the nucleation pathway in the free energy surface, and correlate these regimes to the phase diagram of the bulk system.  Our results clarify the origin of TSN, and elucidate a wide variety of phenomena associated with TSN, including the Ostwald step rule.
\end{abstract}

\date{\today}
\maketitle

\section{introduction}

Fluctuations play a central role in many liquid state phenomena.
For example, it has long been appreciated that fluctuations dominate the physics of critical phenomena and second-order phase transitions~\cite{Stanley:1971}.
Similarly, in the study of supercooled liquids and the origin of the glass transition,
local fluctuations that deviate from the average properties of the bulk liquid phase (e.g. dynamical heterogenerities and locally favored structures) continue to be the focus of much work to explain the complex dynamics observed as a liquid transforms to an amorphous solid~\cite{Berthier:2011hs,Royall:2015if,Turci:2017bp}.
For network-forming liquids such as water, ``two-state" models that assume the occurrence of two distinct, transient local structures have been proposed to explain thermodynamic and dynamic anomalies occurring in both the stable and supercooled liquid~\cite{Anisimov:2018ki}.  
The central role of fluctuations is perhaps most obvious in nucleation phenomena, where a bulk metastable phase decays to a stable phase via the formation of a local fluctuation (the critical nucleus) of sufficient size to be able to grow spontaneously to macroscopic scale~\cite{,Debenedetti:1996,Kelton:2010}.

The behavior of local fluctuations is particularly complex in the case of ``two-step nucleation" (TSN)~\cite{Wolde:1997ez,Tavassoli:2002ey,Vekilov:2004jc,Pan:2005hy,Lutsko:2006cq,vanMeel:2008hb,Chen:2008dl,Li:2008js,Pouget:2009dg,Duff:2009p6360,Vekilov:2010gm,Iwamatsu:2011if,Demichelis:2011ela,Sear:2012ji,Iwamatsu:2012jr,Iwamatsu:2012hy,Santra:2013kn,Wallace:2013df,Peng:2014is,Salvalaglio:2015iv,Kratzer:2015hg,Malek:2015gq,Qi:2015iea,Sosso:2016bda,Ishizuka:2016ep,Guo:2016fo,Bi:2016ek,Ishizuka:2016ep,Iwamatsu:2017ke,Lee:2017fw,Zhang:2017ju,Yamazaki:2017jb,Santra:2018wl}.  In TSN, the
first step in the phase transformation process consists of the appearance in 
the bulk metastable phase of a local fluctuation that resembles an intermediate phase distinct from the stable phase.  In the second step of TSN, this intermediate fluctuation undergoes a transition in which the stable phase emerges from within the intermediate phase.
Evidence for TSN has been observed experimentally in a wide range of molecular and colloidal systems~\cite{Vekilov:2004jc,Vekilov:2010gm,Peng:2014is,Ishizuka:2016ep}, including important cases relevant to understanding protein crystallization~\cite{Zhang:2017ju,Yamazaki:2017jb} and biomineralization~\cite{Li:2008js,Pouget:2009dg}.  
Due to the involvement of an intermediate phase, TSN is poorly described by classical nucleation theory (CNT), in which it is assumed that a nucleus of the stable phase appears directly from the metastable phase~\cite{Debenedetti:1996,Kelton:2010}.  Large deviations from CNT predictions are thus associated with TSN~\cite{Sear:2012ji}.  Given these challenges, an understanding of TSN is required to better control and exploit complex nucleation phenomena.  For example, significant questions remain concerning the nature of long-lived ``pre-nucleation clusters" that have been reported in some TSN processes~\cite{Li:2008js,Pouget:2009dg,Demichelis:2011ela,Wallace:2013df}.  
Control of polymorph selection during nucleation and the origins of the Ostwald step rule~\cite{Kelton:2010} are also facilitated by a better understanding of TSN~\cite{Santra:2013kn,VanDriessche:2018fy,Santra:2018wl}.

A number of theoretical and simulation studies have investigated TSN~\cite{Wolde:1997ez,Tavassoli:2002ey,Pan:2005hy,Lutsko:2006cq,vanMeel:2008hb,Chen:2008dl,Duff:2009p6360,Iwamatsu:2011if,Iwamatsu:2012jr,Iwamatsu:2012hy,Santra:2013kn,Salvalaglio:2015iv,Malek:2015gq,Qi:2015iea,Sosso:2016bda,Bi:2016ek,Iwamatsu:2017ke,Santra:2018wl}.
These works highlight the role of metastable phase transitions involving competing bulk phases, and their connection to the transition from the intermediate to the stable fluctuation that occurs in TSN.
In addition, several works have examined TSN in terms of the two dimensional (2D) free energy surface (FES) that quantifies the nucleation pathway as a function of the size of the nucleus and its degree of similarity to the stable phase~\cite{Wolde:1997ez,Duff:2009p6360,Iwamatsu:2011if,Qi:2015iea,Salvalaglio:2015iv,Malek:2015gq}.  For example, Duff and Peters demonstrated the existence of two distinct regimes of TSN, in which the transition of the nucleus to the stable phase occurs either before or after the formation of the critical nucleus, located at the saddle point of the FES~\cite{Duff:2009p6360}.  
Iwamatsu further showed that the FES for TSN may contain two distinct nucleation pathways, each with its own saddle point~\cite{Iwamatsu:2011if}.  
These works illustrate the complexity of TSN, and help to explain the non-classical phenomena attributed to TSN in experiments.  

Despite the insights obtained to date from experiments, theory and simulation, our understanding of TSN would benefit from a clearer understanding of the relationship between a bulk phase transition and the transition that occurs in the growing nucleus from the intermediate to the stable phase.  This latter transition occurs in a finite-sized system (the fluctuation) and is controlled not only by the chemical potential difference between the intermediate and stable phases, but also by strong surface effects at the interface with the surrounding bulk metastable phase.  In the following, we refer to the transition that occurs in a finite-sized fluctuation as a ``fluctuation phase transition" (FPT) to distinguish it from a bulk phase transition, for which surface effects play no role in determining the thermodynamic conditions of the equilibrium transition.
It would be particularly useful to know how to predict the conditions at which a FPT will occur, and how these conditions are related to the thermodynamic conditions at which bulk phase transitions, both stable and metastable, occur in the same system.  

The present paper has two aims:  
First, we seek to clarify the general thermodynamic behavior of local fluctuations, regardless of whether these fluctuations are involved in a nucleation process, to better understand the properties of fluctuations in their own right.
Second, we wish to specifically elucidate TSN via a detailed examination of the local fluctuations that appear during TSN, and how the behavior of these fluctuations varies over a wide range of thermodynamic conditions.  

To achieve these aims, we first present in Section~\ref{simp} a simple theoretical analysis of fluctuations. 
This analysis uses the assumptions of classical nucleation theory (CNT) to make some general predictions on the nature of local fluctuations in either a stable or metastable phase when more than one type of fluctuation is possible.  This analysis identifies a number of distinct thermodynamic scenarios for how fluctuations behave as a function of their size, including predicting the conditions at which a FPT will occur.  

In Sections~\ref{latt}-\ref{bulk} we then describe simulations of a 2D lattice model, which provides a case study in which our analytical predictions can be tested.  In particular, the model is simple enough to provide a complete thermodynamic description of the fluctuations, in the form of a FES which characterizes the fluctuations in terms of their size and phase composition.  We locate and characterize the FPT as it is observed in the features of the FES for both a stable and metastable phase.  Our lattice model results demonstrate that TSN occurs when a FPT is superimposed on a nucleation process occurring in a metastable phase.  We are thus able to provide a comprehensive perspective on the origins of TSN, clarify its relationship to bulk phase behavior, and elucidate the non-classical nature of TSN.  

In Section~\ref{disc} we discuss the connections between our results and previous work on TSN, such as Refs.~\cite{Duff:2009p6360} and \cite{Iwamatsu:2011if}.  Our results reproduce a number of observations made previously in separate works, as well as identifying new behavior that underlies these previous observations, thus unifying our understanding of the origins of TSN and related phenomena.
At the same time, our results demonstrate that complex thermodynamic behavior is an intrinsic property of fluctuations, independent of metastability and nucleation.  As discussed in Section~\ref{conc}, our findings thus have wider implications for understanding liquid state phenomena that are dominated by the behavior of fluctuations.

\section{CNT analysis of fluctuations}
\label{simp}

We begin with an idealized analysis of the fluctuations in a bulk phase when there are two other bulk phases that may occur in the system.  As we will see, this analysis suggests the existence of several distinct regimes of fluctuation behavior depending on the thermodynamic conditions, including regimes in which a FPT occurs.  Since some of these regimes also correspond to TSN processes, this analysis provides an idealized framework for understanding the origins TSN.  Furthermore, the analysis predicts a regime in which a FPT occurs in a stable phase where nucleation is not possible, demonstrating that a FPT and nucleation can be regarded as independent phenomena.

Consider the free energy cost $G$ to create a fluctuation of size $n$ molecules within a bulk phase $\ca$.  
We assume that any such fluctuation can be associated with one of two other bulk phases $\cb$ or $\cc$.
We also assume that the free energy cost to create a fluctuation of $\cb$ or $\cc$ within $\ca$ is given respectively by the CNT expressions,
\begin{eqnarray}
G_{\ca\cb}&=&n^\alpha\, \phi\, \sigma_{\ca\cb} +  n\, \Delta \mu_{\ca\cb} \cr
G_{\ca\cc}&=&n^\alpha\, \phi\, \sigma_{\ca\cc} +  n\, \Delta \mu_{\ca\cc}
\label{gsimple}
\end{eqnarray}
where $\sigma_{\ca\cb}$ is the $\ca\cb$ surface tension, and $\Delta \mu_{\ca\cb}=\mu_\cb-\mu_\ca$ is the difference in the chemical potential between the bulk phases $\cb$ and $\ca$, and where $\sigma_{\ca\cc}$ and $\Delta \mu_{\ca\cc}$ are similarly defined~\cite{,Debenedetti:1996,Kelton:2010}.  Here we assume that the surface area of the fluctuation is $n^\alpha\, \phi$, where $\alpha=(D-1)/D$ depends on the dimension of space $D$, and $\phi$ is a shape factor.  
For circular fluctuations in $D=2$, $\alpha=1/2$ and $\phi=(4\pi v)^{1/2}$, where $v$ is the area per molecule.  
For spherical fluctuations in $D=3$, $\alpha=2/3$ and $\phi=(36\pi v^2)^{1/3}$, where $v$ is the volume per molecule.  

\begin{figure}
\includegraphics[scale=0.45]{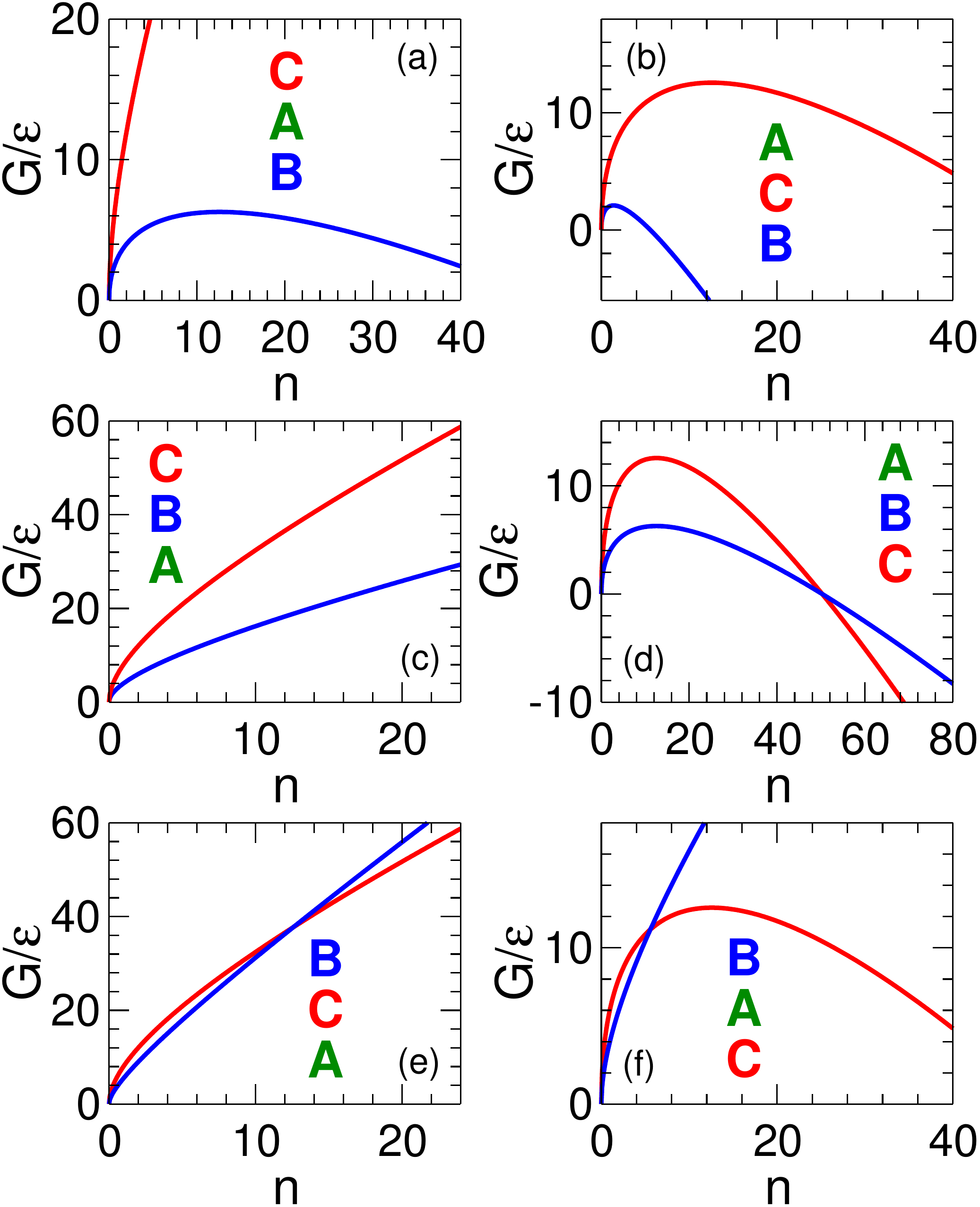}
\caption{
$G_{\ca\cb}$ (blue) and $G_{\ca\cc}$ (red) versus $n$ for $D=2$ in various thermodynamic regimes.  
To plot these curves we scale energies by $\varepsilon=|\Delta \mu_{\ca\cc}|$ and surface tensions by
$\varepsilon v^{-\alpha}$.
In all panels $\sigma_{\ca\cc}=2\sigma_{\ca\cb}=2\varepsilon v^{-\alpha}$.
In the lefthand panels $\Delta \mu_{\ca\cc}=\varepsilon$ and 
in the righthand panels $\Delta \mu_{\ca\cc}=-\varepsilon$.
From top to bottom within each column $\Delta \mu_{\ca\cb}$ increases:  
(a) $\Delta \mu_{\ca\cb}=-\varepsilon/2$;
(b) $\Delta \mu_{\ca\cb}=-3\varepsilon/2$;
(c) $\Delta \mu_{\ca\cb}=\varepsilon/2$;
(d) $\Delta \mu_{\ca\cb}=-\varepsilon/2$;
(e) $\Delta \mu_{\ca\cb}=2\varepsilon$;
(f) $\Delta \mu_{\ca\cb}=\varepsilon/2$.
To indicate the relative stability of the three bulk phases in each panel, the phases are listed vertically according to their value of $\mu$, with $\mu$ increasing from bottom to top in each list.}
\label{simple}
\end{figure}

Since $\alpha<1$, the variation of $G$ with $n$ is always dominated by the surface contribution as $n \to 0$; see Fig.~\ref{simple}.
As a result, the most probable small fluctuations occurring in phase $\ca$ (i.e. the small fluctuations with the lowest free energy)
will always correspond to the phase $\cb$ or $\cc$ that has the lower surface tension with $\ca$, regardless of the values of $\Delta \mu_{\ca\cb}$ or $\Delta \mu_{\ca\cc}$.   Although this result is apparent from the assumptions of CNT, it has important consequences that, to our knowledge, have not been explicitly recognized in previous work.  In particular, our analysis predicts that the initial fluctuations of a bulk phase always favor the local structure that has the lowest surface tension, and that the bulk chemical potential for this structure is irrelevant.

Furthermore, our analysis predicts the conditions at which an abrupt change in the structure of the most probable local fluctuation may occur.
Let us assume that $\sigma_{\ca\cb}< \sigma_{\ca\cc}$, 
in which case $\cb$ fluctuations dominate at small $n$.
If $G_{\ca\cb}$ and $G_{\ca\cc}$ 
intersect at $n>1$, then the most probable fluctuation in $\ca$ will undergo a FPT from $\cb$-like to $\cc$-like as $n$ increases.  Assuming the validity of Eq.~\ref{gsimple}, the value of $n=n_c$ at which the FPT occurs is given by,
\begin{eqnarray}
n_c^{1/D}
               &=&\phi\, \frac{\sigma_{\ca\cc}-\sigma_{\ca\cb}}{\Delta \mu_{\cc\cb}}.
\label{nc}
\end{eqnarray}
For $n_c$ to be non-zero, positive and real, the quotient in Eq.~\ref{nc} must be non-zero and positive.  
Assuming that $\sigma_{\ca\cb}< \sigma_{\ca\cc}$, and if $\cc$ is more stable than $\cb$
(i.e. $\Delta \mu_{\cc\cb}>0$), 
a fluctuation of $\ca$ will undergo a FPT from $\cb$ to $\cc$ at $n=n_c$ as it grows.  
When $\Delta \mu_{\cc\cb}>0$ and when approaching the conditions where $\cb$ and $\cc$ coexist, then 
$\Delta \mu_{\cc\cb}\to 0^+$, guaranteeing the existence of a range of states at which the FPT occurs with $n_c\gg 1$.
On the other hand, if $\cb$ is more stable than $\cc$
(i.e. $\Delta \mu_{\cc\cb}<0$), then $n_c$ is undefined and no FPT occurs; that is, the fluctuations of $\ca$ remain $\cb$-like for all $n$.  
Notably, the above reasoning does not depend on the value of $\mu_\ca$
and thus applies to the behavior of the fluctuations of $\ca$ regardless of whether $\ca$ is stable or metastable with respect to either or both of the bulk $\cb$ and $\cc$ phases. 

In Fig.~\ref{simple} we show schematically all possible relationships between $G_{\ca\cb}$ and $G_{\ca\cc}$ when 
$\sigma_{\ca\cb}< \sigma_{\ca\cc}$.  
In Fig.~\ref{simple}(a,b,c) no FPT occurs because $\Delta \mu_{\cc\cb}<0$.
In Fig.~\ref{simple}(e) a FPT occurs in the fluctuations of the stable $\ca$ phase.
In Fig.~\ref{simple}(d,f), a FPT occurs in the fluctuations of the metastable $\ca$ phase.  In these two latter cases, a nucleation process from $\ca$ to $\cc$ occurs in concert with a FPT from $\cb$ to $\cc$. 
The cases in Fig.~\ref{simple}(d,f) may thus be expected to correspond to TSN.  From Eq.~\ref{nc} we also predict that $n_c$ diverges on approach to the $\cb\cc$ coexistence line (or its metastable extension within the stability field of $\ca$) since $\Delta \mu_{\cc\cb}=0$ on this line.

\section{lattice model simulations}
\label{latt}

Next we present results obtained from a lattice model to test and elaborate on the predictions of the previous section.  As shown below, this model provides a simple example of a system having a triple point at which three distinct bulk phases coexist.  Furthermore, within any one phase, fluctuations corresponding to the other two phases are easily identified.

We conduct Monte Carlo (MC) simulations of a 2D Ising model, with nearest-neighbor (nn) and next-nearest-neighbor (nnn) interactions, on a square lattice of $N=L^2$ sites with periodic boundary conditions.  Each site $i$ is assigned an Ising spin $s_i=\pm 1$. 
The energy $E$ of a microstate is given by,
\begin{equation}
\frac{E}{J} = \sum_{\langle \rm nn \rangle} s_i s_j
- \frac{1}{2} \sum_{\langle \rm nnn \rangle} s_i s_j 
 -H\sum_{i=1}^N s_i
-H_s\sum_{i=1}^N \sigma_i s_i,
\label{ham}
\end{equation}
where $J$ is the magnitude of the nn interaction energy.  
Interactions between nn sites are antiferromagnetic, while nnn interactions are ferromagnetic.
The first (second) sum in Eq.~\ref{ham} is carried out over all distinct nn (nnn) pairs of sites $i$ and $j$.  
The third and fourth terms in Eq.~\ref{ham} specify the influence of the direct magnetic field $H$ and the staggered field $H_s$.  
We define $\sigma_i=(-1)^{x_i+y_i}$, 
where $x_i$ and $y_i$ are integer horizontal and vertical coordinates of site $i$, so that 
the sign of $H_s\sigma_i$ alternates in a checkerboard fashion on the lattice.
We sample configurations using Metropolis single-spin-flip MC dynamics~\cite{Binder:2009vp}.  

This model has been studied previously to model metamagnetic systems exhibiting a tricritical point, for which it provides a prototypical example in 2D~\cite{Landau:1972mt,Landau:1981bi,Rikvold:1983fs,Herrmann:1984fs,Landau:1986wi}.  
At $T=0$, four stable phases are observed, depending on the values of $H$ and $H_s$.
We label these phases so as to maintain the ``$\ca \cb \cc$" notation used in the previous section.
There are two ferromagnetic phases which we label
$\cb$ (all $s_i=1$) and 
$\cal\bar B$ (all $s_i=-1$);
and two antiferromagnetic phases labelled
$\cc$ (all $s_i=\sigma_i$) and 
$\ca$ (all $s_i=-\sigma_i$).
Since the topology of the phase diagram is unchanged when $H\to -H$, we only consider $H>0$ here.  Consequently the $\cal \bar B$ phase will not appear in our analysis.  
All our simulations are carried out at temperature $T$ such that $J/\beta=1$, where $\beta=1/kT$ and $k$ is Boltzmann's constant.  
This is well below the $T$ for the
N\' eel transition ($kT/J=3.802$) and the tricritical point ($kT/J=1.205$)~\cite{Rikvold:1983fs}.  Thus the phase diagram in the plane of $H$ and $H_s$ at fixed $kT/J=1$ contains only first-order phase transitions, 
arranged as three coexistence lines meeting at a triple point located at $H_s=0$ and $H=3.9876$, as shown in Fig.~\ref{pd1}.  {\color{black} 
See Supplementary Materials (SM) Sections~S1-S3 
and Refs.~\cite{Borgs:1990ig,Binder:2008p5313,Tuckerman:2010,Binder:2009vp}
for the details of our phase diagram calculation}.

\begin{figure}
\includegraphics[scale=0.45]{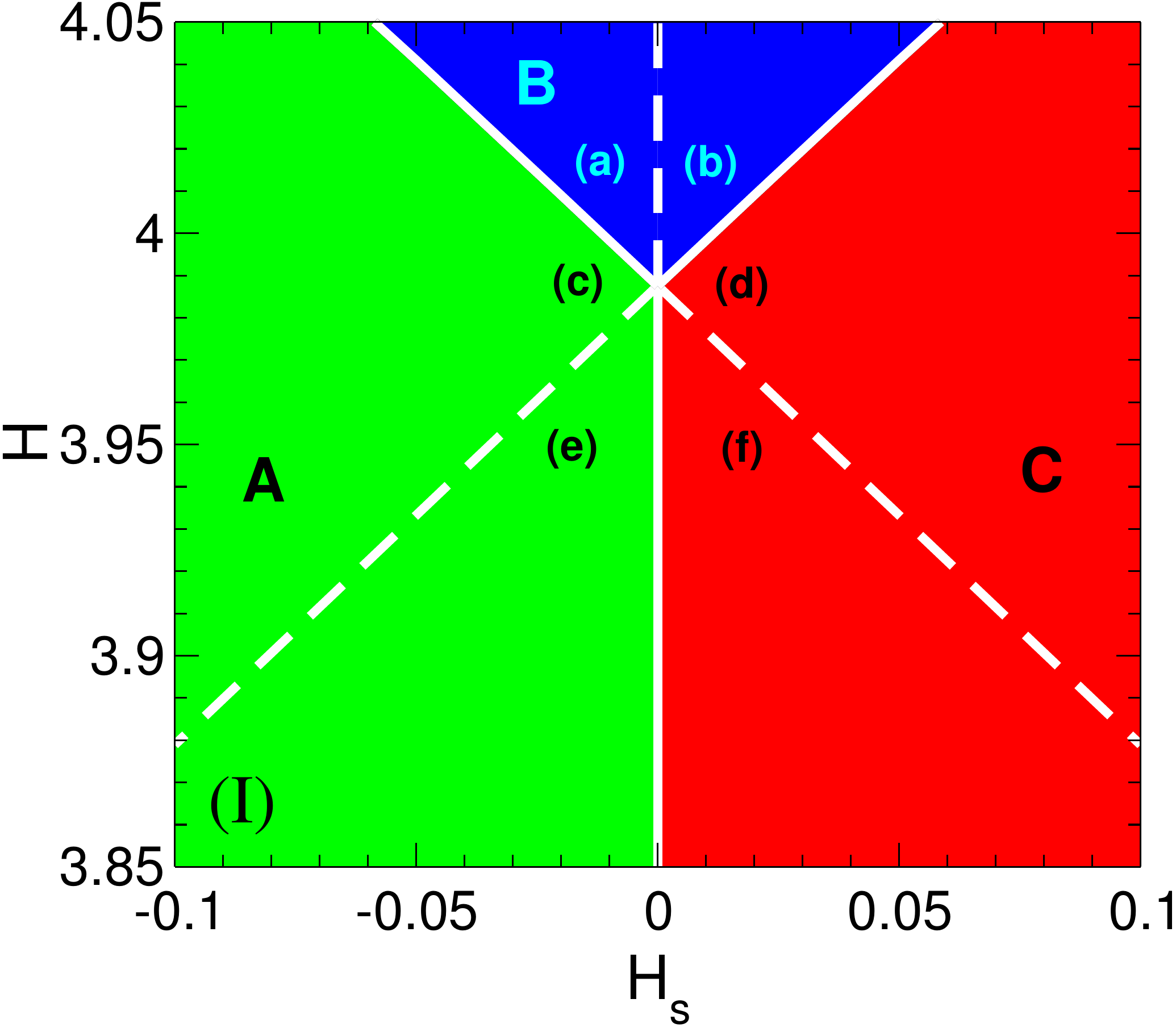}\\
\vspace{0.8cm}
\includegraphics[scale=0.45]{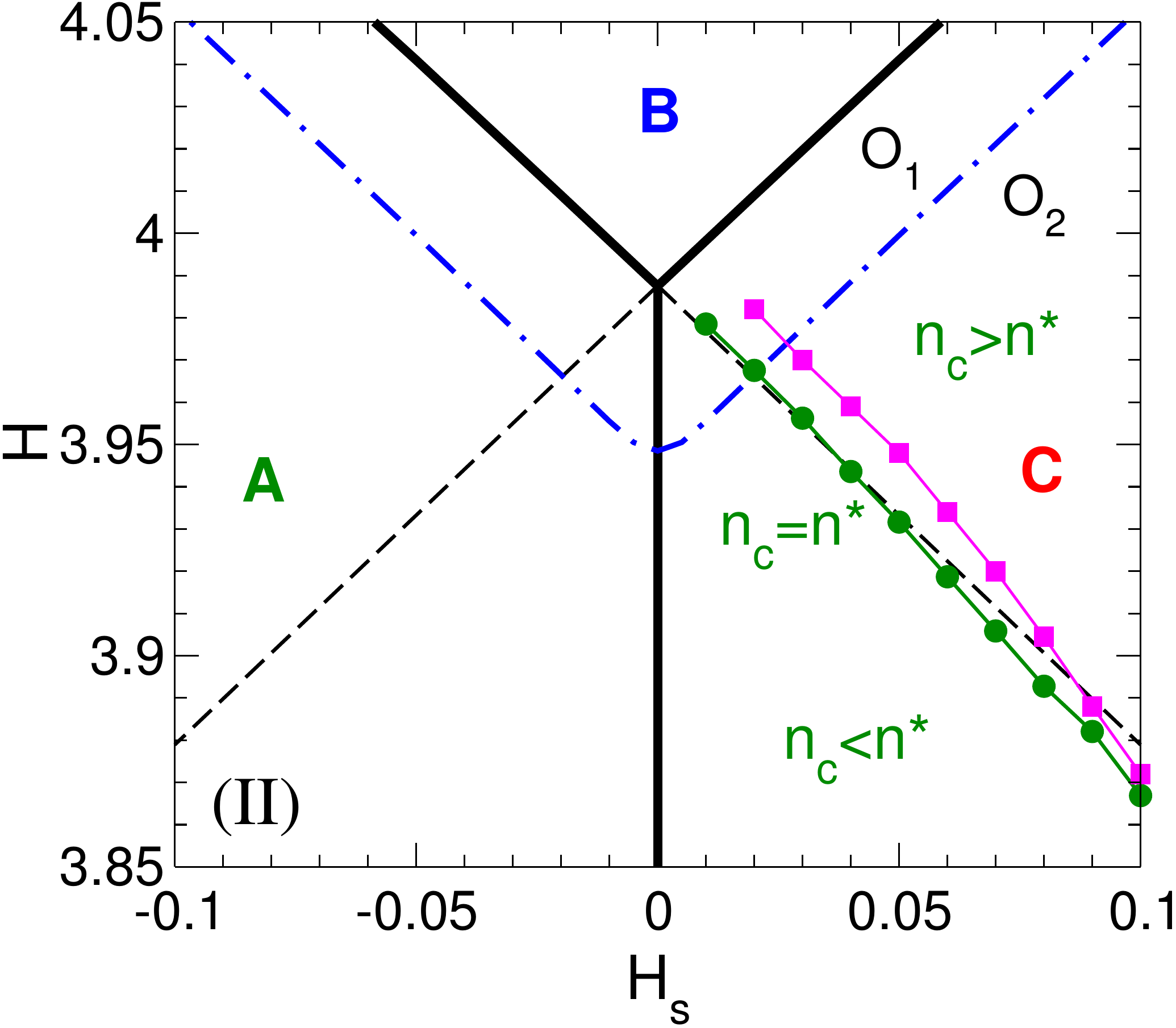}
\caption{Phase diagram for $kT/J=1$.  Solid lines are coexistence lines and dashed lines are metastable extensions of coexistence lines.  
In panel (I), the stability field of each phase is indicated with a solid color, green for $\ca$, blue for $\cb$, and red for $\cc$.  The six regions demarcated by the solid and dashed lines are labelled (a)-(f) and correspond to the six identically labelled cases shown in Fig.~\ref{simple}.
In panel (II), the dot-dashed line is the limit of metastability (LOM) of the bulk $\cb$ phase for a system of size $L=64$.  Green circles locate points on the line ${\cal L}_n$ at which $n_c=n^*$.  
Magenta squares locate points on the line at which $\beta G^*=20$; below this line $\beta G^*>20$ and above it $\beta G^*<20$.  O$_1$ labels the region bounded by the $\ca\cb$ and $\cb\cc$
coexistence lines 
and the LOM of the $\cb$ phase.  O$_2$ labels the region between the $\ca\cb$ and $\cb\cc$
coexistence lines and which lies beyond the LOM of the $\cb$ phase.}
\label{pd1}
\end{figure}

The phase diagram presented in Fig.~\ref{pd1} 
also includes the metastable extensions of the $\ca\cc$, $\cb\cc$ and $\ca\cb$ coexistence lines.  
The phase diagram is thereby divided into six distinct regions each corresponding to a unique ordering of the chemical potentials $\mu_\ca$, $\mu_\cb$ and $\mu_\cc$.  
Furthermore, as shown in 
{\color{black} 
SM Section~S4}, 
we find that 
$\sigma_{\ca\cb}<\sigma_{\ca\cc}$ 
throughout the region of the $(H_s,H)$ plane explored here~\cite{Binder:2008p5313,Binder:2011du,Binder:2012ey}.
Our lattice model thus realizes the relationships between the surface tensions and chemical potentials considered in the previous section:  The six regions of the phase diagram in Fig.~\ref{pd1} correspond to the six panels of Fig.~\ref{simple}.
If we choose the $\ca$ phase of the lattice model to correspond to the $\ca$ phase of Section~\ref{simp},
then the predictions of Section~\ref{simp} can be tested by examining the behavior of the fluctuations of $\ca$ in the lattice model in the various regions of the model phase diagram.

\begin{figure*}
\newcommand\x{0.22}
\includegraphics[scale=\x]{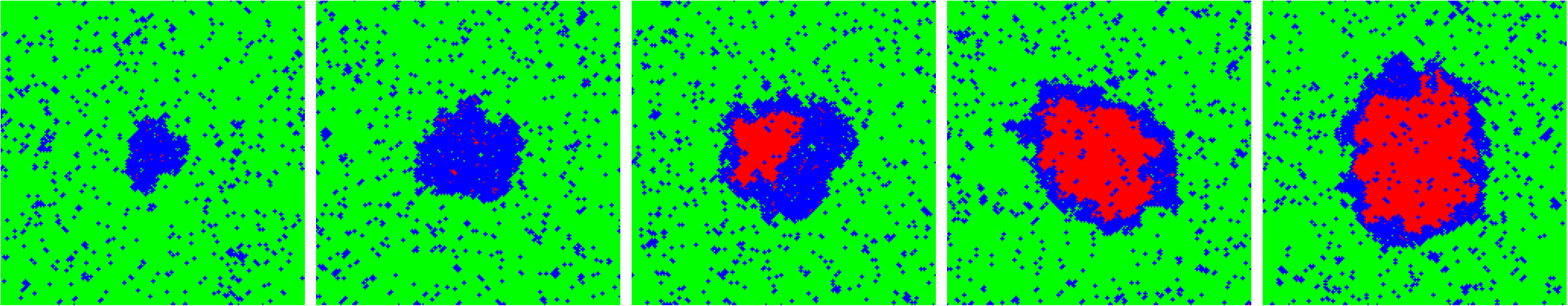}
\caption{System configurations containing fluctuations of different sizes and compositions obtained from 2D umbrella sampling simulations carried out at 
$H_s=0.01$, 
$H=3.981$, 
and $L=200$.  As described in 
{\color{black}
SM Section~S5,} 
green sites have a local structure corresponding to the $\ca$ phase, blue to the $\cb$ phase, and red to the $\cc$ phase.  From left to right $(\nmax,f)=(1393,0.020)$,
$(3476,0.040)$,
$(5474,0.303)$,
$(7492,0.583)$ and
$(9658,0.633)$.  This sequence of microstates approximately follows the path of $\langle f \rangle$ shown in Fig.~\ref{FES1}(b) and illustrates the two-step nature of the nucleation process under these conditions.}
\label{pic1}
\end{figure*}

To explore the scenarios predicted in Section~\ref{simp}, we must characterize the 
fluctuations that appear in the bulk $\ca$ phase.
Due to the simplicity of our lattice model, we show in 
{\color{black} 
SM Section~S5} that 
it is straightforward 
to identify all local fluctuations as clusters of size $n$ that deviate from the structure of $\ca$.
All sites within a given cluster can further be classified according to their correspondence to either $\cb$ or $\cc$.  We thereby define the phase composition of each cluster as $f={\bar n} /n$, where $\bar n$ is the number of sites in the cluster that are classified as $\cc$.  
Fig.~\ref{pic1} shows example clusters of various $n$ and $f$, from mostly $\cu$-like ($f\to 0$) to mostly $\cc$-like ($f\to 1$).

To quantify the thermodynamic behavior of the fluctuations that occur in $\ca$, we measure $G(n_{\rm max},f)$, 
the FES of the bulk $\ccb$ phase in which the largest non-$\ca$ cluster in the system is of size $n_{\rm max}$ and has composition $f$~\cite{Duff:2009p6360}.
We obtain the FES from umbrella sampling MC simulations at fixed $(N,H_s,H,T)$~\cite{Tuckerman:2010}.
We compute $G(n_{\rm max},f)$ from,
\begin{equation}
\beta G(\nmax,f)=-\log [P(\nmax,f)] + C,
\label{gnf}
\end{equation}
where $P(\nmax,f)$ 
is proportional to the probability to observe a microstate with values $n_{\rm max}$ and $f$. 
The value of the arbitrary constant $C$
is chosen so that the global minimum of $G(\nmax,f)$ is zero.
We estimate
$P(\nmax,f)$ from 2D umbrella sampling simulations using a biasing potential that depends on both $\nmax$ and $f$,
\begin{equation}
U_B=\kappa_n(\nmax-\nmax^*)^2 + \kappa_f(f-f^*)^2,
\label{us}
\end{equation}
where $\nmax^*$ and $f^*$ are target values of $\nmax$ and $f$ to be sampled in a given umbrella sampling simulation, and 
$\kappa_n$ and $\kappa_f$ control the range of sampling around $\nmax^*$ and $f^*$.  
{\color{black} 
See SM Section~S6 
for details of our 2D umbrella sampling simulations.}
Results from multiple umbrella sampling runs conducted at fixed $(N,H_s,H,T)$
are combined using the weighted histogram analysis method (WHAM) to estimate the full $G(n_{\rm max},f)$ FES at a given state point~\cite{Kumar:1992,Tuckerman:2010,Grossfield:2018}.
Once $G(n_{\rm max},f)$ has been calculated, we can also compute
the one dimensional (1D) free energy as a function of $\nmax$ alone, defined as,
\begin{equation}
\beta G_1=-\log \int_0^1 \exp[-\beta G(\nmax,f)]\,df.
\end{equation}

\section{Fluctuation phase transition in a stable phase}

In this section we analyze the behavior observed in the fluctuations of $\ca$ when $\ca$ is stable and no nucleation process is possible.  As an example, we focus on the state point $(H_s,H)=(0,3.9)$.  
This point is on the $\ca\cc$ coexistence line, and so the nucleation barrier to convert $\ca$ to $\cc$ is infinitely large.  
In terms of the analysis of Section~\ref{simp}, this point is on the boundary of the regions described by panels (e) and (f) of Fig.~\ref{simple}.  We thus expect that the fluctuations of $\ca$ are $\cb$-like at small $\nmax$ and then undergo a FPT to $\cc$ at $n_c$.

\begin{figure}
\includegraphics[scale=0.4]{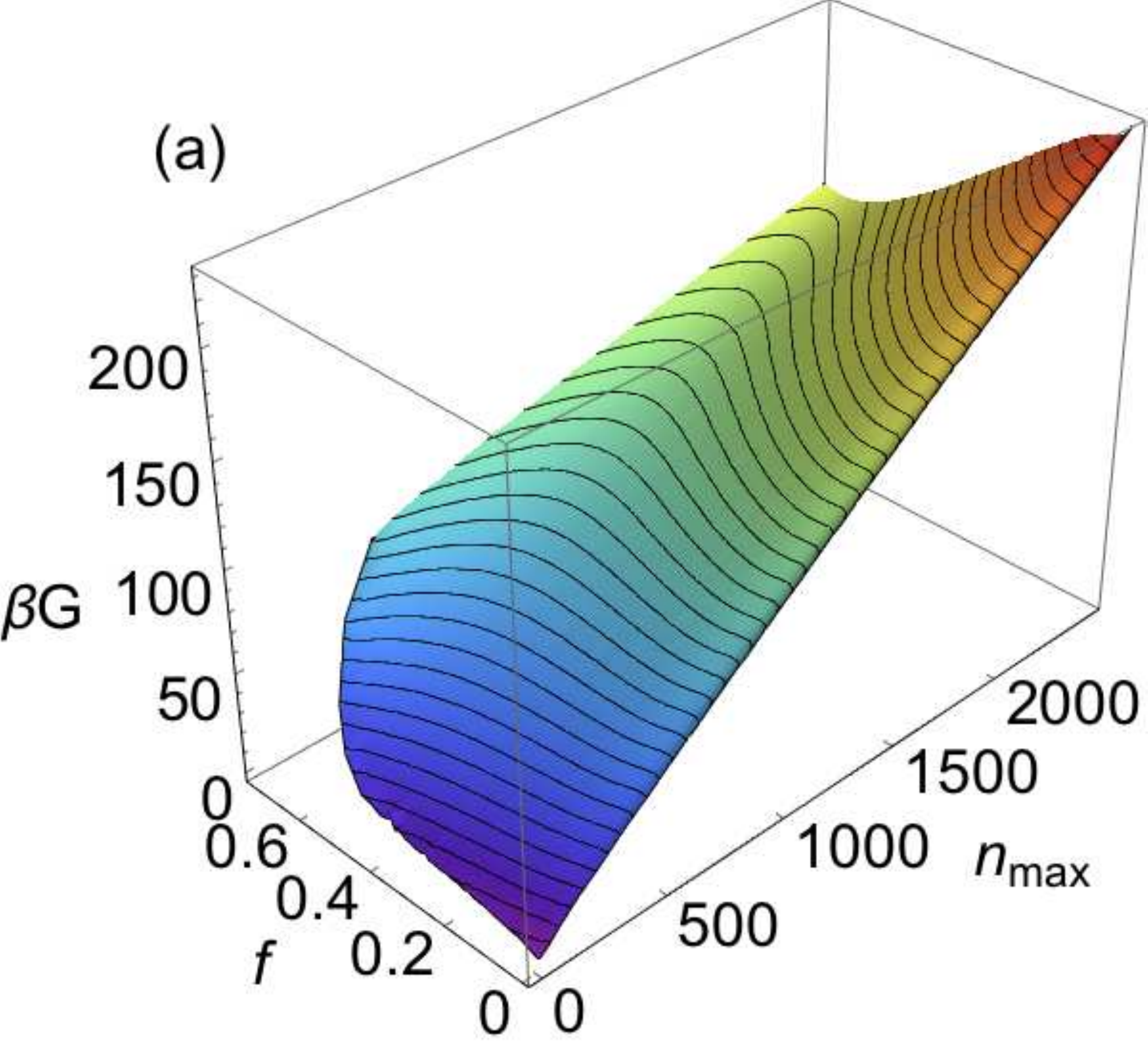}\\
\vspace{0.8cm}
\includegraphics[scale=0.35]{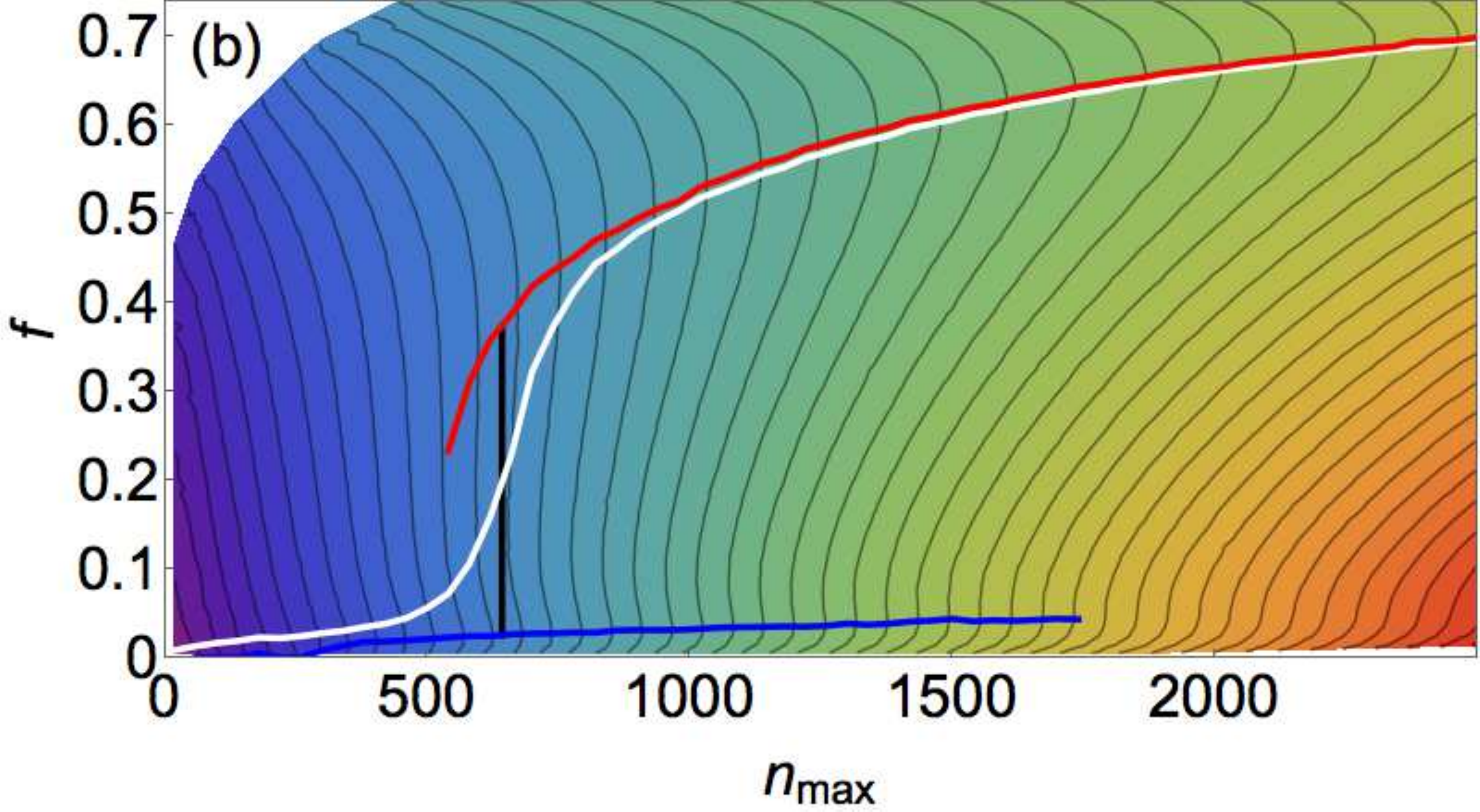}
\caption{(a) Surface plot and (b) contour plot of $G(\nmax,f)$ 
for $(H_s,H)=(0,3.9)$ and $L=128$.  Contours are $5kT$ apart in both (a) and (b).  In (b) we also plot $f_\cb$ (blue line), 
$f_\cc$ (red line) and
$\langle f \rangle$ (white line).  The black vertical line is located at $\nmax=n_c$.}
\label{noFPT}
\end{figure}

We present the FES describing the fluctuations of $\ca$ at this state point in Fig.~\ref{noFPT}.
Under these conditions, all local fluctuations that induce a deviation from the most probable state of the $\ca$ phase increase the system free energy, regardless of their size or composition.  The FES therefore exhibits only one basin with a minimum in the lower-left corner of Fig.~\ref{noFPT}(b) associated with the bulk $\ca$ phase, and no transition states (i.e. saddle points) occur in the surface.  However, the FES is not featureless.  It contains two channels, indicated by the red and blue lines in Fig.~\ref{noFPT}(b).  These lines locate 
the values of $f$ at which a local minimum occurs in 
$G(n_{\rm max},f)$ 
at a fixed value of $\nmax$.
Along the low-$f$ channel $\cb$ fluctuations dominate, while the high-$f$ channel corresponds to fluctuations in which the core is $\cc$, wetted by a surface layer of $\cb$.  
We define 
$f_\cb$ as the values of $f$ along the minimum of the low-$f$ channel, and $f_\cc$ for the high-$f$ channel.

\begin{figure}
\includegraphics[scale=0.4]{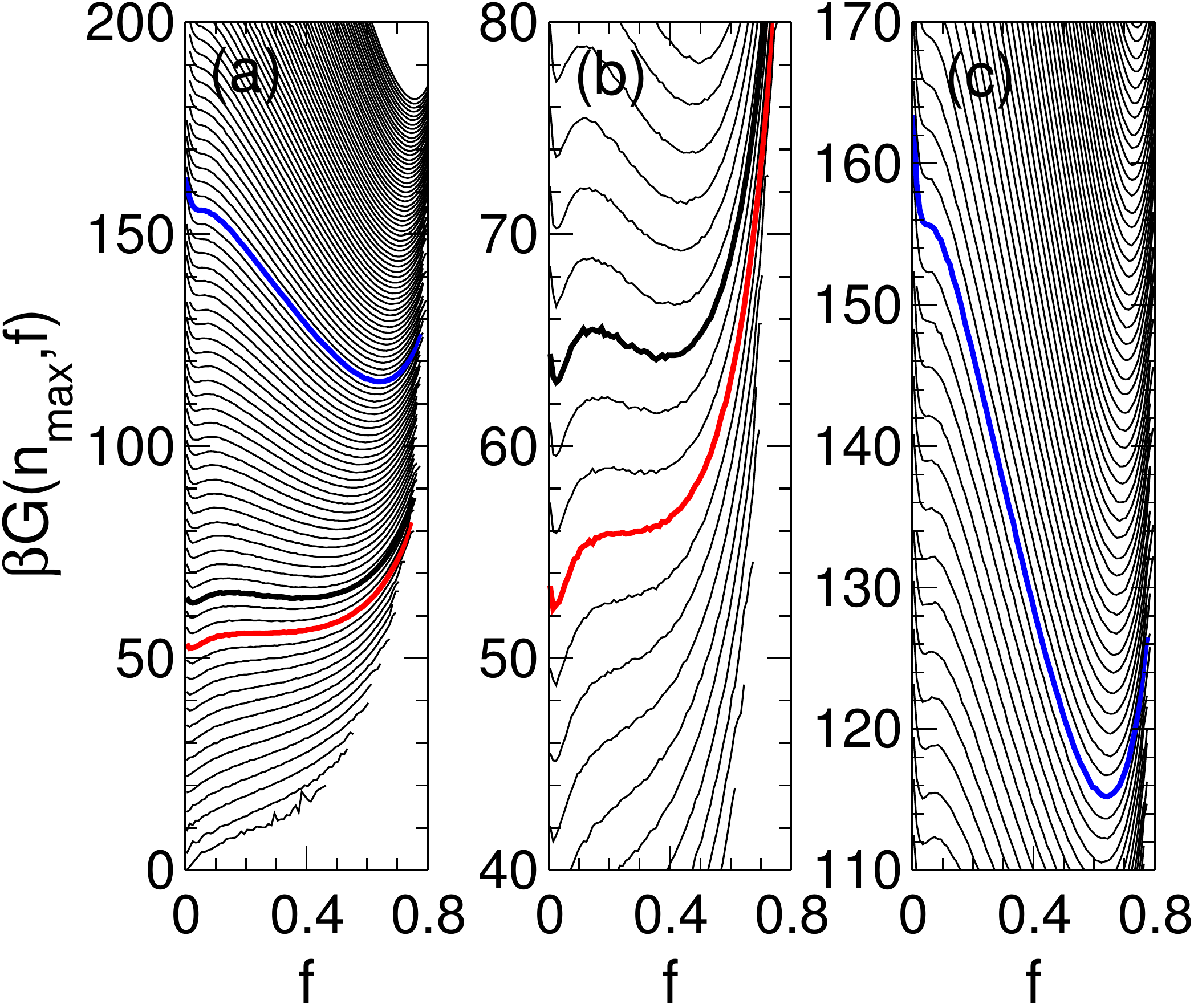}
\caption{Cuts through the $G(\nmax,f)$ surface at fixed $\nmax$ for
$(H_s,H)=(0,3.9)$ and $L=128$.  
$\nmax$ changes by 40 from one cut to to next.  Several cuts are highlighted:  
The red curve is for $\nmax=539$ and corresponds to the spinodal endpoint of the $\cc$ channel.
The blue curve is for $\nmax=1739$ and corresponds to the spinodal endpoint of the $\cb$ channel.
The thick black curve is for $\nmax=659$ and corresponds to the point of coexistence between the $\cb$ and $\cc$ channels at $\nmax=n_c$.}
\label{noFPT1}
\end{figure}

It is notable that the two channels are unconnected, and that neither channel is defined for all $\nmax$.  
At small $\nmax$ only the $\cb$ channel exists, while only the $\cc$ channel exists at large $\nmax$.  This behavior is highlighted in Fig.~\ref{noFPT1}, where we show cuts through the FES at fixed $\nmax$.  For a finite range of $\nmax$, $G$ versus $f$ exhibits two minima with a maximum in between.  At small $\nmax$ the high-$f$ minimum disappears, and at large $\nmax$ the low-$f$ minimum disappears.

To quantify the relative free energies associated with these two channels we define,
for a fixed value of $\nmax$,
\begin{eqnarray}
\beta G_{\cb}&=&-\log \int_0^{f_{\rm max}} \exp[-\beta G(\nmax,f)]\,df \cr
\beta G_{\cc}&=&-\log \int_{f_{\rm max}}^1 \exp[-\beta G(\nmax,f)]\,df,
\end{eqnarray}
where $f_{\rm max}$ is the value of $f$ at which a local maximum occurs in $G(n_{\rm max},f)$ as a function of $f$ at fixed $\nmax$, 
if the maximum exists.
If $f_{\cb}$ is defined but $f_{\cc}$ is not, then $f_{\rm max}$ is set to 1.
If $f_{\cc}$ is defined but $f_{\cb}$ is not, then $f_{\rm max}$ is set to 0.
So defined, $G_\cb$ and $G_\cc$ decompose $G_1$ into contributions associated with the respective $\cb$ and $\cc$ channels.

\begin{figure}
\includegraphics[scale=0.4]{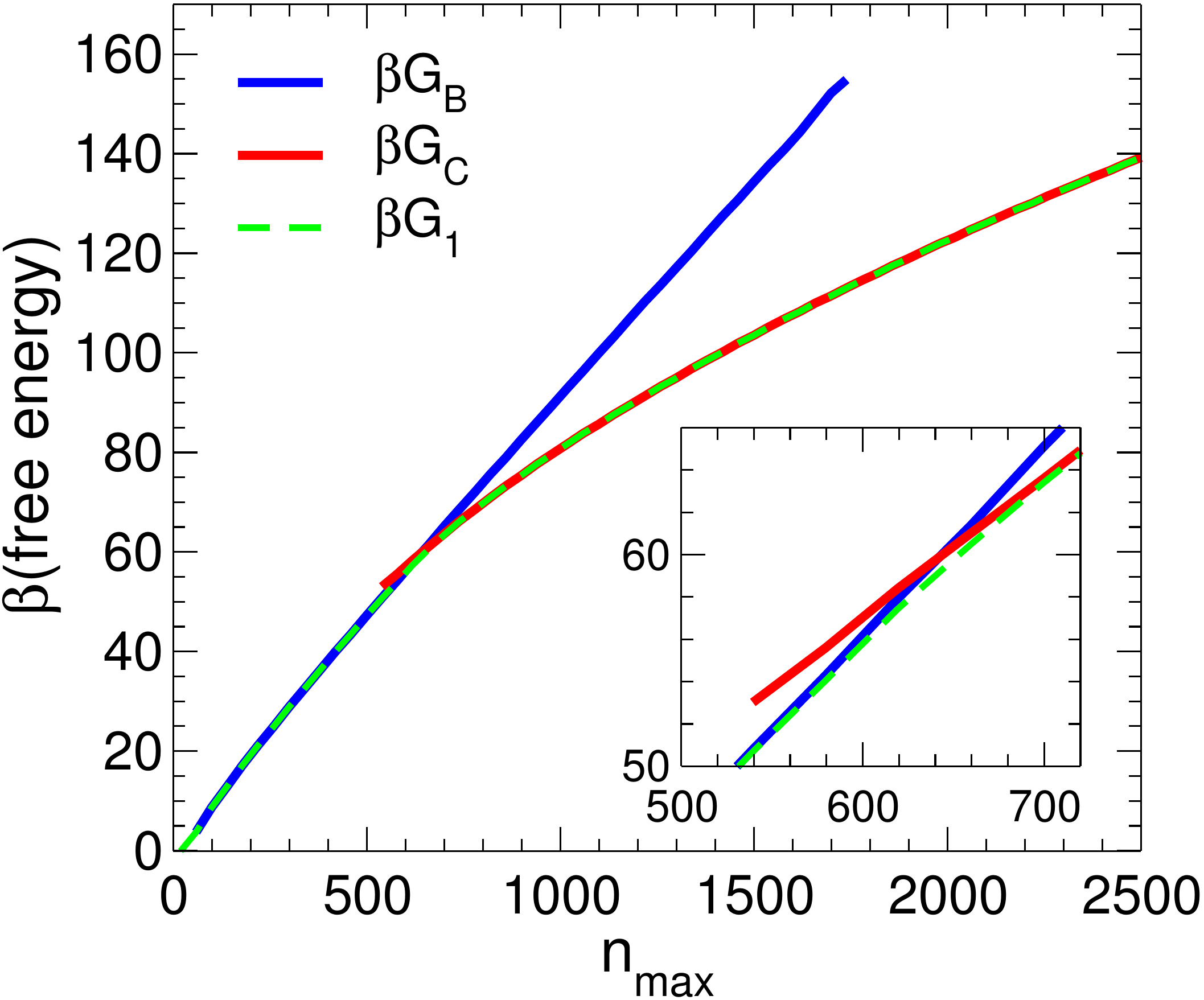}
\caption{$G_\cb$, $G_\cc$ and $G_1$ for
$(H_s,H)=(0,3.9)$ and $L=128$.  The insert shows a closeup of the same data as in the main panel highlighting the intersection of
$G_\cb$ and $G_\cc$ at $\nmax=n_c$.}
\label{noFPT2}
\end{figure}

We plot $G_\cb$, $G_\cc$
and $G_1$ in Fig.~\ref{noFPT2}.  
We see that the $\cb$ channel makes the dominant contribution to the total free energy at small $\nmax$, while the $\cc$ channel dominates at large $\nmax$.  A well-defined FPT is identified by the intersection of $G_\cb$ and $G_\cc$, and corresponds to a ``kink" in the $G_1$ curve.
The value of $\nmax$ at this intersection defines $n_c$, and identifies the coexistence condition where distinct $\cb$-dominated and $\cc$-dominated fluctuations of equal size are equally probable.  We further see that both channels have metastable extensions beyond $n_c$ that end at well-defined limits of metastability, occurring at the values of $\nmax$ where the low-$f$ and high-$f$ minima disappear, as highlighted in  Fig.~\ref{noFPT1}.
In the following, we use the term ``spinodal" to refer to the limit of metastability that terminates a channel in the FES, in analogy to the use of this term when referring to the limit of metastability of a bulk phase in a mean-field system.
The behavior shown in Fig.~\ref{noFPT2}, where both coexistence and metastability are observed,
demonstrates that the FPT is a first-order phase transition occurring in a finite-sized system (the fluctuation) as the system size ($\nmax$) increases.  

An appropriate order parameter for the FPT is $f$.  
For a fixed value of $\nmax$ we define,
\begin{equation}
\langle f \rangle=\frac{\int_0^1 f\exp[-\beta G(\nmax,f)]\,df} {\int_0^1 \exp[-\beta G(\nmax,f)]\,df}.
\end{equation}
We plot $\langle f \rangle$ 
in Fig.~\ref{noFPT}(a).  
The variation of $\langle f \rangle$ is steepest at $\nmax=n_c$.
Even though the FPT is first-order, $\langle f \rangle$ 
does not jump discontinuously at $n_c$ because of the finite size of the system.  However, the most probable value of $f$ does have a jump discontinuity at $n_c$.
We also note that the fluctuations of $f$ at fixed $\nmax$ can be defined as,
\begin{equation}
\chi=\langle f^2 \rangle - \langle f \rangle^2.
\end{equation}
{\color{black}
As shown in SM Section~S7}, 
$n_c$ can be accurately estimated as the value of $\nmax$ at which a maximum occurs in $\chi$.  This procedure allows $n_c$ to be evaluated without having to separately compute $G_\cb$ and $G_\cc$.

The spinodal endpoints that terminate the $G_\cb$ and $G_\cc$ curves
are a significant difference between the behavior plotted in Fig.~\ref{noFPT2} and that predicted in Fig.~\ref{simple}(e).  In our lattice model, a thermodynamic distinction between the $\cb$ and $\cc$ fluctuations only exists for $\nmax$ between these spinodals, where both channels are observed.
These spinodals have important physical consequences.  At small $\nmax$, the most probable fluctuations are always $\cb$, and $\cc$ fluctuations, although they may occur, have no local stability relative to changes in $f$.  This occurs in spite of the fact that $\cc$ has a lower bulk-phase chemical potential than $\cb$ under these conditions.  Thus the prediction made in Section~\ref{simp}, that the most probable small fluctuation corresponds to the phase with the lowest surface tension, becomes even stronger in our lattice model:  Not only is this small fluctuation most probable, it is also the {\it only} fluctuation that is stable with respect to changes in composition.
This observation is in line with a similar conclusion obtained by Harrowell, who predicted that sub-critical clusters in a supercooled liquid are not stable as crystal-like clusters below a threshold size~\cite{Harrowell:2010jt}.  The same is true here for our $\cc$ fluctuations. 

Conversely, at sufficiently large $\nmax$, only $\cc$ fluctuations are stable with respect to changes in composition.
That is, even though a fluctuation is most likely to start out as a $\cb$ fluctuation, and even if it persists as a $\cb$ fluctuation in the metastable portion of the $\cb$ channel when $\nmax>n_c$, it cannot remain a $\cb$ fluctuation at arbitrarily large $\nmax$.  It must eventually convert to $\cc$.

Most of the features of the FES discussed here, including the FPT, occur for $G\gg kT$.  As a consequence, the most commonly observed fluctuations of $\ca$ are entirely dominated by $\cb$, despite the lower bulk chemical potential of $\cc$.  
Nonetheless, observable effects associated with the FPT can be observed in this system during non-equilibrium processes.  For example, if a large nucleus of $\cc$ is inserted into the bulk $\ca$ phase under these conditions, it will spontaneously shrink in size along the $\cc$ channel of the FES.  If the degrees of freedom associated with changes in $f$ relax quickly relative to the rate at which the nucleus size decreases, the shrinking nucleus will then undergo a FPT from $\cc$ to $\cb$ at some size between $n_c$ and the spinodal of the $\cc$ channel.
This case illustrates the differences between the FPT described here and a conventional first-order phase transition occurring between bulk phases.  The FPT occurs in a finite-sized system (the fluctuation), which arises as a departure from the most probable state of the surrounding system (the homogeneous bulk phase), and the parameter that drives the system through the phase transition is the size of the fluctuation.  The size of a fluctuation will normally be subject to strong thermodynamic driving forces that cause it to spontaneously increase or decrease in size.  
The dynamics of the system and its preparation history will therefore have a significant influence on if and how a FPT manifests itself in a particular case.

\section{Two-step nucleation:  fluctuation phase transition in a metastable phase}

We now focus on state points where $\ca$ is metastable and $\cc$ is stable, i.e. regions (d) and (f) in the phase diagram of Fig.~\ref{pd1}.  Based on the predictions of Section~\ref{simp}, we expect in regions (d) and (f) that small $\cb$ fluctuations appear first and then convert to $\cc$ at larger size via a FPT, just like in region (e).  However, in regions (d) and (f) we should also observe a transition state in the FES that is absent in region (e).  For $\nmax$ beyond this transition state, the fluctuation will grow in size spontaneously, leading ultimately to the formation of the bulk $\cc$ phase.  In this case, $\cc$ is formed from $\ca$ via a TSN process, in which the $\cb$ fluctuations that occur initially play the role of the intermediate phase.

\begin{figure}[t]
\includegraphics[scale=0.35]{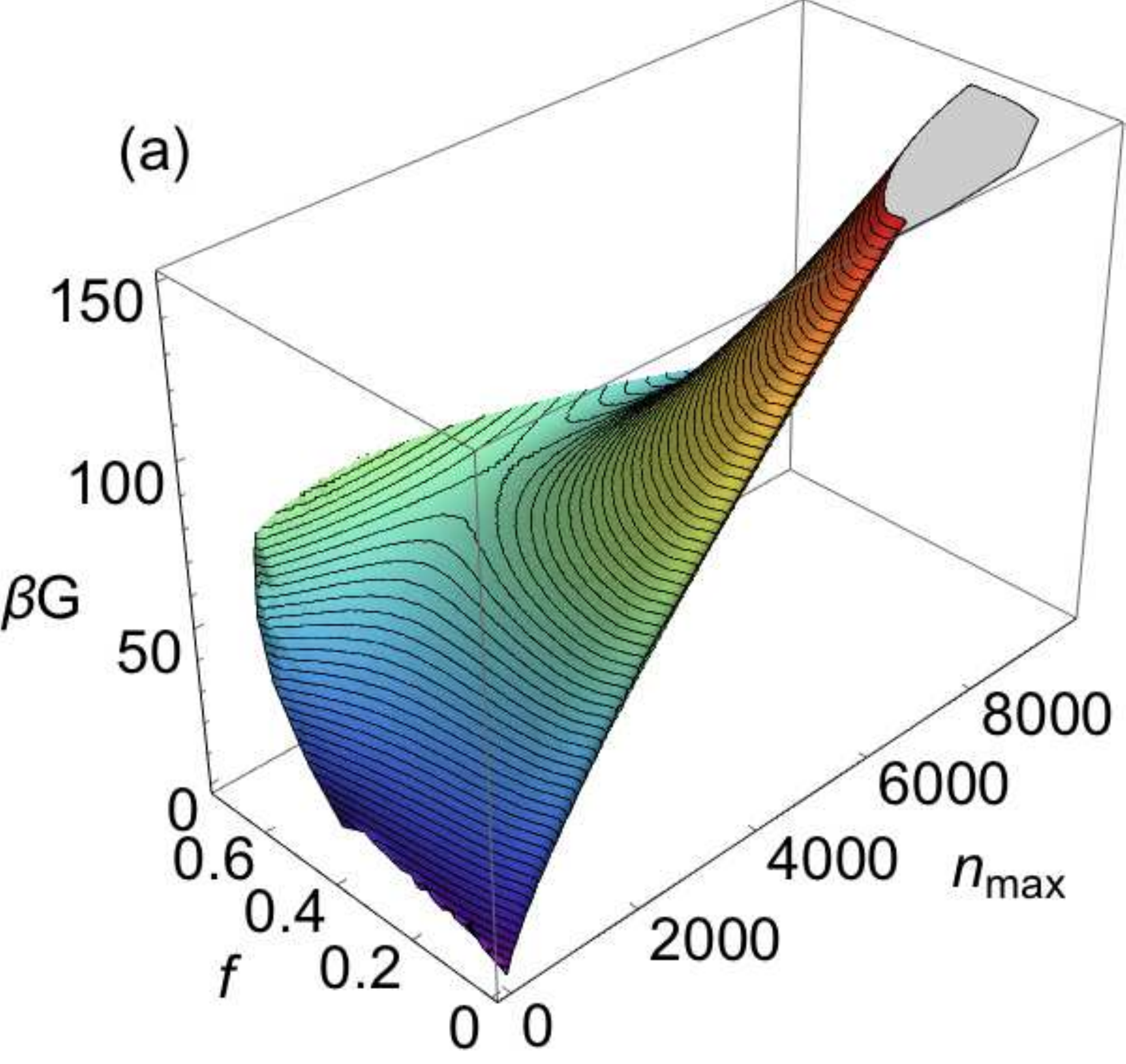}
\hspace{0.5cm}
\includegraphics[scale=0.35]{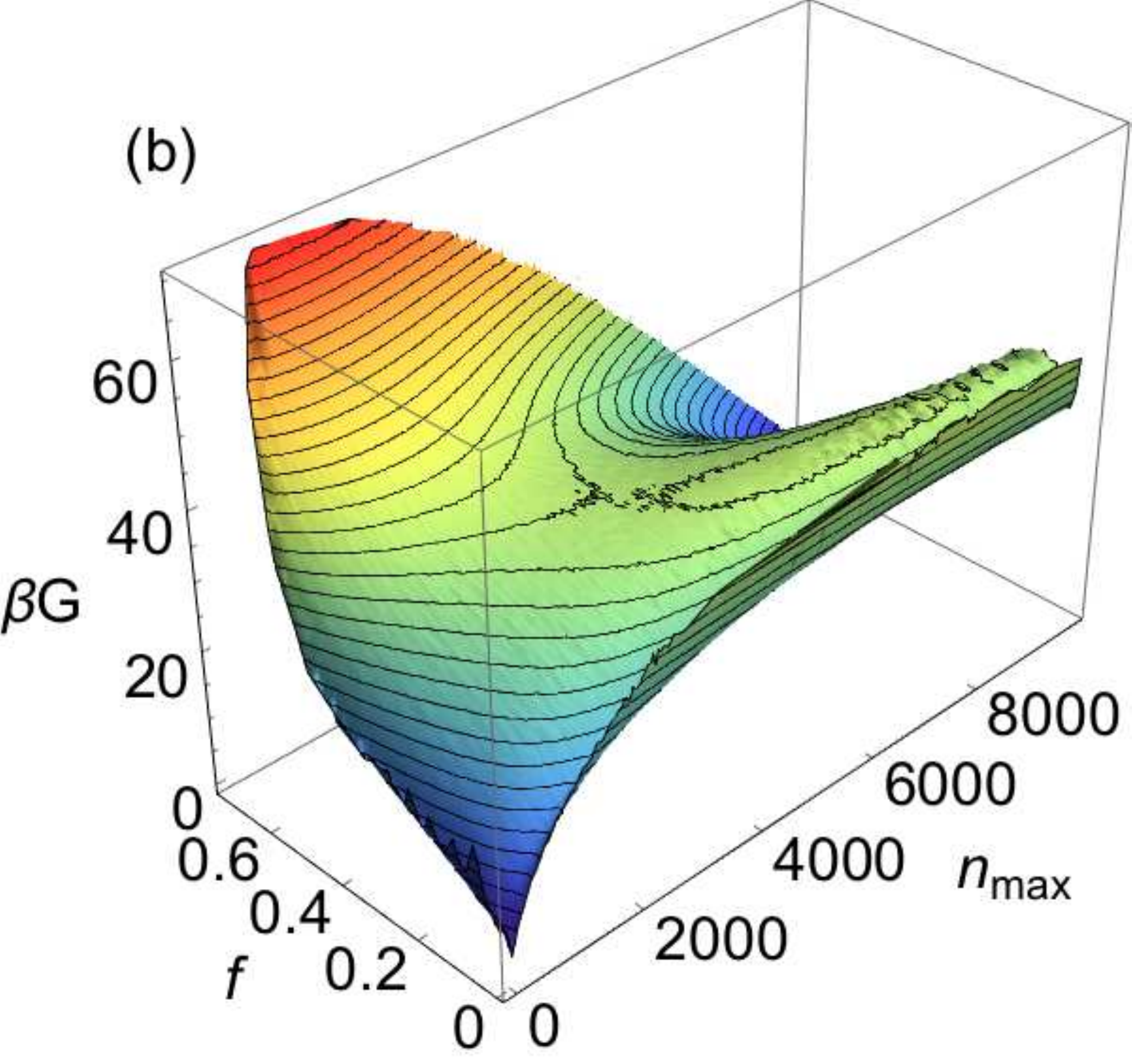}\\
\vspace{0.5cm}
\includegraphics[scale=0.35]{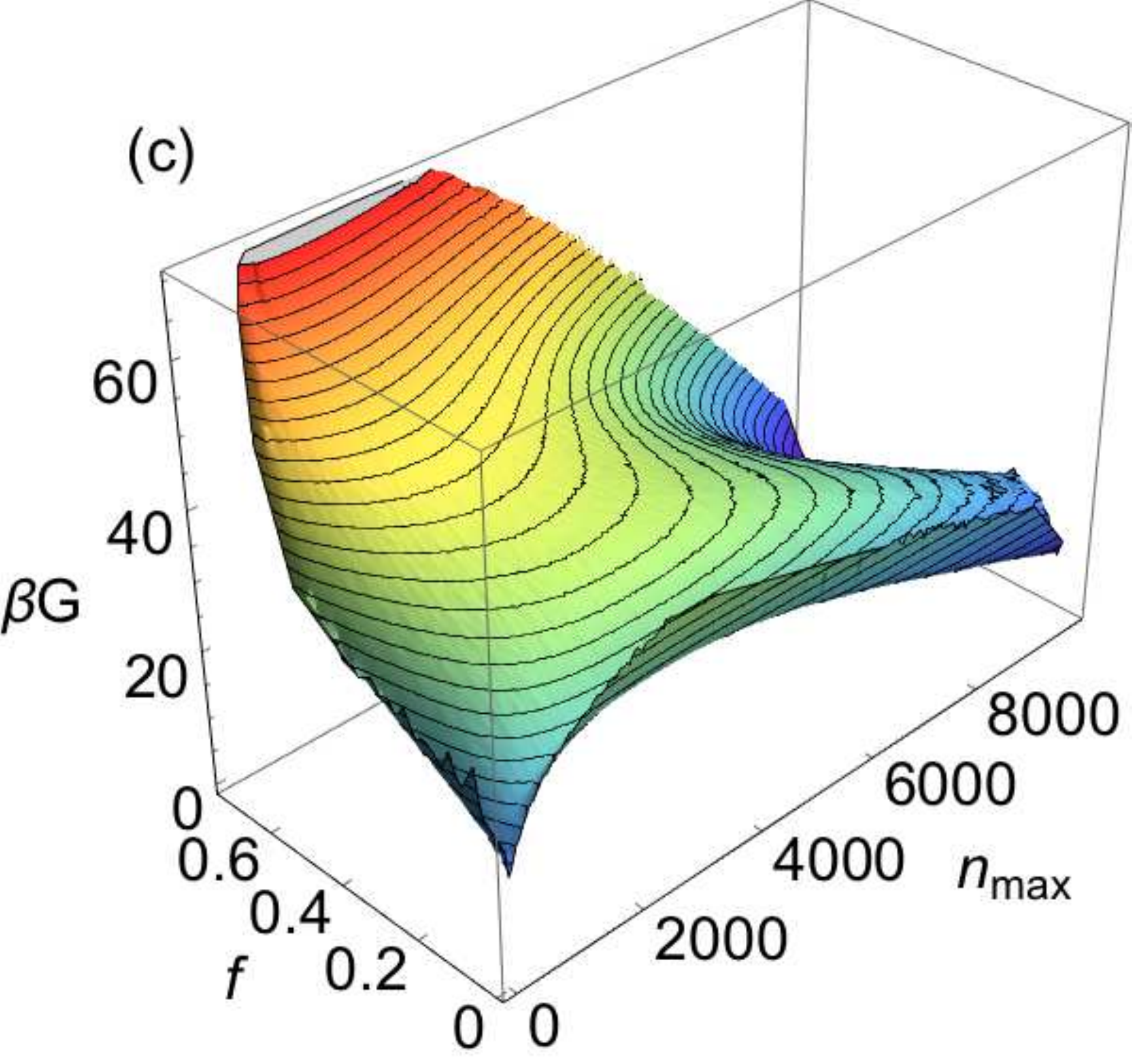}
\caption{Surface plots of $G(n_{\rm max},f)$ for $H_s=0.01$ and $L=200$.  
Panels (a), (b) and (c) correspond respectively to 
$H=\{3.96,3.981,3.985\}$. Contours are $2kT$ apart.}
\label{FES2}
\end{figure}

\begin{figure}[t]
\includegraphics[scale=0.35]{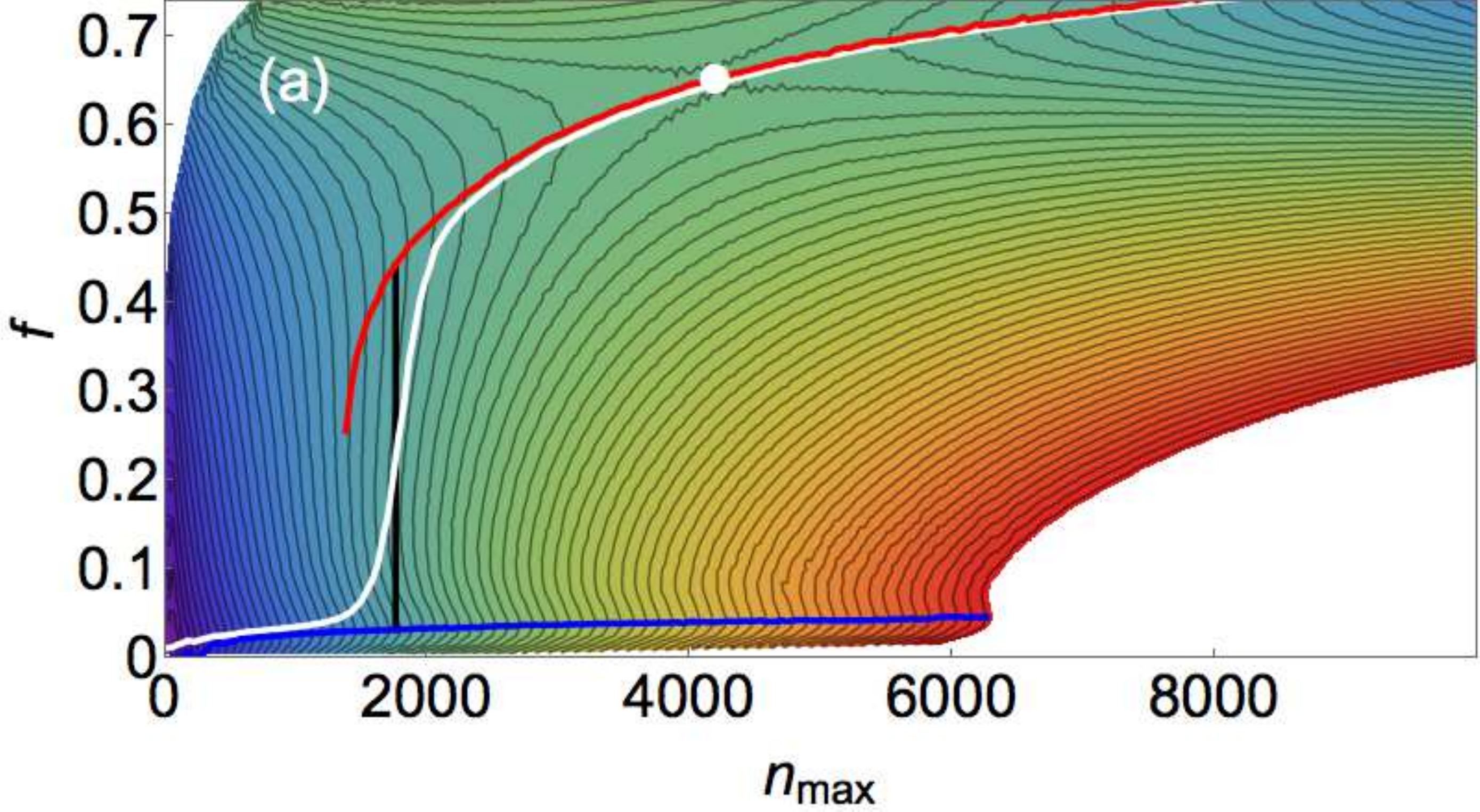}
\includegraphics[scale=0.35]{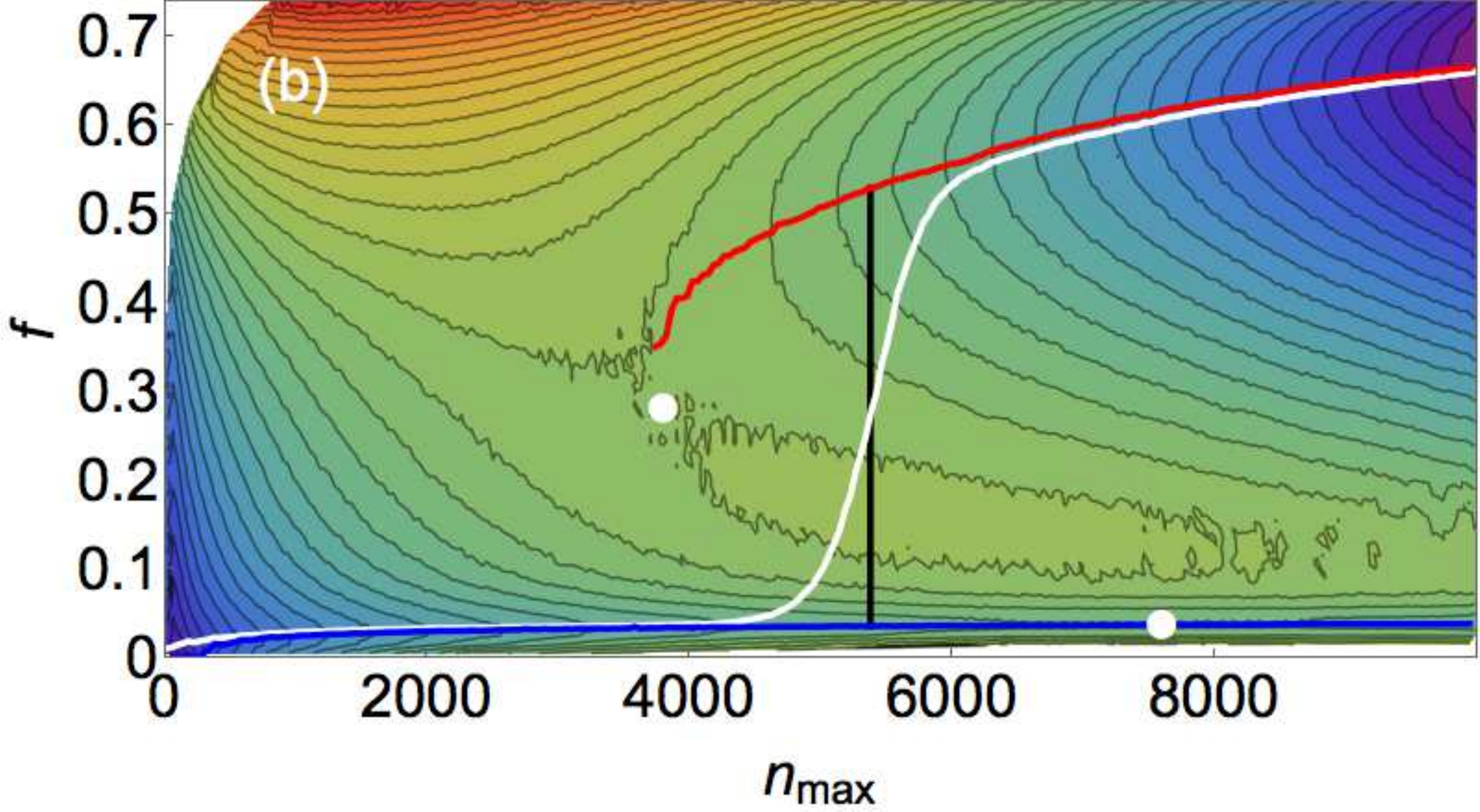}
\includegraphics[scale=0.35]{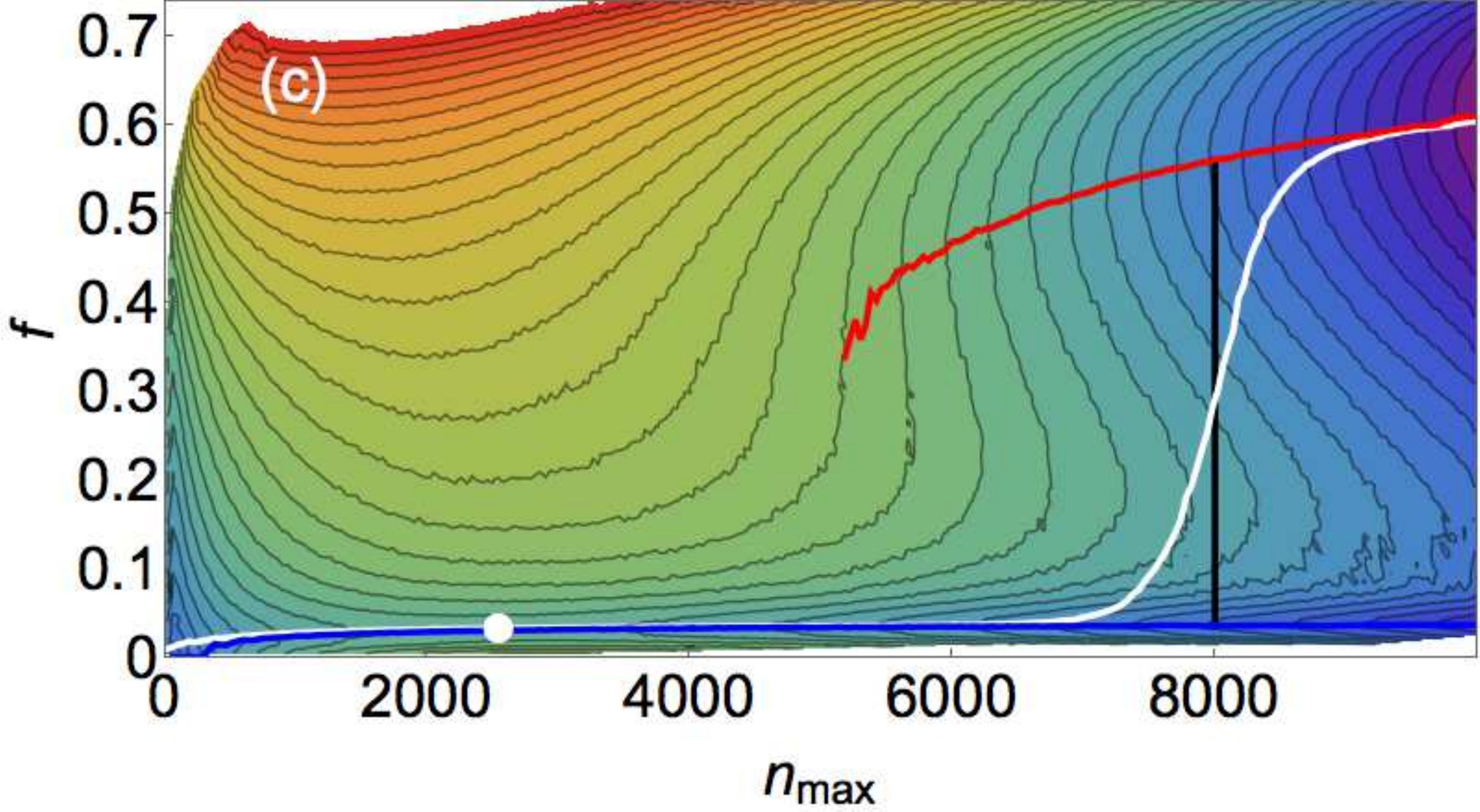}
\caption{Contour plots of $G(n_{\rm max},f)$ for $H_s=0.01$ and $L=200$.  
Panels (a), (b) and (c) correspond respectively to 
$H=\{3.96,3.981,3.985\}$. 
Contours are $2kT$ apart.
In each panel we also plot $f_\cb$ (blue line), 
$f_\cc$ (red line) and
$\langle f \rangle$ (white line).  The black vertical line is located at $\nmax=n_c$.
The white dots locate saddle points in the FES.}
\label{FES1}
\end{figure}

\begin{figure}
\includegraphics[scale=0.4]{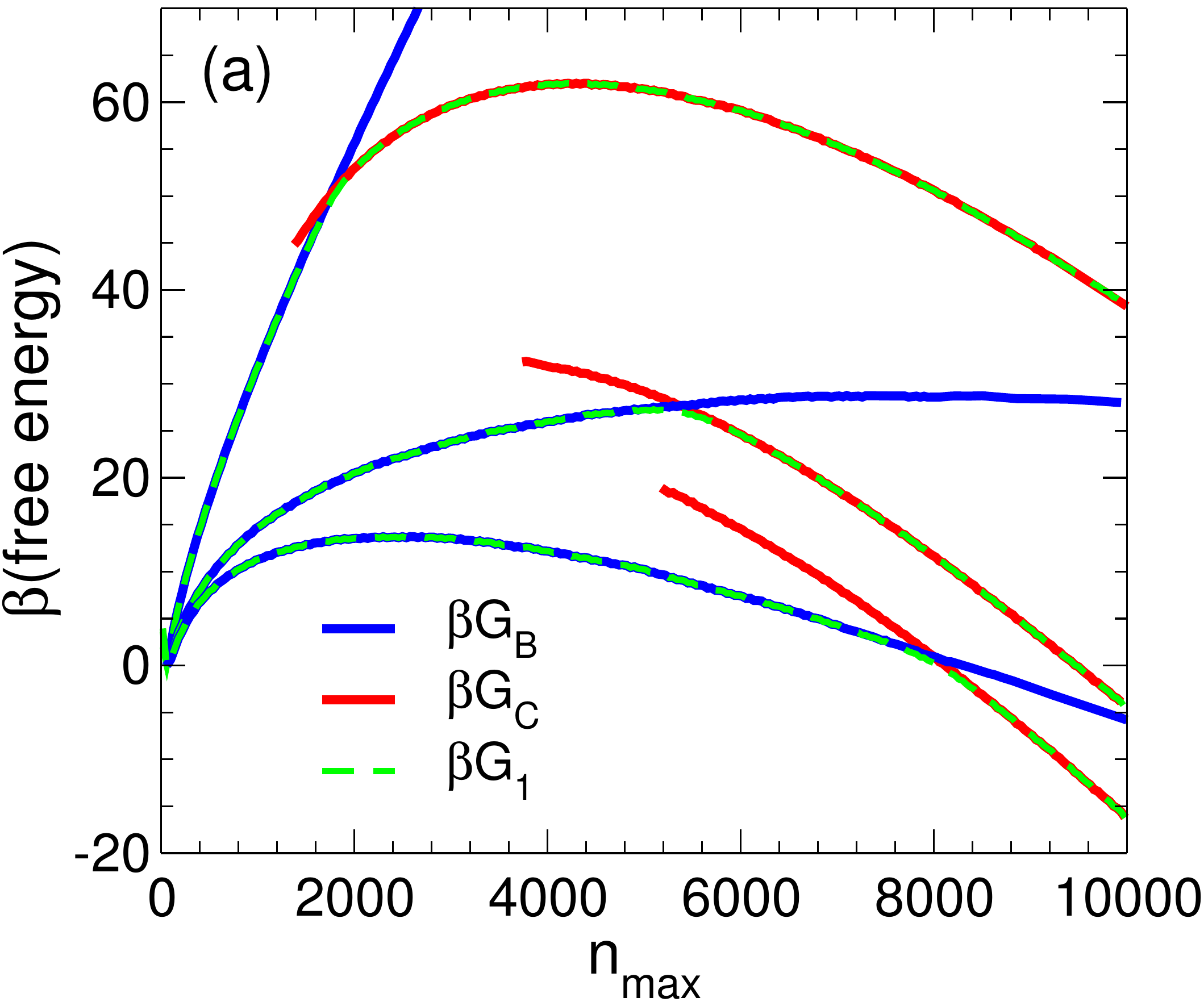}\\
\vspace{1cm}
\includegraphics[scale=0.4]{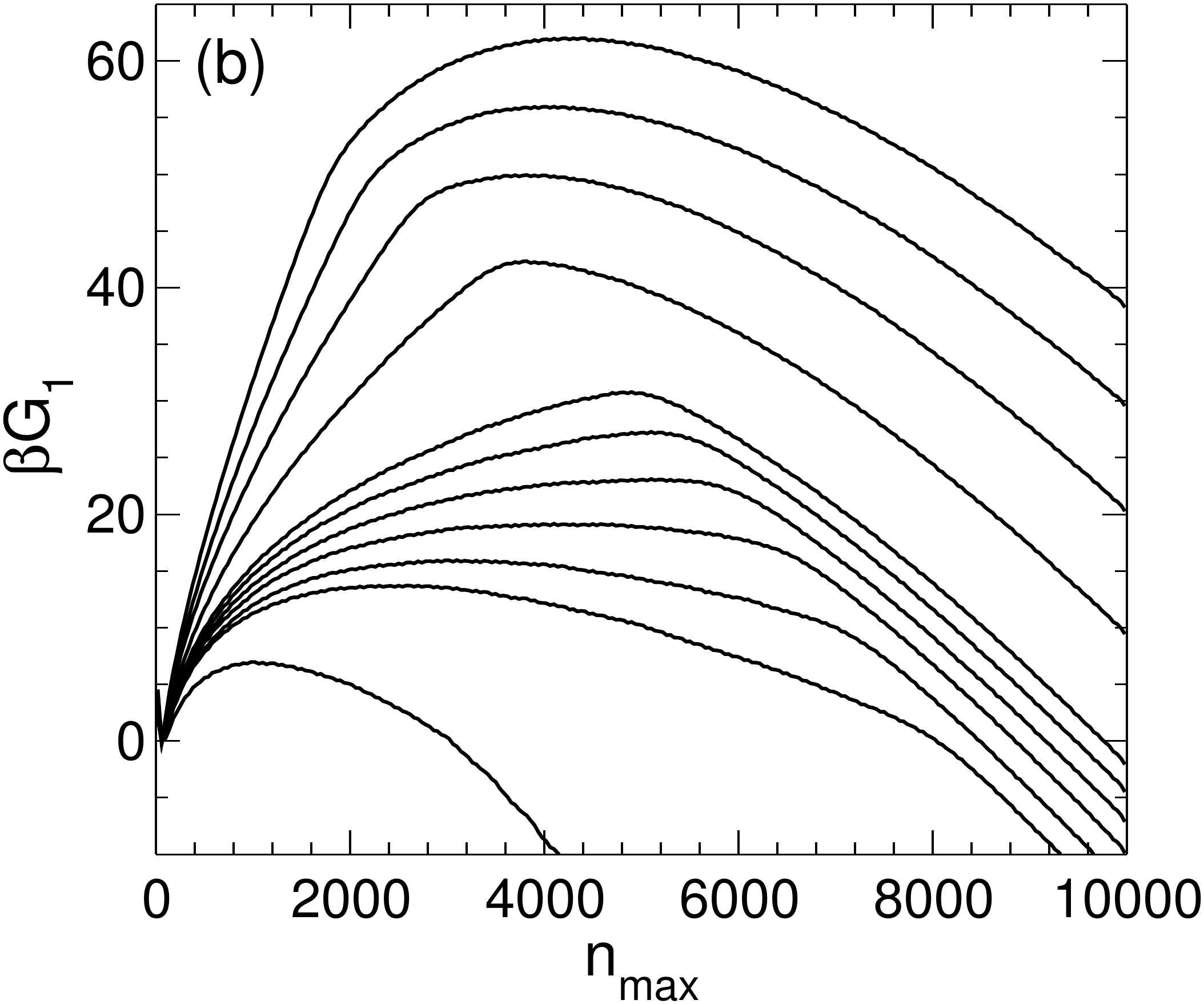}
\caption{(a) $G_\cb$, $G_\cc$ and $G_1$ for $H_s=0.01$ and $L=200$.  
From top to bottom $H=\{3.96,3.981,3.985\}$.
(b) $G_1$ for $H_s=0.01$ and $L=200$.  
From top to bottom $H=\{3.960$, 3.965, 3.970, 3.975, 3.980, 3.981, 3.982, 3.983, 3.984, 3.985, 3.990$\}$.
}
\label{G1D}
\end{figure}

\begin{figure}
\includegraphics[scale=0.4]{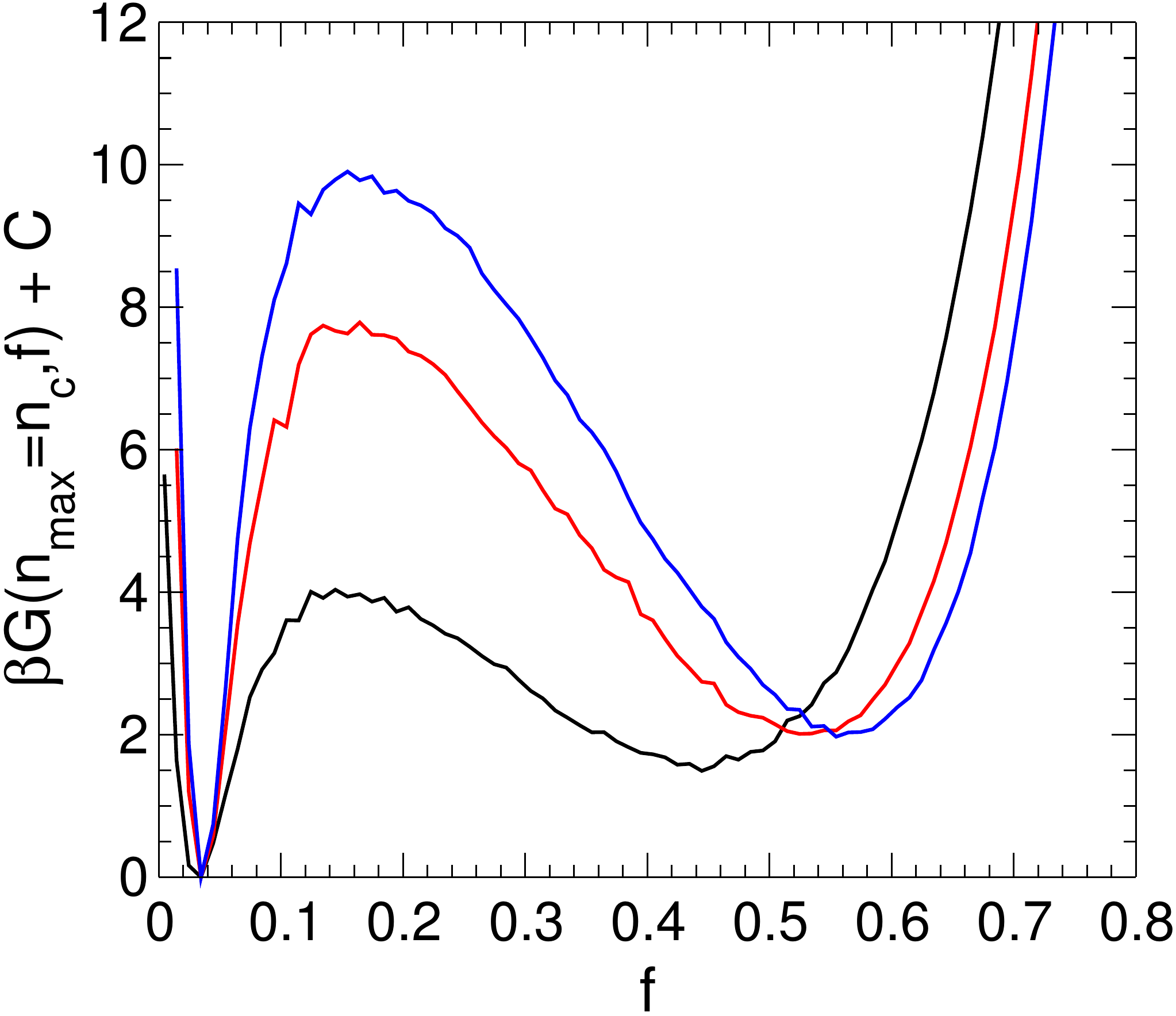}
\caption{$G(n_{\rm max}=n_c,f)$ for $H_s=0.01$ and $L=200$.  
The black curve corresponds to $H=3.96$ and $\nmax=1779$.
The red curve corresponds to $H=3.981$ and $\nmax=5379$.
The blue curve corresponds to $H=3.985$ and $\nmax=8019$.  Each curve has been shifted by a constant $C$ so that the minimum value is zero.}
\label{Gcuts}
\end{figure}

Figs.~\ref{FES2} and \ref{FES1}
show the FES at fixed $H_s=0.01$ for three values of $H$ 
within the stability field of $\cc$.  $G_1$, $G_\cb$ and $G_\cc$ are shown for each of these cases in Fig.~\ref{G1D}.
{\color{black}
In SM Section~S8, 
we provide additional plots of the FES for other values of $H$ between 3.960 and 3.985.}
The free energy basin in the lower left corner of each surface in Fig.~\ref{FES1}
now corresponds to the metastable bulk $\ccb$ phase, and the channel in the upper right corner leads to the stable $\cc$ phase.  In all cases, we observe a FPT with the same set of features found when $\ca$ is stable:  There are two distinct, unconnected channels in the FES. As shown in Fig.~\ref{G1D}, the coexistence value of $n_c$ is well defined at the crossing of $G_\cb$ and $G_\cc$, and is coincident with a kink in $G_1$.  The variation of $\langle f \rangle$ with $\nmax$ is steepest in the vicinity of $n_c$ (Fig.~\ref{FES1}).
We also observe the lower spinodal limit for the $\cc$ channel;  the upper spinodal limit for the $\cb$ channel is beyond the range of $\nmax$ accessible to our simulations for this system size ($L=200$).  

In Fig.~\ref{Gcuts} we show $G(\nmax=n_c,f)$, the cut through the $G(\nmax,f)$ surface at the point of coexistence between the $\cb$ and $\cc$ channels, for each FES plotted in Figs.~\ref{FES2} and \ref{FES1}.  We see that the height of the free energy barrier between the two channels at the coexistence condition increases with $n_c$.  This is in line with the expectation for a first-order phase transition occurring in a finite-sized system (i.e. the fluctuation)~\cite{Binder:2011du,Binder:2012ey}.
As the size of the fluctuation increases, the phase transition within it must surmount a larger barrier because of the larger interface that must be created between the $\cc$-like core and the wetting layer of $\cb$ that surrounds the core.

In addition to the FPT, each FES in Figs.~\ref{FES2} and \ref{FES1}
exhibits features associated with the nucleation process by which the metastable $\ca$ phase converts to the stable $\cc$ phase.  Fig.~\ref{G1D} shows that $G_1$ in this regime exhibits a maximum at $\nmax=n^*$ corresponding to the size of the critical nucleus. We also observe that the kink in $G_1$ corresponding to  
$\nmax=n_c$ may occur either before of after $n^*$.
Figs.~\ref{FES1}(a) and (c) thus typify two distinct regimes of behavior:
In Fig.~\ref{FES1}(a) $n_c<n^*$ 
while in 
Fig.~\ref{FES1}(c) $n_c>n^*$.  
We also see in Figs.~\ref{FES1}(a) and (c) that the value of $n^*$ corresponds closely to $\nmax$ at which a saddle point occurs in the FES.  This saddle point locates the most probable transition state at which the system exits the basin of the metastable phase.   
Thus the FPT can occur either before or after the transition state.  We also find that all the qualitative features of the FPT occur in exactly the same way regardless of whether the FPT occurs before or after the transition state.  This behavior emphasizes that the FPT is an independent phenomenon from the nucleation process.

Our results thus demonstrate that the superposition of a FPT on the nucleation process generates the characteristic signatures of TSN.  When $n_c<n^*$ [Fig.~\ref{FES1}(a)], the FPT occurs in the sub-critical nucleus.  In this case, the most probable small nucleus resembles the $\cb$ phase, and a small $\cc$ nucleus is unstable with respect to fluctuations in $f$.  Then as it grows larger the most probable nucleus switches to the $\cc$ phase (via the FPT) prior to reaching the critical size, and so the structure of the critical nucleus reflects the structure of the bulk stable phase that will ultimately form.  

When $n_c>n^*$ [Fig.~\ref{FES1}(c)], the FPT occurs in the post-critical nucleus.  In this regime the most probable nucleus resembles the $\cb$ phase all the way up to and beyond the size of the critical nucleus.  Indeed, in Fig.~\ref{FES1}(c) we see that a $\cc$ nucleus is unstable at $n^*$.
As a consequence, the structure of the most probable critical nucleus bears no resemblance to the stable $\cc$ phase, and cannot do so, even as a metastable nucleus.
The post-critical $\cb$ nucleus then grows spontaneously along the $\cb$ channel in the FES.  The transition of this growing post-critical $\cb$ nucleus to the $\cc$ channel only becomes thermodynamically possible for $\nmax$ greater that the lower spinodal for the $\cc$ channel, and is only likely to occur for $\nmax\ge n_c$.  
These observations emphasize that the transition state (saddle point) is not necessarily the entrance to the basin of the stable phase.  Rather, it only identifies the exit from the basin of the metastable phase.

The topology of the FES in Fig.~\ref{FES1}(c) exposes the difference between the first and second ``steps" of TSN when $n_c>n^*$.  The first step is a conventional barrier-crossing process where the transition state corresponds to a well-defined saddle point.  The size and composition of the critical nucleus at this step is defined solely by the thermodynamic features encoded in the FES.  The second step is associated with the FPT, and is a process where the system does not pass through a saddle point but rather crosses over an extended ridge in the FES~\cite{Iwamatsu:2011if}.  Consequently, even though $n_c$ is defined by the properties of the FES, the average size of the post-critical nucleus when it crosses the ridge may not be determined solely by the FES.  For example, 
if the degrees of freedom associated with changes in $f$ relax much faster than those associated with changes in the size of the nucleus, then we can expect the FPT to occur close to $n_c$.
In this case, the average path of the system on the FES will follow closely the curve for $\langle f \rangle$.
However, if the relaxation of $f$ is comparable to or slower than for $\nmax$, then it is likely that the growing nucleus will significantly ``overshoot" the coexistence condition at $n_c$ and continue to grow in size along the (now metastable) $\cb$ channel.  In this case, the average path of the system on the FES will not follow $\langle f \rangle$.  Thus when $n_c>n^*$, the second step of TSN is qualitatively different from the first:  The nucleation pathway for the first step is entirely controlled by thermodynamics, whereas the pathway for the second step depends on both thermodynamic and dynamic factors.  This distinction can help explain the wide variety of behavior observed in TSN in different systems.

Fig.~\ref{FES1}(b) corresponds to the case when $n_c\sim n^*$, and displays complex behavior.  We observe two saddle points on either side of an unusually flat region of the FES.  
Although $n_c$ and $n^*$ are close in value, $n^*$ is not close to the value of $\nmax$ of either saddle point.  In this case, the transition state in the FES by which the system leaves the metastable state is not sharply defined.  As a consequence, we can expect particularly strong deviations from CNT in this regime.  

\begin{figure}
\includegraphics[scale=0.4]{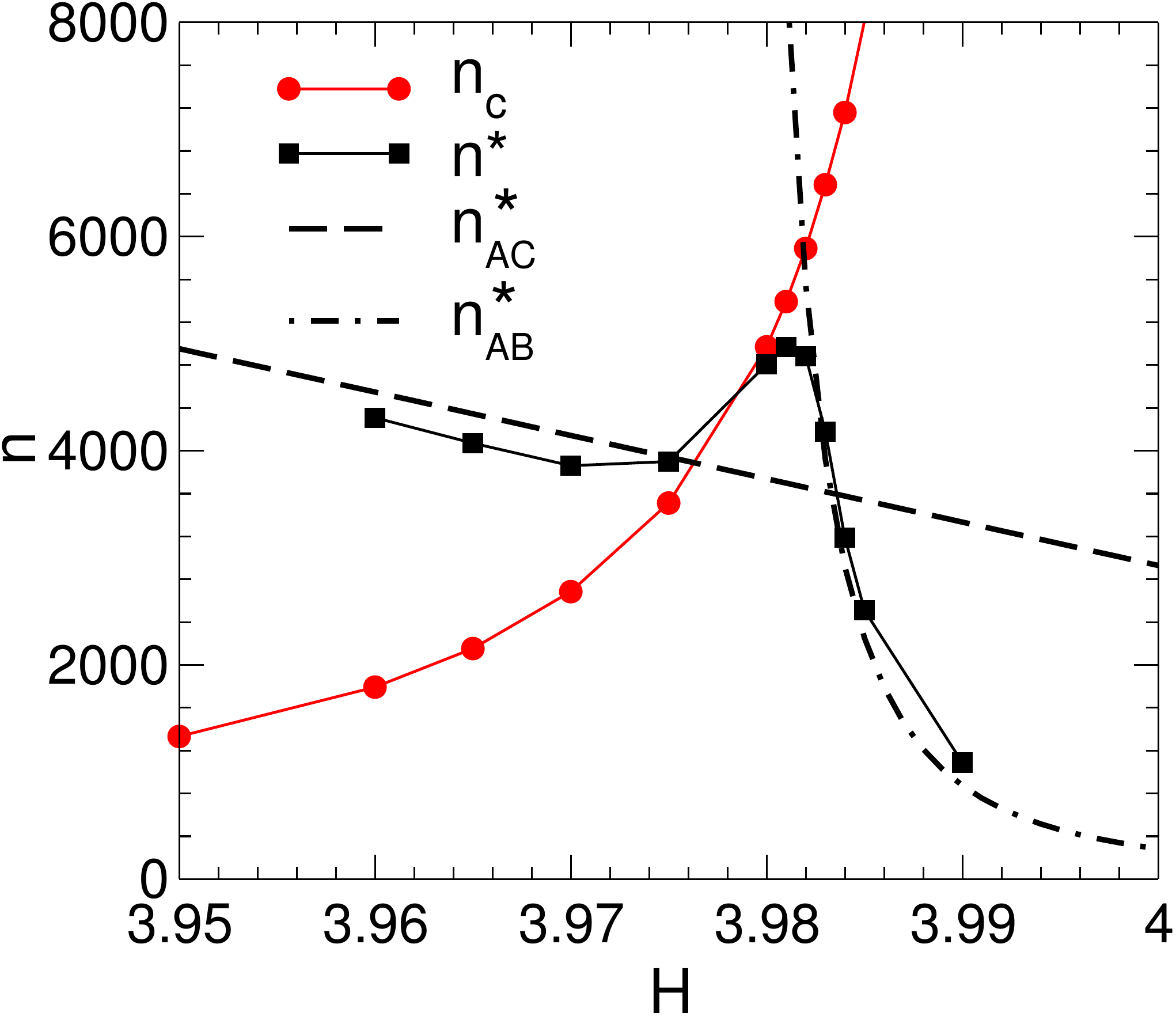}
\caption{Plot of $n_c$ and $n^*$ versus $H$,
for $H_s=0.01$ and $L=200$ as obtained from our 2D umbrella sampling simulations.  Also shown are the CNT predictions for 
$n^*_{\ca\cb}$ and $n^*_{\ca\cc}$ computed as described in the text.
}
\label{nonCNT}
\end{figure}

An example of the unusual behavior occurring when $n_c\sim n^*$ 
is shown 
in Fig.~\ref{nonCNT}, where we plot $n_c$ and $n^*$ as a function of $H$ at fixed $H_s=0.01$.  As shown in Fig.~\ref{G1D}, the height of the nucleation barrier decreases monotonically as $H$ increases.  However, in Fig.~\ref{nonCNT} we observe that $n^*$ does not decrease monotonically with $H$, but rather exhibits a minimum and a maximum in the vicinity of the value of $H$ at which $n_c = n^*$.  
This complex and highly non-classical behavior arises from the crossover from the $n_c <n^*$ 
to the $n_c >n^*$ regimes as $H$ increases.  
To understand this effect, we consider the CNT expression~\cite{,Debenedetti:1996,Kelton:2010} for $n^*$ in $D=2$:
$n^*=\pi \sigma^2/(\Delta \mu)^2.$
{\color{black} As described in SM,}
we have obtained approximate expressions to describe the dependence on $H$ and $H_s$ of the 
chemical potentials 
{\color{black}(see SM Section~S3)}
and
surface tensions 
{\color{black}(see SM Section~S4)}
for all three phases in our lattice model.
We use these expressions to calculate the CNT prediction for the variation of $n^*$ with $H$, and compare this with the observed behavior.
In Fig.~\ref{nonCNT} we show that 
for $n_c <n^*$, the $H$-dependence of $n^*$ approximately follows that expected for the nucleation of $\cc$ directly from $\ca$:  $n^*_{\ca \cc}=\pi \sigma^2_{\ca \cc}/(\Delta \mu_{\ca\cc})^2$.  
While $\Delta \mu_{\ca\cc}$ is constant for all $H$ at fixed $H_s$, $n^*_{\ca\cc}$ decreases gradually with $H$ due to the approximately linear decrease of $\sigma^2_{\ca \cc}$ with $H$.  
However, when $n_c >n^*$, the $H$-dependence of $n^*$ switches to follow the prediction for the nucleation of 
$\cb$ directly from $\ca$:  $n^*_{\ca \cb}=\pi \sigma^2_{\ca \cb}/(\Delta \mu_{\ca\cb})^2$.  In this expression, $\sigma_{\ca \cb}$ is constant with $H$, but the magnitude of $\Delta \mu_{\ca\cb}$ increases linearly as $H$ increases at fixed $H_s$, resulting in a fast decrease of $n^*$.  The crossover in the behavior of $n^*$ occurs when the barrier for the $\ca\to\cb$ process becomes smaller than the $\ca\to\cc$ process.  Interestingly, at the point of this crossover, the critical nucleus along the $\cb$ channel is larger than that on the $\cc$ channel, resulting in the maximum in $n^*$ observed in Fig.~\ref{nonCNT}.  


\section{Two-step nucleation and bulk phase behavior}
\label{bulk}


 \begin{figure*}
\includegraphics[scale=0.36]{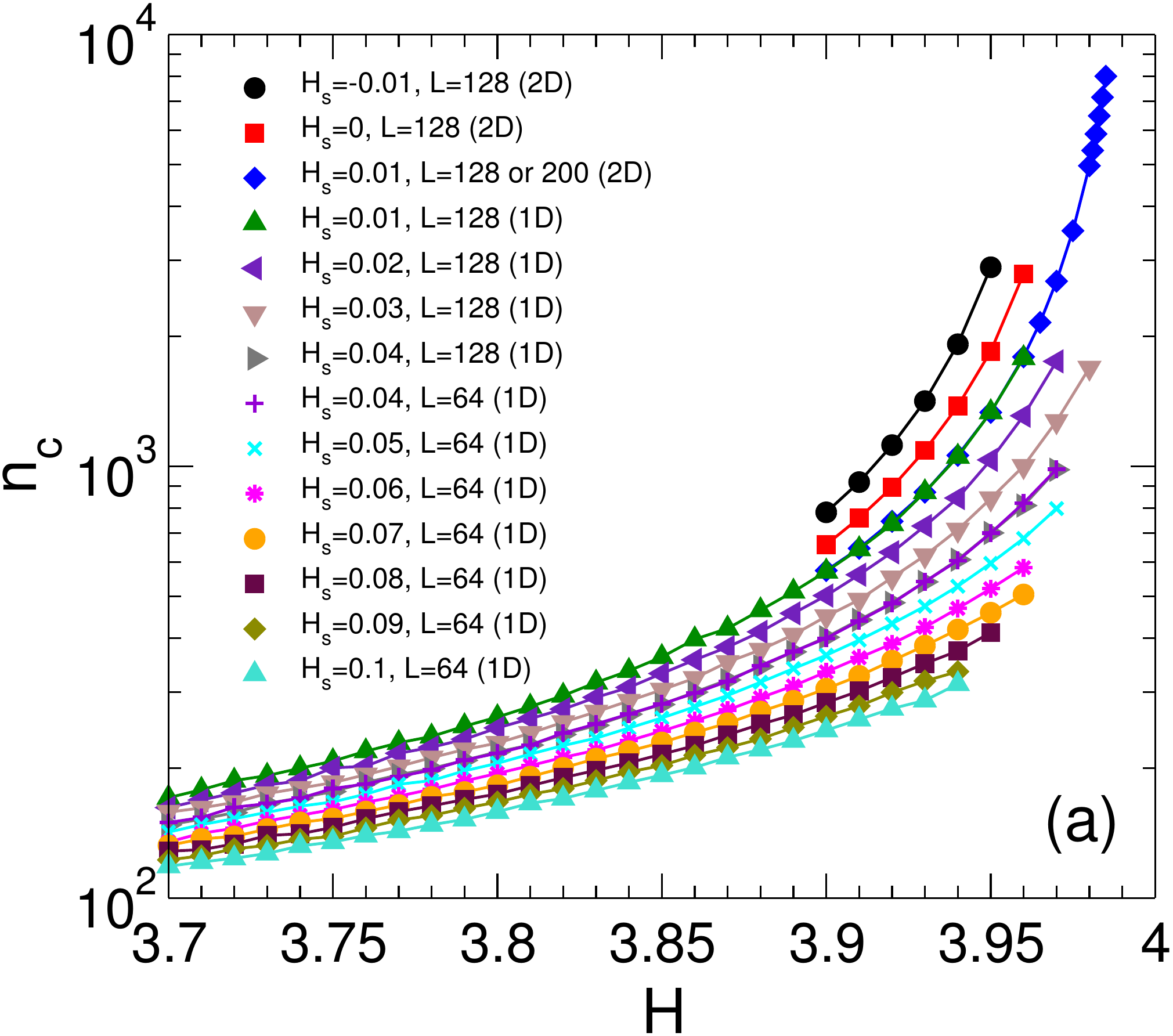}
\hspace{1cm}
\includegraphics[scale=0.36]{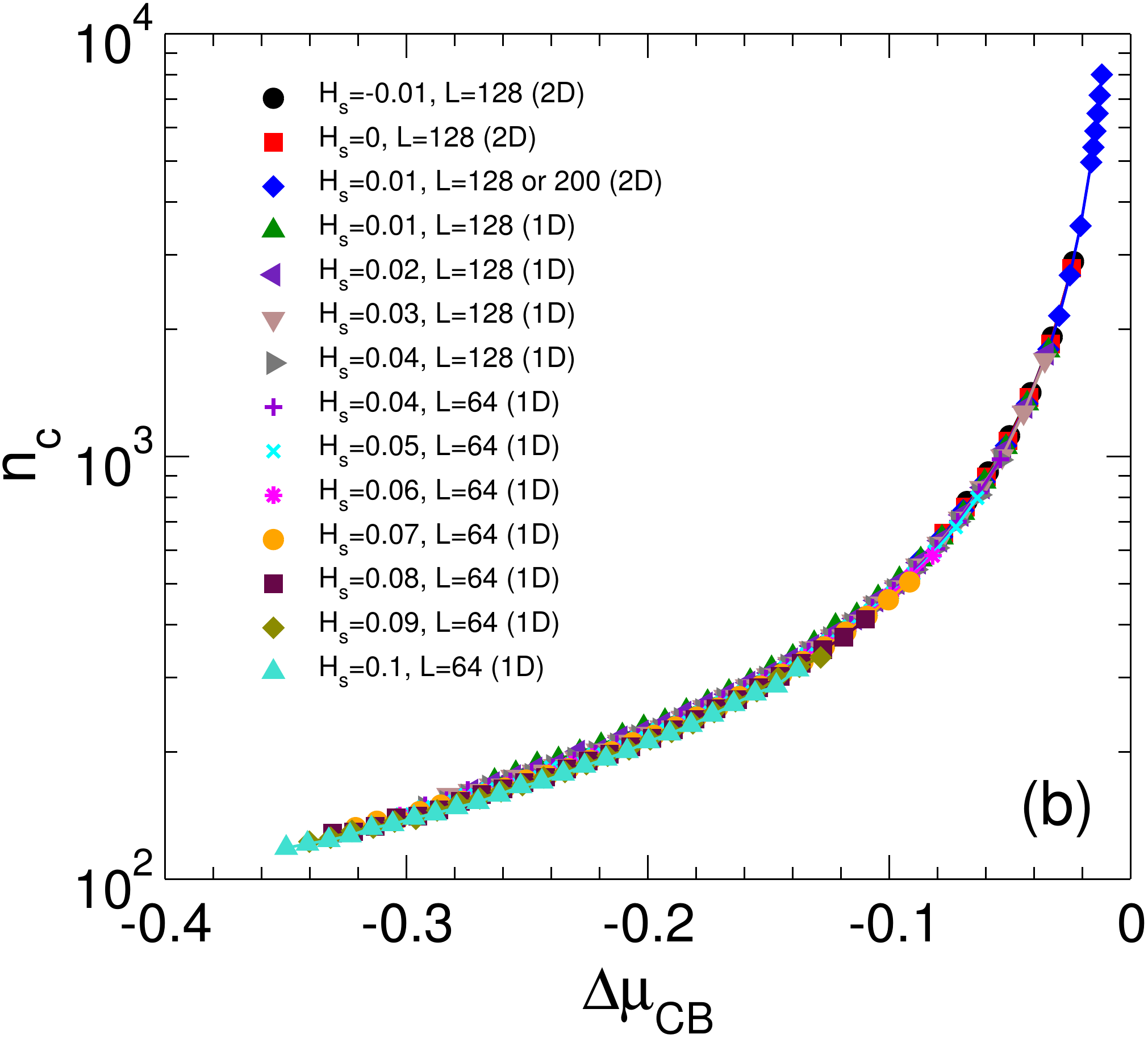}\\
\vspace{1cm}
\includegraphics[scale=0.36]{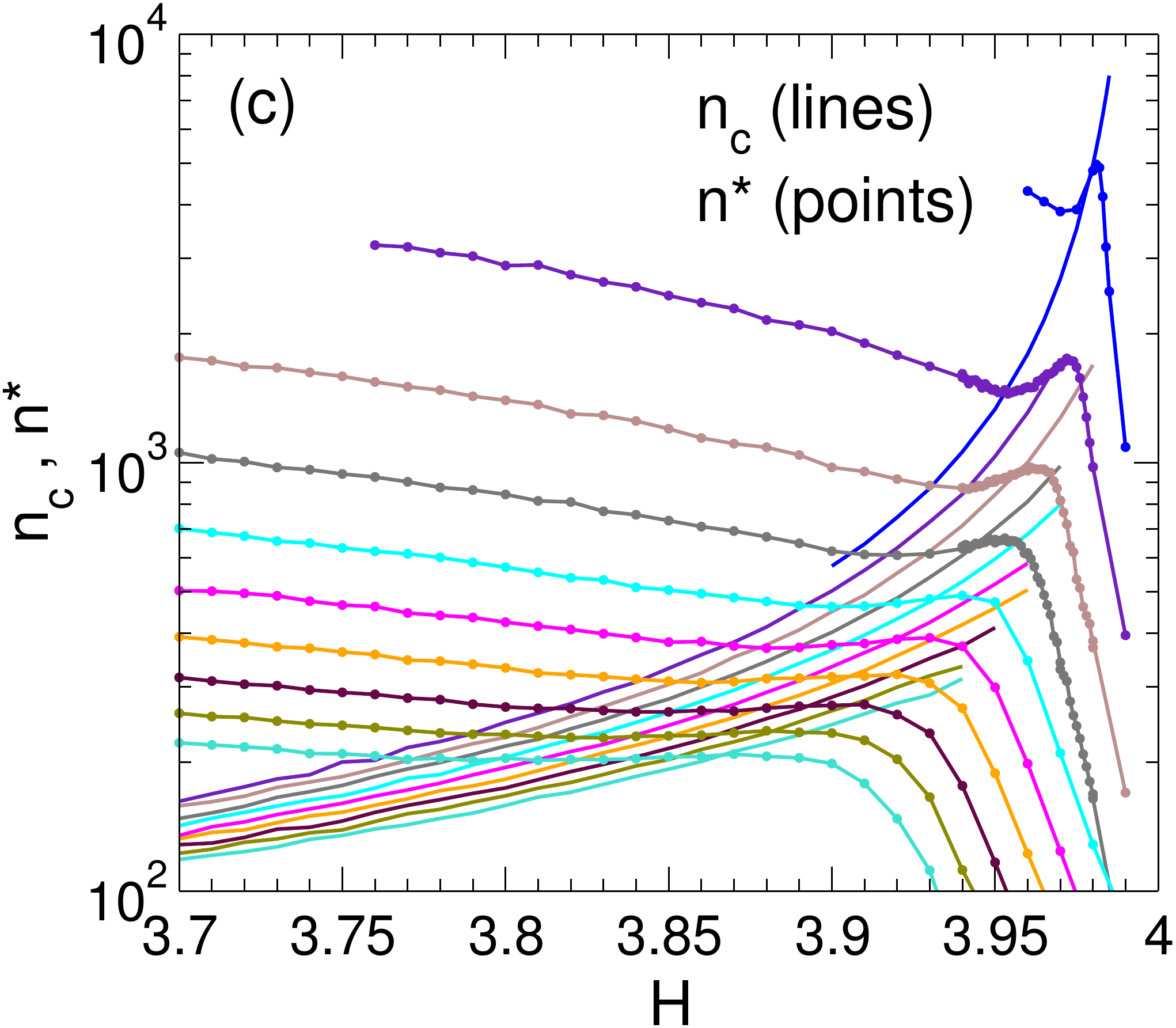}
\caption{(a) $n_c$ versus $H$ for various $H_s$.  Data are obtained from either 1D or 2D umbrella sampling simulations, as indicated in the legend.  For data obtained from 2D umbrella sampling runs at $H_s=0.01$, we use a system of size $L=128$ for $H<3.96$ and $L=200$ for $H\ge 3.96$. 
(b) Same data as in (a), but plotted versus $\Delta \mu_{\cc\cb}$.
(c) Lines without symbols plot $n_c$ versus $H$, for $H_s$ from 0.01 to 0.10 in steps of 0.01, from top to bottom.  Lines with dots plot our data for $n^*$ versus $H$, for the same set of $H_s$ values, from top to bottom.  
Data for $n^*$ at $H_s=0.01$ are obtained from 2D umbrella sampling runs with $L=200$.  
Data for $n^*$ at $H_s=\{0.02,0.03,0.04\}$ are obtained from 1D umbrella sampling runs with $L=128$.  
All other $n^*$ data are obtained from 1D umbrella sampling runs with $L=64$.  
}
\label{nn}
\end{figure*}

We next seek to identify where in the phase diagram the different regimes of TSN occur.  To do so, we quantify the variation of $n_c$ and $n^*$ over a wide range of $H$ and $H_s$ within the stability fields of $\ca$ and $\cc$. 
We achieve this efficiently by augmenting the results obtained from our 2D umbrella sampling runs with 1D umbrella sampling simulations, as described in 
{\color{black} SM Section~S9}.  
As shown in Fig.~\ref{nn}(a), 
we find that the variation of $n_c$ with $H$ and $H_s$ is relatively simple:  For fixed $H_s$, $n_c$ grows and diverges as $H$ approaches 
the $\cu \cc$ coexistence line ${\cal L}_{\cu \cc}$ 
from below.  This behavior is anticipated by the form of Eq.~\ref{nc} and confirmed in Fig.~\ref{nn}(b), where we plot $n_c$ as a function of $\Delta\mu_{\cc\cb}$ 
for various values of $H_s$, both positive and negative.  The data for all values of $H_s$ fall on a single master curve.
Fig.~\ref{nn}(b) confirms the prediction of Eq.~\ref{nc} that $n_c$ diverges as $\Delta\mu_{\cc\cb}\to 0$ (i.e. approaching ${\cal L}_{\cb\cc}$), and does not depend on $\mu_{\ca}$, which changes as $H_s$ changes.  That is, the value of $n_c$ is unaffected by the presence or absence of a nucleation process, and is an intrinsic property of the fluctuations of $\ca$.

The variation of $n^*$ for various $H_s$ is shown in Fig.~\ref{nn}(c).  We find that the maximum of $n^*$ noted in Fig.~\ref{nonCNT} is sharpest at small $H_s$, and fades in prominence as $H_s$ increases.  
For each $H_s$ we locate the intersection of $n_c$ and $n^*$, and plot the locus of points ${\cal L}_n$ at which $n_c=n^*$ in Fig.~\ref{pd1}.
For $H$ less than ${\cal L}_n$, 
TSN will be observed where $n_c<n^*$; 
for $H$ greater than ${\cal L}_n$, TSN with $n_c>n^*$ will occur.  

Notably, we find that ${\cal L}_n$ is nearly coincident with the $\ca\cb$ coesixtence line $\cal L_{\ca \cb}$, especially as $H_s\to 0$.  This correspondence occurs in our model due to a combination of influences.  
Based on Eq.~\ref{simple}, we would expect that ${\cal L}_n$ should occur for $H$ above ${\cal L}_{\ca \cb}$, 
since $\Delta \mu_{\ca \cb}<0$ is required to form a $\cu$-like nucleus that grows spontaneously within $\ca$.  
However, Fig.~\ref{nonCNT} 
shows that the intersection of $n_c$ and $n^*$ occurs at lower $H$ than predicted by our simple CNT analysis, due to the complexity of the FES when $n_c\sim n^*$.  In addition, we see in Fig.~\ref{nn}(c) that $n^*$ drops very quickly for $H$ above ${\cal L}_n$.
Related to this behavior, we also observe 
that the height of the nucleation barrier $G^*$ decreases rapidly for $H$ above ${\cal L}_n$.  
See
{\color{black}SM Section~S10} 
for details of our calculation of $G^*$~\cite{Wolde:1996p3069,Auer:2004db,Lundrigan:2009p5256}.
In Fig.~\ref{pd1} we plot the locus along which $\beta G^*=20$, which we find lies close to and just above ${\cal L}_n$.  For $H$ above the $\beta G^*=20$ locus, the basin of metastability for $\ccb$ quickly becomes poorly defined, and the TSN process consists of an almost barrierless decay to a spontaneously growing nucleus of $\cu$, which eventually converts to $\cc$ via the FPT.  
As a result of these effects, ${\cal L}_n$ 
is on the one hand unlikely to occur much below ${\cal L}_{\ca\cb}$, and on the other hand is unlikely to occur much above it.  These factors effectively constrain ${\cal L}_n$ 
to lie very close to ${\cal L}_{\ca\cb}$.
If this behavior proves to be common in other systems, it provides a simple way to predict the crossover from the 
$n_c<n^*$ to the $n_c>n^*$ 
regime, by locating the metastable extension of the bulk phase coexistence line for the two phases involved in the FPT.

We have also assessed the limits of metastability (LOM) of the bulk $\cb$ phase, 
{\color{black}as described in SM Section~S2}.  
In practice, the LOM of a homogeneous bulk phase depends on the system size~\cite{Binder:2012ey}.  For a small system, the LOM for a given bulk phase may occur significantly outside the stable phase boundary of that phase, but as $L\to \infty$ the LOM approaches the stable phase boundary.  We show an example in Fig.~\ref{pd1}(b), where we plot the LOM for the bulk $\cb$ phase for $L=64$.  The LOM for each of our system sizes with $L>64$ lies between the boundary shown in Fig.~\ref{pd1}(b) and the $\ca\cb$ and $\cb\cc$ coexistence curves.
Therefore most of our results for $n_c$ in Fig.~\ref{nn} are obtained beyond the LOM of the bulk $\cb$ phase for the system sizes studied here, demonstrating that the FPT remains well-defined even when the bulk $\cb$ phase is unstable.
Furthermore, small fluctuations occurring within the $\ca$ phase always resemble the $\cb$ phase, even when the bulk $\cb$ phase is unstable, in both the $n_c<n^*$ and the $n_c>n^*$ regimes.  These observations emphasize that a local structure that is unstable as a bulk phase can still play a significant role, both as the dominant small fluctuation, and as an ``intermediate phase" in a TSN process.

\section{Relationship to previous modelling of two-step nucleation}
\label{disc}


As indicated in the Introduction, a number of previous simulation and theoretical studies have examined behavior related to TSN.
Previous work, particularly by Sear, has also demonstrated that many complex nucleation phenomena can be elucidated by studying lattice models similar to the one employed here~\cite{Sear:2005p6366,Page:2006p6353,Sear:2007p6363,Sear:2008p6361,Sear:2008p7131,Sear:2009p6355,Duff:2009p6360,Sear:2011dv}.
Notably, the present work reproduces several phenomena first identified by Duff and Peters~\cite{Duff:2009p6360} who also used a lattice model.
Their work introduced the FES in the specific form that we use, and showed that the conversion of the nucleus to the stable phase can occur before or after the transition state, although in the latter case the conversion of the nucleus was not explicitly observed.  Their simulations also did not allow the calculation of a FES of sufficient resolution to resolve the first-order character of the FPT as observed here, and they did not identify the spinodals that bracket the FPT.  Ref.~\cite{Duff:2009p6360} presents a CNT-based analysis of TSN, although the implications for the nature of fluctuations in general was not explored.  

The analytical study of TSN by Iwamatsu~\cite{Iwamatsu:2011if} identified the thermodynamic conditions for the conversion of the nucleus to the stable phase, pointed out its first-order character, and noted that this conversion crosses a ridge in the FES.  This work also showed that the FES may display two distinct saddle points, as observed here.  However, Ref.~\cite{Iwamatsu:2011if} argued that there were cases where the FES has two independent channels leading out of the metastable phase, in contrast to our results, where we find only one.
Further, Ref.~\cite{Iwamatsu:2011if} 
did not identify cases in which the conversion of the nucleus to the stable phase occurs before reaching the transition state, nor did it identify spinodal endpoints along any channel in the FES.
These differences may arise from fundamental differences between our modelling and that in Ref.~\cite{Iwamatsu:2011if}.  However, it is also possible that a higher resolution analysis of the cases presented in Ref.~\cite{Iwamatsu:2011if} might reveal the same pattern of behavior observed here.
Such a test to see if Ref.~\cite{Iwamatsu:2011if} can be reconciled with our results merits investigation, as this would clarify the possible topologies of the nucleation pathway in TSN.
As mentioned earlier, our observation of spinodals bracketing the FPT is consistent with the analysis of Harrowell on the stability of sub-critical crystal nuclei~\cite{Harrowell:2010jt}, and so it would be useful to assess the generality of this result by re-examining the features of the FES as presented in both Refs.~\cite{Duff:2009p6360} and \cite{Iwamatsu:2011if}. 

A recent example consistent with the pattern of behavior shown here is the simulation study by Santra, Singh and Bagchi~\cite{Santra:2018wl}, which focusses on the competition between BCC and FCC crystal nucleation in a hard-core repulsive Yukawa system.  They showed that a post-critical BCC nucleus forms and grows spontaneously even under conditions where bulk FCC is the most stable phase.  This case corresponds to the $n_c>n^*$ regime identified here.  
Ref.~\cite{Santra:2018wl} also evaluates 1D ``cuts" through the FES, which in the terminology of the present work correspond  
approximately to $f=0$ (BCC-like) and $f=1$ (FCC-like).  Although the behavior of the 1D nucleation barriers so obtained is consistent with the 2D surfaces studied here, it would be useful to confirm this correspondence by computing the full 2D FES for the system studied in Ref.~\cite{Santra:2018wl}.

While several of the phenomena reported here have been documented in prior work, these observations are fragmented across separate studies, and are also limited in the range of thermodynamic conditions examined.  
The general pattern of behavior presented here captures the key features of these earlier studies, clarifies the interrelationships of these findings, and also reveals important details of the FES not previously appreciated.  We further show how the properties of the FES evolve over a wide range of thermodynamic conditions and correlate these changes to stable and metastable coexistence boundaries in the bulk phase diagram.  Notably, no previous work has to our knowledge pointed out that the FPT is an intrinsic property of fluctuations, and is distinct and independent from nucleation phenomena.  Our work thus broadens, clarifies, and hopefully simplifies, the conceptual framework for understanding the many phenomena associated with TSN.

\section{Discussion}
\label{conc}


In summary, we have attempted to clarify TSN by first disentangling the physics of the FPT from the nucleation process itself, and showing that these are indeed distinct phenomena.
We then examine how these two phenomena combine to produce TSN via a high-resolution study of the nucleation FES for a prototypical lattice model, conducted over a wide range of thermodynamic conditions.

Our results demonstrate that regardless of the thermodynamic conditions under which nucleation occurs, the initial fluctuation of the system away from its equilibrium state always takes the form of a local structure with the lowest surface tension with the surrounding phase.  
In other words, polymorph selection, at least at the local level, is controlled entirely by surface tension.
When the lowest-surface-tension structure does not correspond to the most stable phase, 
then the initial stage of nucleus growth will not resemble the stable phase, and the result is TSN.
The conversion of the fluctuation to the stable phase is a first-order FPT, which occurs by traversing a ridge in the FES.  
The transition state by which the system exits the metastable phase occurs at a saddle point in the FES and may occur before ($n_c<n^*$) 
or after ($n_c>n^*$) 
the FPT.

The Ostwald step rule (OSR) states that the bulk phase that forms first from a metastable phase is not the most stable phase, but the phase with the chemical potential that is below but closest to the metastable phase~\cite{Tavassoli:2002ey,Kelton:2010,Santra:2013kn,Santra:2018wl}.  Our findings are consistent with the OSR but also provide a modified and more general way of understanding it.  In our lattice model, when metastable $\ca$ transforms to stable $\cc$, the $\cb$ phase always appears first at the local level, regardless of whether $\cb$ has a higher or lower chemical potential.  It is the low $\ca\cb$ surface tension that ensures that $\cb$ forms first within $\ca$;  their relative chemical potentials are initially irrelevant.  
When $n_c<n^*$,
the initial sub-critical nucleus resembles the $\cb$ phase, but because the FPT occurs prior to the transition state, there is no indication in the post-critical nucleus that the $\cb$ phase was initially dominant.
However, when $n_c>n^*$, the post-critical nucleus resembles the $\cb$ phase, creating the conditions in which the OSR may be realized.
The OSR is formally obeyed in our model phase diagram in the region bounded by ${\cal L}_{\ca\cb}$, 
${\cal L}_{\cb\cc}$ 
and the LOM of the $\cb$ phase 
(region ``O$_1$" in Fig.~\ref{pd1})
because this is the region in which $\cb$ is both observable as a 
bulk metastable phase and also has a lower chemical potential than $\ca$.  
At points in the phase diagram between ${\cal L}_{\ca\cb}$ and
${\cal L}_{\cb\cc}$ but 
beyond the LOM of the $\cb$ phase
(region ``O$_2$" in Fig.~\ref{pd1})
the bulk $\cb$ phase is unstable.  In this region, the post-critical nucleus will resemble $\cb$, but must convert to the stable $\cc$ phase at a finite size that is smaller than the system size.  Thus the observation of behavior that obeys the OSR depends on an interplay of system-size effects (which control the location of the LOM) and the range of $\nmax$ for the growing post-critical nucleus over which the low-$f$ channel in the FES remains well-defined (which is controlled by the location of the spinodal on the low-$f$ channel).
In the present study we have restricted our evaluation of the FES to the range of $\nmax$ in which the largest fluctuation does not approach the system size.
It would be useful for future work to extend the FES to larger $\nmax$ to further clarify the behavior related to the OSR.  

As noted, the two steps in TSN are qualitatively different.  One crosses a saddle point in the FES and the other crosses a ridge.
As such, both steps are activated processes.  At the same time, the process that takes the system over the saddle point is similar to that in conventional (i.e. one-step) nucleation, while the ridge-crossing process of the FPT is more complex.  The FPT occurs in a system (the fluctuation) the size of which is spontaneously increasing or decreasing, depending on the shape of the FES.  Furthermore, there is a well-defined critical size for the nucleus of the stable phase to form inside the fluctuation, and until the fluctuation itself reaches this size, the stable phase will not be observed.  This interplay of size effects is consistent with the existence of the spinodal limit of the high-$f$ channel on the FES that prevents a $\cc$-like fluctuation from being stable at small $\nmax$, and suggests that this phenomenon is indeed general~\cite{Harrowell:2010jt}.  In addition, the existence of a spinodal limit at large $\nmax$ along the low-$f$ channel means that if the growing $\cb$ nucleus does not cross the ridge to the $\cc$ phase, then it will ultimately do so via a barrierless process at the spinodal.  That is, the activated nature of the ridge-crossing FPT process is lost if the $\cb$ nucleus grows sufficiently large. 

We also note that all of our analysis concerning TSN assumes that the intermediate phase ($\cb$) completely wets the stable phase ($\ca$), a condition realized in our lattice model.  An important direction for future work is to generalize these considerations to cases in which incomplete wetting occurs.  
In addition, the lattice model results presented here are all obtained at fixed $T$.  While this constraint has simplified our analysis, now that the characteristics of the model FES have been described in detail, it will be interesting in subsequent studies to explore the $T$ dependence of these features, especially for the FPT itself.

 
Our work also has a number of practical implications for simulation studies of nucleation.  
It is widely appreciated that care must be taken when choosing a local order parameter to define the nucleus, the size of which serves as the reaction coordinate in many studies which evaluate the nucleation barrier~\cite{Wolde:1996p3069,Auer:2004db,Jacobson:2011jx,Reinhardt:2012ey}.  Our results show that when this order parameter recognizes both the intermediate and the stable phase contributions to the nucleus, the ``kink'' in $G_1$ is a characteristic signature of TSN, which also locates $n_c$.  Such a kink may be discerned in previous work;  see e.g. Fig.~2 of Ref.~\cite{Qi:2015iea}.  Conversely, caution must be exercised when conducting 1D umbrella sampling with respect to $\nmax$:  If $n_c$ is large, then the barrier between the low-$f$ and high-$f$ channels will also be large, and so a series of simulations of progressively larger $\nmax$ may become trapped in the low-$f$ channel even when $\nmax>n_c$.  When practical, 2D umbrella sampling to compute the full FES should be carried out, to ensure that the complete nucleation pathway is observed.
It is also common to choose a local order parameter which only detects a nucleus of the stable phase.  Our work demonstrates that when TSN occurs, the initial nuclei generated by this approach will not correspond to the most probable initial nuclei, resulting in a distortion in the shape of $G_1$ at small $n$.  
Where possible, a local order parameter should be chosen that identifies all structures that deviate from the metastable phase, not just those that resemble the stable phase.  Recent work suggests that such an approach is feasible in molecular systems~\cite{Krebs:2018ib}.  
Finally, we note the challenges that will be associated with estimating the nucleation rate from transition state theory when the transition state is not sharply defined in the FES, as in Fig.~\ref{FES1}(b)~\cite{Auer:2004db}.  This difficulty may help explain the large deviations between estimated and observed nucleation rates noted for many systems exhibiting complex nucleation processes~\cite{Kelton:2010,Sear:2012ji}.

In addition, it is notable that in our lattice model we observe conditions where the most probable fluctuation of a given size does not correspond to a stable bulk phase under the same conditions, i.e. conditions beyond the LOM of the bulk phase.  Sear noted a similar effect in a simulation study of heterogeneous nucleation~\cite{Sear:2009p6355}.
It is therefore conceivable that, in other systems, a local fluctuation that never corresponds to a bulk phase might play an important role in the growth of the nucleus.  Such fluctuations might include amorphous solid clusters or spatially limited structures such as icosohedra~\cite{Royall:2015if}.  
This possibility, combined with the activated nature of the FPT, could account for long-lived metastable ``prenucleation clusters" that grow to mesoscopic size before conversion to the stable phase, as has been reported e.g. in crystallization of CaCO$_3$~\cite{Li:2008js,Pouget:2009dg,Demichelis:2011ela,Wallace:2013df}. 

We have shown that the dominant contribution of the surface tension to the free energy of small fluctuations underlies and explains the complexities of TSN in our model system.  Recent work on the competition between glass and crystal formation in supercooled liquids also points to the central role of low-surface-tension fluctuations~\cite{Russo:2018hl}, which if different from the stable crystal can promote glass formation.
The controlling influence of the surface tension in determining the most probable initial deviation from equilibrium may therefore be a principle with wide ranging implications for the behavior of metastable systems.

\bigskip
\begin{acknowledgments}
ISV, RKB and PHP thank NSERC for support.  
Computational resources were provided by ACEnet and Compute Canada.
We thank K.~De'Bell, D. Eaton. K.M.~Poduska, F.~Sciortino and R.~Timmons for helpful discussions.
\end{acknowledgments}



%




\clearpage

\centerline{\bf SUPPLEMENTAL MATERIAL}
\bigskip
\centerline{\bf Phase transitions in fluctuations and their role in two-step nucleation}
\bigskip
\centerline{D. James, S. Beairsto, C. Hartt, O. Zavalov, I. Saika-Voivod, R.K. Bowles and P.H. Poole}
\bigskip
\centerline{\today}

\setcounter{equation}{0}
\setcounter{figure}{0}
\setcounter{section}{0}
\setcounter{table}{0}
\setcounter{page}{1}
\makeatletter
\renewcommand{\theequation}{S\arabic{equation}}
\renewcommand{\thefigure}{S\arabic{figure}}
\renewcommand{\thesection}{S\arabic{section}}
\section{phase diagram}
\label{diag}

To evaluate the phase diagram of our lattice model, we use umbrella sampling MC simulations to compute the system free energy as a function of 
two bulk order parameters, 
the magnetization $m$ and the staggered magnetization $m_s$~\cite{Binder:2009vp,Tuckerman:2010}.  These are defined as,
\begin{eqnarray}
m&=&\frac{1}{N}\sum_{i=1}^N s_i \\
m_s&=&\frac{1}{N}\sum_{i=1}^N \sigma_i s_i.
\label{m}
\end{eqnarray}
Note that $m$ and $m_s$ are subject to the constraints,
\begin{eqnarray}
m+m_s&\le&1\cr
m-m_s&\le&1.
\label{m1}
\end{eqnarray}

The order parameters $m$ and $m_s$ can be used to identify each of the phases in our system.
At $T=0$, four phases are observed, depending on the values of $H$ and $H_s$: two ferromagnetically ordered  phases with $(m,m_s)=(-1,0)$ and $(m,m_s)=(1,0)$, denoted respectively as 
$\bar\cb$ and $\cb$;
and two antiferromagnetically ordered phases with $(m,m_s)=(0,-1)$ and $(m,m_s)=(0,1)$, denoted respectively as
$\ca$ and $\cc$.  Since we only consider $H>0$, the $\bar\cb$ phase does not appear in our analysis.

\begin{figure}
\includegraphics[scale=0.4]{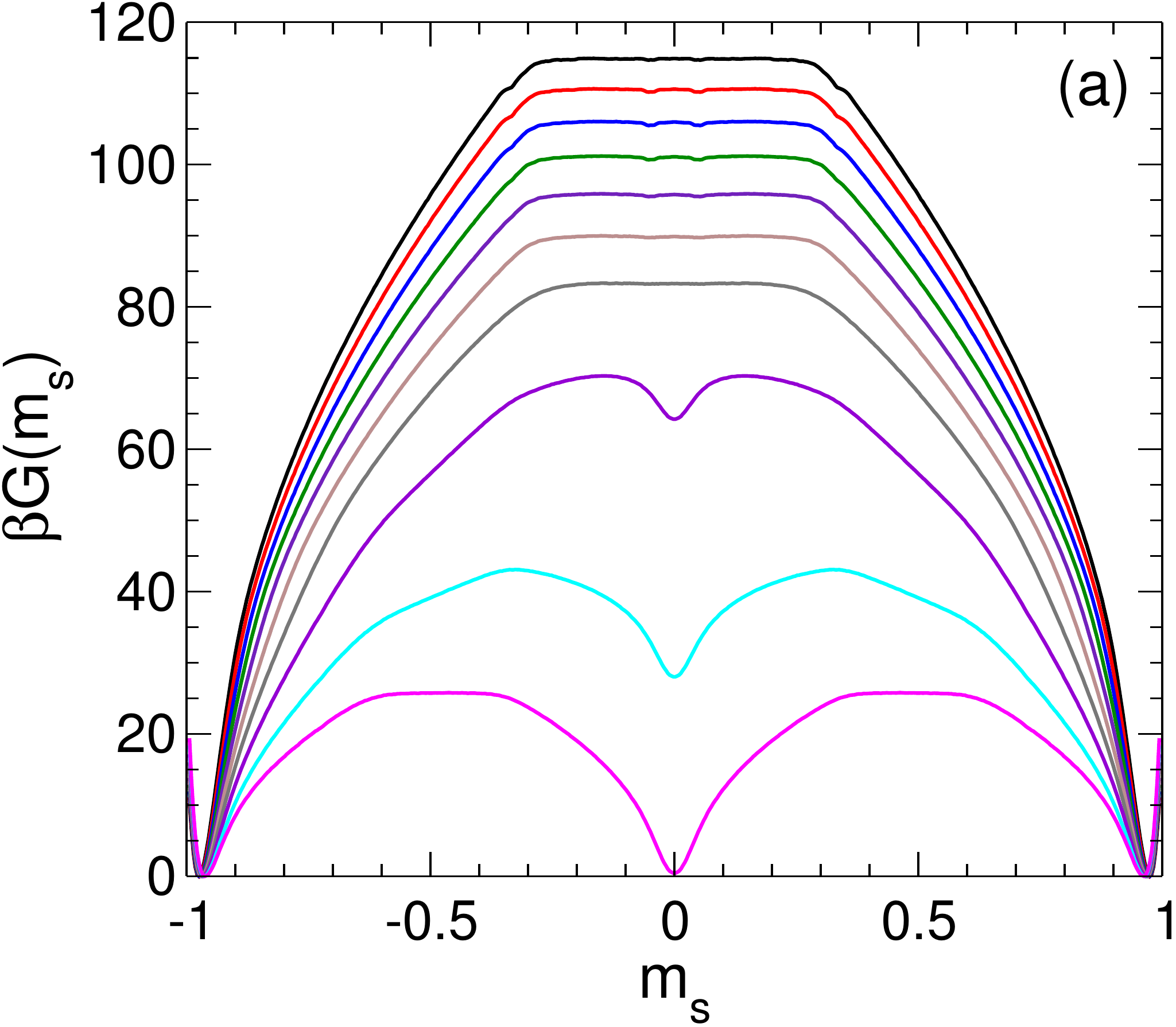}\\
\vspace{0.8cm}
\includegraphics[scale=0.4]{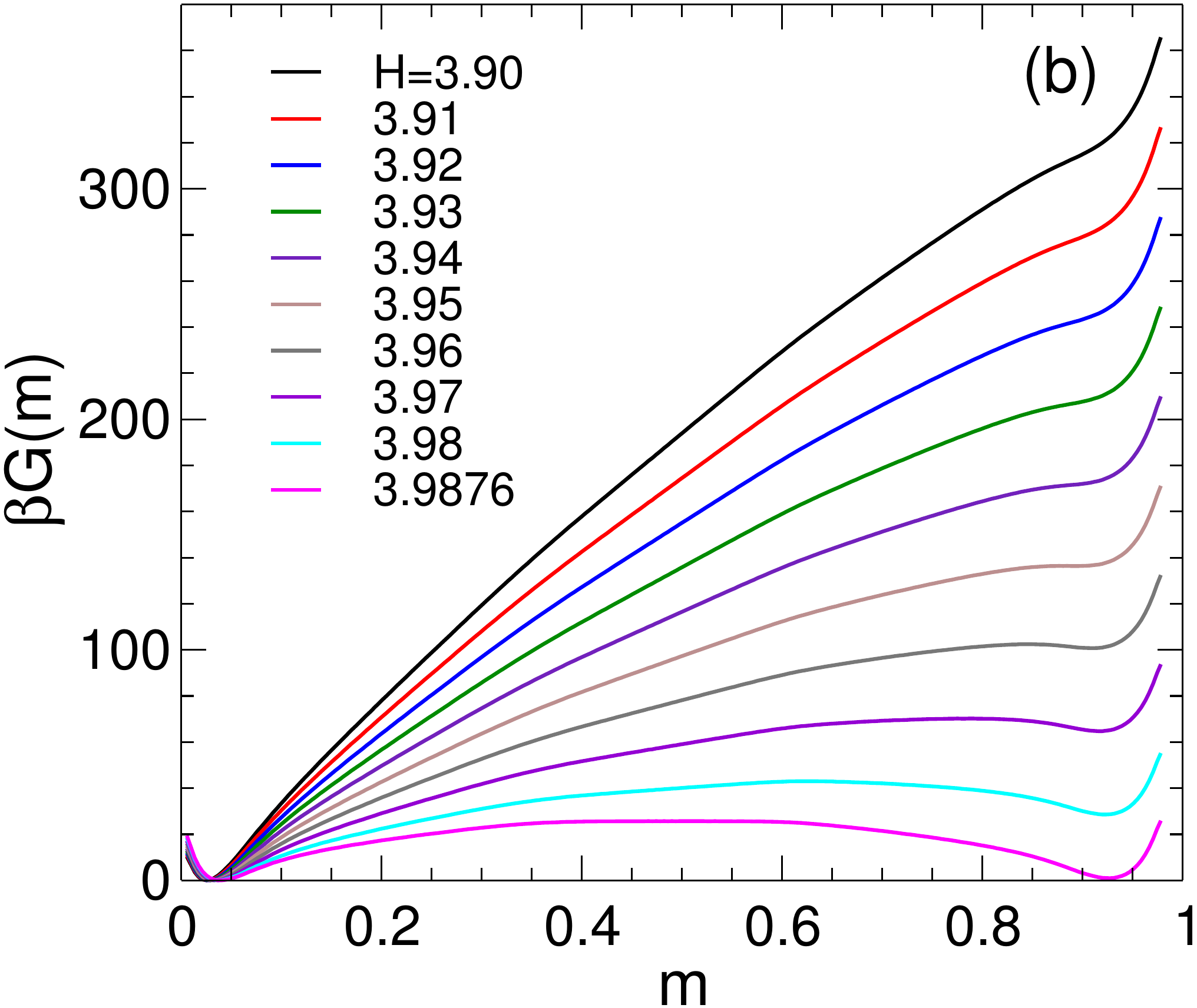}
\caption{(a) $G(m_s)$ and (b) $G(m)$ along the $\ca\cc$ coexistence curve (where $H_s=0$) for a system with $L=64$.  The legend given in (b) applies to both panels.}
\label{sigmaCC}
\end{figure}

To locate stability fields and phase boundaries for each phase, we evaluate, 
\begin{equation}
\beta G(m_s,m)=-\log [P( m_s,m)] + C,
\label{ggg}
\end{equation}
where $G(m_s,m)$ 
is the conditional free energy of the system at fixed $(H_s, H, T)$ as a function of $m_s$ and $m$.  
$P(m_s,m)$ is a function that is proportional to
the probability of observing a given value of $m_s$ and $m$ under the same conditions of $(H_s, H, T)$.
The value of the arbitrary constant $C$ in Eq.~\ref{ggg}, and in all subsequent equations in which it occurs, is 
always chosen so that the global minimum of the corresponding free energy is zero.
We also define,
\begin{eqnarray}
\beta G(m_s)&=&-\log [P( m_s)]+C\cr
P(m_s)&=&\int_0^1 P(m_s,m)\, dm,
\end{eqnarray}
and,
\begin{eqnarray}
\beta G(m)&=&-\log [P( m)]+C\cr
P(m)&=&\int_{-1}^1 P(m_s,m)\, dm_s.
\end{eqnarray}

\begin{figure}
\includegraphics[scale=0.4]{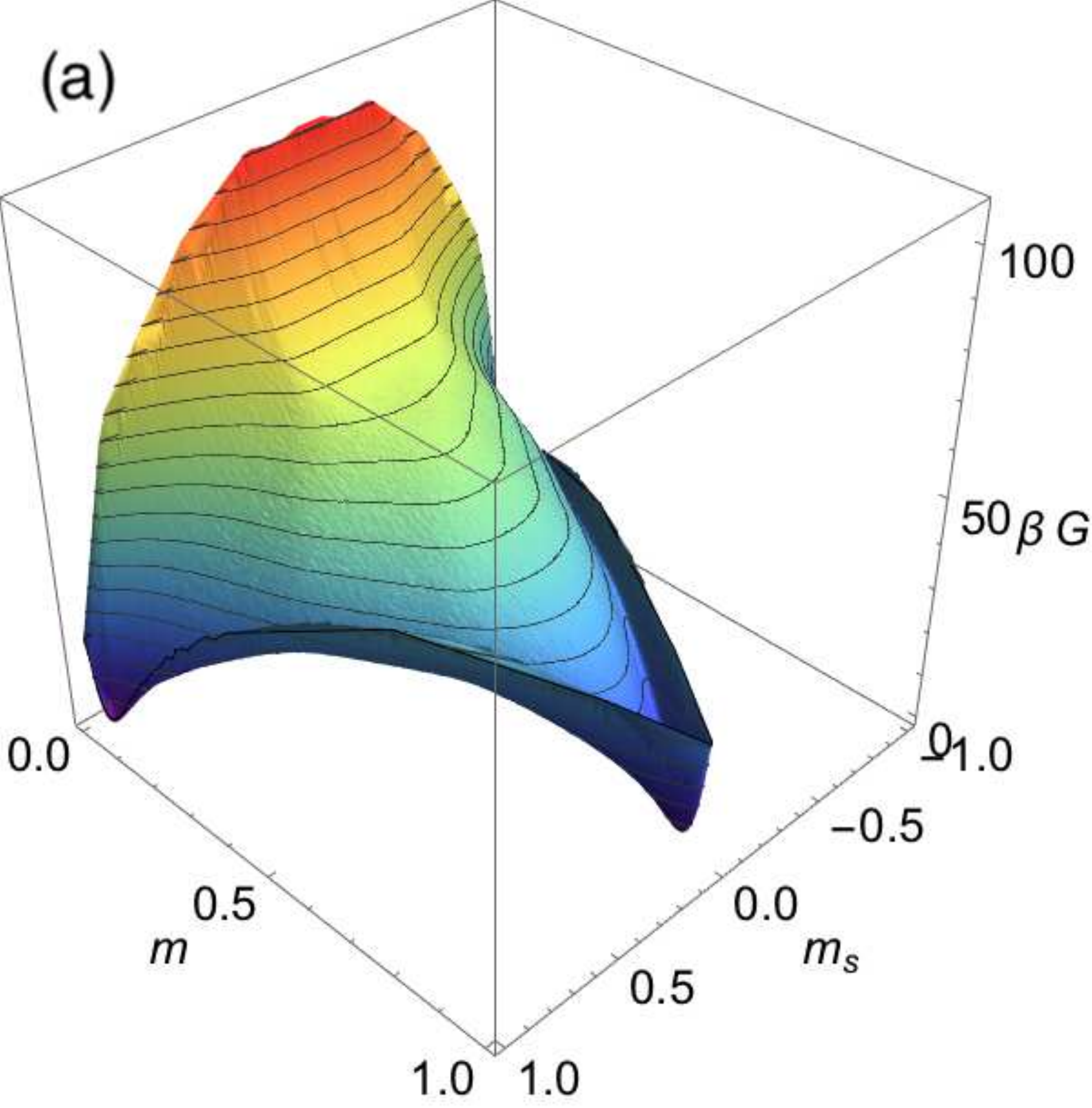}\\
\vspace{0.8cm}
\includegraphics[scale=0.4]{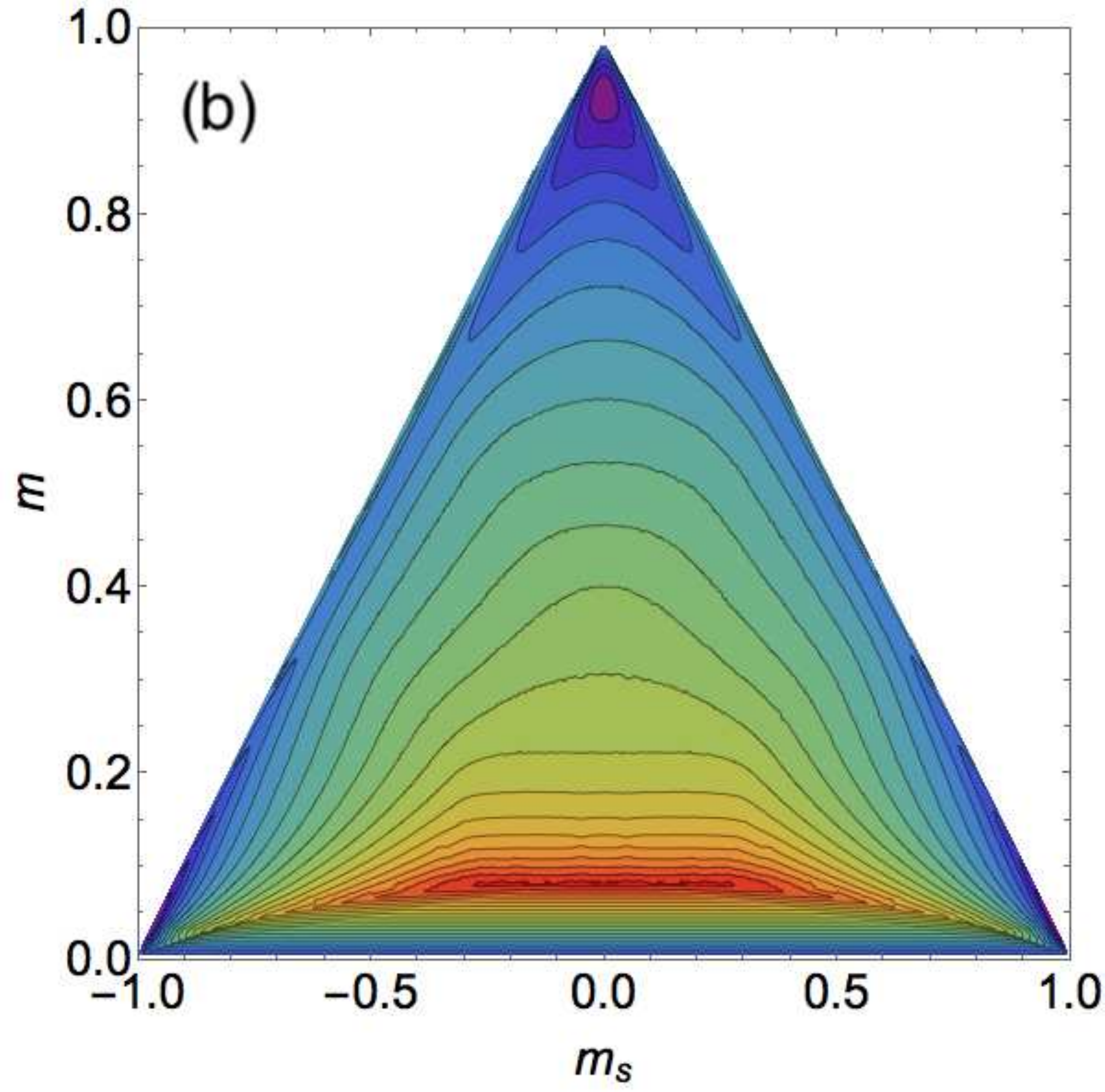}
\caption{$G(m_s,m)$ at the triple point for a system with  $L=64$.  Contours are $5kT$ apart.  Panel (a) is a surface plot of $G(m_s,m)$ and panel (b) is a contour plot of the same data.}
\label{GTP}
\end{figure}

We estimate $P(m_s,m)$ from umbrella sampling simulations using a biasing potential,
\begin{equation}
U_B'=\kappa_s(m_s-m_s^*)^2 + \kappa_m(m-m^*)^2,
\end{equation}
where $m_s^*$ and $m^*$ specify the target values of $m_s$ and $m$ to be sampled, and 
$\kappa_s$ and $\kappa_m$ control the range of sampling around the target values.
All our umbrella sampling simulations using $U_B'$ are carried out 
with a system size of $L=64$ and at the state point $(H_s^0, H^0, kT/J)=(0,3.9875,1)$.
As we will see below, this point is on the $\ca\cc$ coexistence line and is very close to the $\ca\cb\cc$ triple point.  Since this point is on the $H_s=0$ axis, we only need to compute $P(m_s,m)$ for $0<m_s<1$ because $P(m_s,m)=P(-m_s,m)$ under these conditions.  Also, since $m_s+m\le1$, the range of $P(m_s,m)$ is further restricted to the triangle-shaped region bounded by $m_s=0$, $m=0$, and $m_s+m=1$.  We cover this region using 903 umbrella windows with $m_s^*=100i/L^2$ where $i$ is an integer in the range $[0, 41]$, and $m^*=100j/L^2$ where $j$ is an integer in the range $[0, 41-i]$.  We choose $\kappa_s=\kappa_m=0.0005L^4J$.  Trial configurations are accepted or rejected using the umbrella potential every 1~MCS.  
In all of our simulations, one MCS corresponds to $L^2$ attempts to flip the spin of a randomly chosen lattice site.
The simulation for each umbrella window is initialized using a perfect $\cc$ configuration.
Each run is equilibrated for $2\times 10^5$~MCS, after which $m_s$ and $m$ are recorded every 400~MCS for the next $4\times 10^6$~MCS.  

The time series of $m_s$ and $m$ for all umbrella simulations are analyzed and combined using WHAM~\cite{Kumar:1992,Grossfield:2018} to generate estimates of $P(m_s,m)$ and $G(m_s,m)$ at $(H_s^0, H^0, kT/J)=(0,3.9875,1)$.
For the WHAM analysis, we exclude every run with an umbrella sampling acceptance rate of less than 0.04.  This reduces the number of windows analyzed to 819.  The windows excluded are:  
$j=0$ and $0\le i \le 33$; 
$j=1$ and $0\le i \le 28$; and
$j=2$ and $0\le i \le 20$.  These windows correspond to regions in which $G(m_s,m)$ is very large and steep, and which make a negligible contribution to the average properties of equilibrium states near the triple point.
We estimate that the error in our computed values for $G(m_s,m)$ is at most $1kT$.  

$P(m_s,m)$ provides the complete density of states as a function of $m_s$ and $m$, and can be used to find $P(m_s,m)$ or $G(m_s,m)$ 
at nearby values of $(H_s, H)$ 
by reweighting according to,
\begin{equation}
G(m_s,m; H_s,H) = G(m_s,m; H_s^0,H^0) - N(H_s-H_s^0)m_s - N(H-H^0)m.
\end{equation}
Having reweighted $G(m_s,m)$ to new values of $(H_s, H)$, we can also obtain 
$G(m_s)$ and $G(m)$ at the same $(H_s, H)$.

\begin{figure}
\includegraphics[scale=0.4]{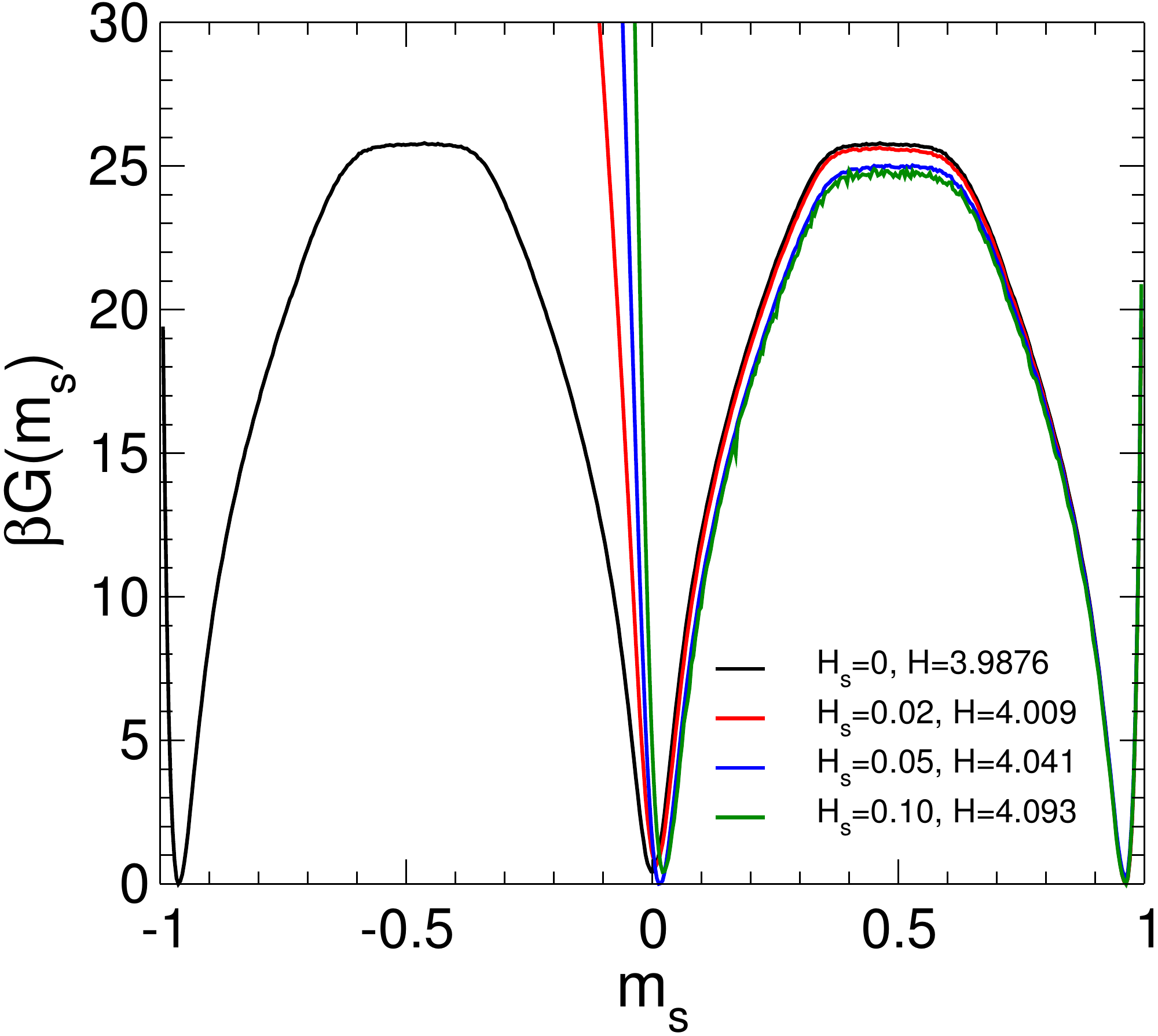}
\caption{$G(m_s)$ along the $\cb\cc$ coexistence curve for an $L=64$ system.} 
\label{sigmaUC}
\end{figure}

Due to the symmetry of the system Hamiltonian, the $\ca\cc$ coexistence line lies on the $H_s=0$ axis and so
the value of $H_s$ at the $\ca\cb\cc$
triple point is $H_s^T=0$.
At the triple point, $G(m_s,m)$ will exhibit three basins of approximately equal depth, one for each of the phases $\ca$, $\cb$ and $\cc$.  We further note that when $H>0$ and the $\bar\cb$ phase can be ignored, $m_s$ by itself serves as an order parameter to distinguish each phase, since $m_s\simeq -1$ in $\ca$, 
$m_s\simeq 0$ in $\cb$, and 
$m_s\simeq 1$ in $\cc$.
At the triple point, $G(m_s)$ 
will therefore also display three minima of approximately equal depth.  In Fig.~\ref{sigmaCC}(a) we show $G(m_s)$ at $H_s=0$ for several $H$ approaching the triple point, where we observe the emergence of these three minima.  To precisely locate
$H^T$, the value of $H$ at the triple point, we
evaluate $P(m_s)$ 
at several values of $H$ and seek conditions where the areas under the three peaks in 
$P(m_s)$ 
corresponding to each phase are equal~\cite{Borgs:1990ig,Binder:2008p5313}.  
We find $H^T=3.9876\pm0.0005$.
Fig.~\ref{sigmaCC}
shows $G(m_s)$ and $G(m)$
at the triple point, and 
Fig.~\ref{GTP} shows $G(m_s,m)$ 
at the triple point.

We locate points on the $\cb\cc$ coexistence curve by examining the behavior of $G(m_s)$ and $P(m_s)$ at several fixed values of  $H_s>0$.  For a given value of $H_s$, we seek the value of $H$ at which the areas under the peaks in $P(m_s)$ for the $\cb$ and $\cc$ phases are equal.  Fig.~\ref{sigmaUC} shows $G(m_s)$ at several points on the $\cb\cc$ coexistence curve determined in this way, confirming that under these conditions the minima for the $\cb$ and $\cc$ phases are of approximately equal depth.  
The result for the $\cb\cc$ coexistence curve is shown in Fig.~\ref{pd1}, for which the statistical error in $H$ is $0.0005$.  The $\ca\cb$ coexistence curve in Fig.~\ref{pd1} is  simply the reflection of the $\cb\cc$ coexistence curve about the $H_s=0$ axis.

\section{Limit of metastability of the $\cb$ phase}
\label{lom}

As shown in Fig.~\ref{sigmaCC}(b), at $H_s=0$ a local minimum corresponding to the $\cb$ phase occurs in $G(m)$ in the vicinity of $m\simeq 0.95$.  This minimum persists even for values of $H$ outside of the stability field of $\cb$, i.e. for values of $H$ below the triple point.  Under these conditions, this minimum of $G(m)$ corresponds to the metastable bulk $\cb$ phase.  As $H$ decreases further this minimum disappears, thus defining the limit of metastability (LOM) of the bulk $\cb$ phase.  We locate the LOM of the $\cb$ phase in the phase diagram
by seeking the value of $H$ at which the local minimum for $\cb$ in $G(m)$ 
disappears with decreasing $H$ at several fixed values of $H_s$.  The result for a system of size $L=64$ is shown in Fig.~\ref{pd1}.  Note the LOM is system-size dependent.  As $L\to \infty$, the LOM approaches the $\cb\cc$ and $\ca\cb$ coexistence curves~\cite{Binder:2012ey}.

\section{Chemical potential of bulk phases and metastable phase boundaries}
\label{mu}

\begin{figure}
\includegraphics[scale=0.4]{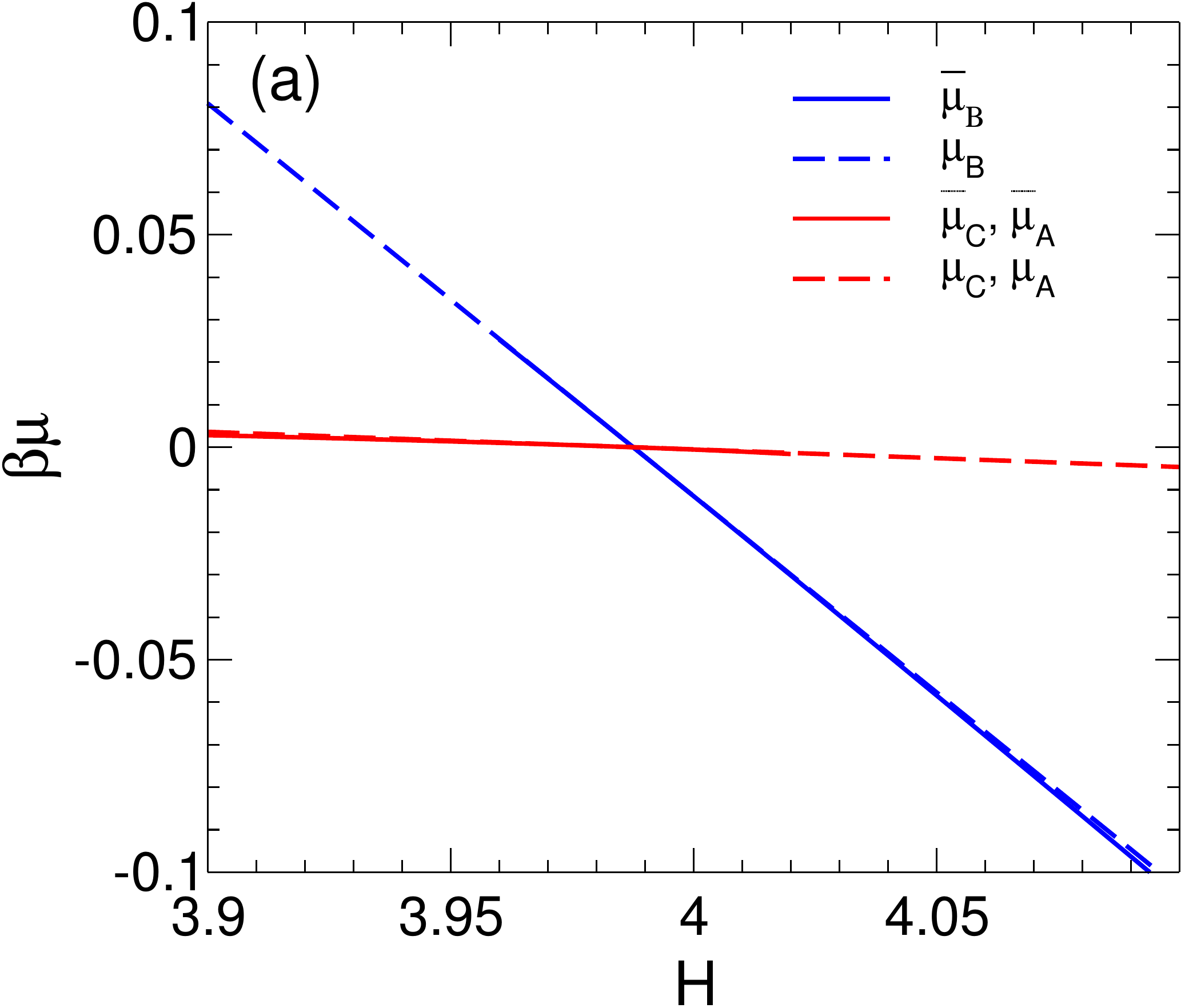}\\
\vspace{0.8cm}
\includegraphics[scale=0.4]{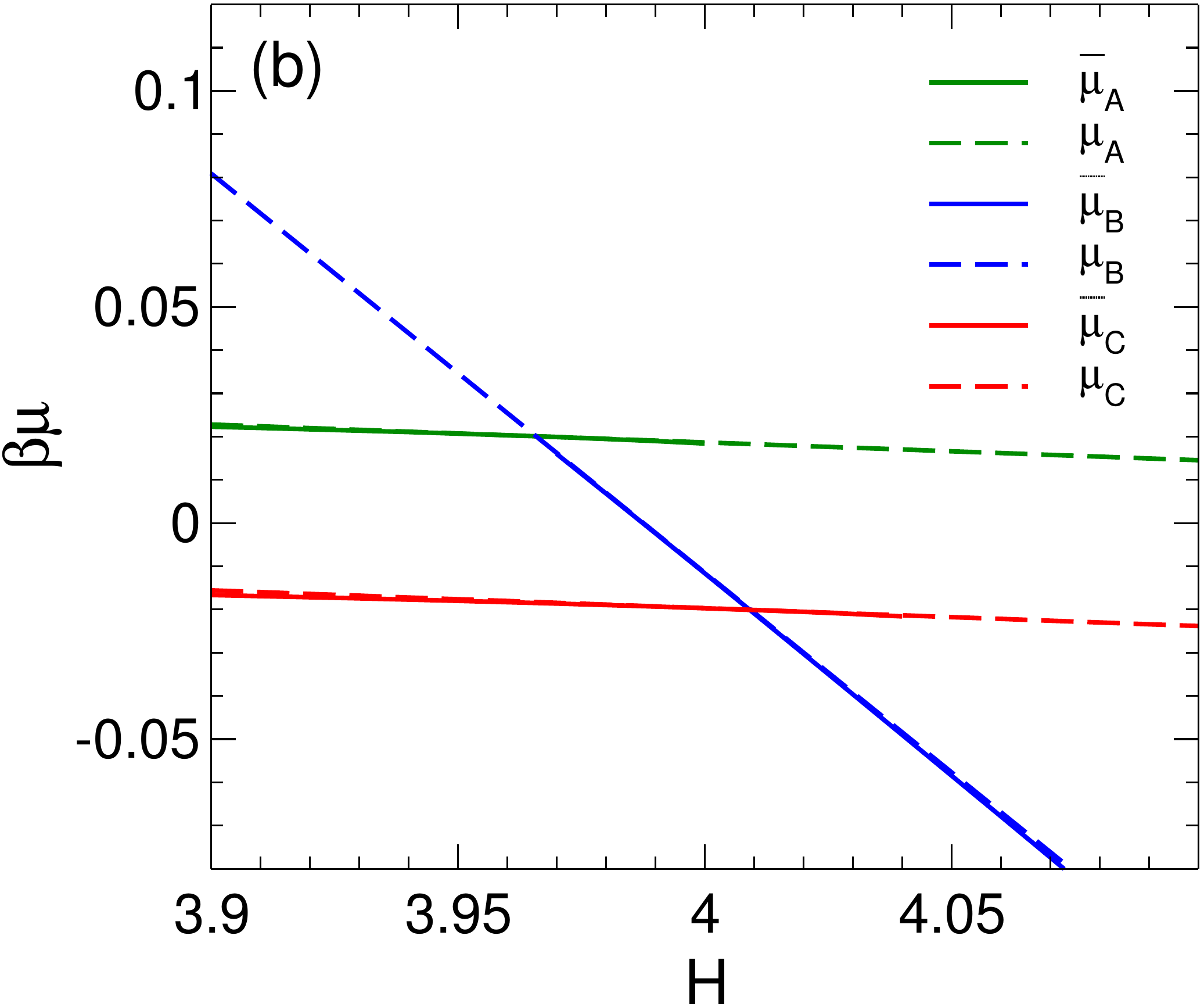}
\caption{
Chemical potentials of bulk phases as a function of $H$ at fixed $H_s$.  Solid lines end at the limit of metastability for the bulk phase in a system with $L=64$. In (a) $H_s=0$, and in (b) $H_s=0.02$.  Note that when $H_s=0$, $\mu_\ca=\mu_\cc$.}
\label{GH}
\end{figure}

We estimate the chemical potential of each bulk phase at a given value of $(H_s,H)$, relative to its value at the triple point, using,

\begin{eqnarray}
\bar \mu_{\ca}(H_s,H)&=&-\frac{kT}{N}\log \Biggl( \int_{-1}^{-0.9} dm_s \int_0^1 dm\, \exp[-\beta G(m_s,m; H_s,H)] \Biggr) - \bar \mu_\ca^0 \\
\bar \mu_{\cb}(H_s,H)&=&-\frac{kT}{N}\log \Biggl( \int_{-1}^1 dm_s \int_{0.9}^1 dm\, \exp[-\beta G(m_s,m; H_s,H)] \Biggr) - \bar \mu_\cb^0\\
\bar \mu_{\cc}(H_s,H)&=&-\frac{kT}{N}\log \Biggl( \int_{0.9}^1 dm_s \int_0^1 dm\, \exp[-\beta G(m_s,m; H_s,H)] \Biggr) -  \bar \mu_\cc^0,
\end{eqnarray}
where $\bar \mu_\ca^0$ 
is chosen such that $\bar \mu_{\ca}(H_s^T,H^T)=0$, and similarly for $\bar \mu_\cb^0$ and $\bar\mu_\cc^0$.  The integrals in the above relations sum over a region of the $(m_s,m)$ plane which encompasses the minimum of the basin for the corresponding phase, and in which the system is a single homogeneous phase.  In Fig.~\ref{GH} we plot $\bar \mu_{\ca}$, $\bar \mu_{\cb}$ and $\bar \mu_{\cc}$ as a function of $H$ at $H_s=0$ and $H_s=0.02$.  These lines terminate at the value of $H$ for the LOM of each phase for our $L=64$ system, i.e. where the local minimum in $G(m_s,m; H_s,H)$ ceases to exist.

Although the chemical potential is always well defined within the stability field of each phase, for a metastable bulk phase it is only defined when a local minimum is observed in $G(m_s,m; H_s,H)$.  However, for the purpose of analyzing the nucleation behavior predicted by Eq.~\ref{simple}, it would be useful to have an approximate way to assign a value to the chemical potential for a phase that is beyond its LOM.  We find that it is possible to do so in our lattice model because the dependence of $\bar\mu$ on $H_s$ and $H$ is very close to linear for all three phases; see Fig.~\ref{GH}.
As a simple approximation, we therefore model the chemical potential for each phase, relative to the triple point, using the expressions:
\begin{eqnarray}
\mu_{\ca}(H_s,H)&=& - (H_s-H_s^T)m_s^{\ca} - (H-H^T)m^{\ca} \label{simpleG1}\\
\mu_{\cb}(H_s,H)&=& - (H_s-H_s^T)m_s^{\cb} - (H-H^T)m^{\cb} \label{simpleG2}\\
\mu_{\cc}(H_s,H)&=& - (H_s-H_s^T)m_s^{\cc} - (H-H^T)m^{\cc}.\label{simpleG3}
\end{eqnarray}
In these relations, we use the value of $m_s$ and $m$ for each phase at the triple point to fix the rate of change of $\mu$ with $H_s$ or $H$, since 
$m_s=-(\partial \mu/\partial H_s)_{H,T}$ and
$m=-(\partial \mu/\partial H)_{H_s,T}$.
At the triple point, 
the average value of $m_s$ and $m$ 
for the $\ca$ phase is
$m_s^{\ca}=-0.959$ and $m^{\ca}=0.0413$; 
for the $\cb$ phase
is $m_s^{\cb}=0$ and $m^{\cb}=0.924$; and
for the $\cc$ phase 
is $m_s^{\cc}=-m_s^\ca$ and $m^{\cc}=m^\ca$.
We compare $\bar\mu$ and $\mu$ for each phase in Fig.~\ref{GH}, and find that these two approaches give nearly indistinguishable results.
The metastable extensions of the coexistence boundaries shown in Fig.~\ref{pd1} are estimated by finding the intersection of the surfaces defined in Eqs.~\ref{simpleG1}-\ref{simpleG3}.

Note that since $m^{\cc}=m^\ca$, then $\Delta \mu_{\ca\cc}=\mu_\cc-\mu_\ca$ does not depend on $H$ and is constant at fixed $H_s$.  In 
contrast, $\Delta \mu_{\ca\cb}=\mu_\cb-\mu_\ca$ decreases linearly with $H$ at fixed $H_s$.
These observations are relevant for understanding the behavior of the CNT estimates for $n^*_{\ca\cb}$  and $n^*_{\ca\cc}$ 
plotted in Fig.~\ref{nonCNT}.

\section{Surface tension}
\label{st}

\begin{figure}
\includegraphics[scale=0.4]{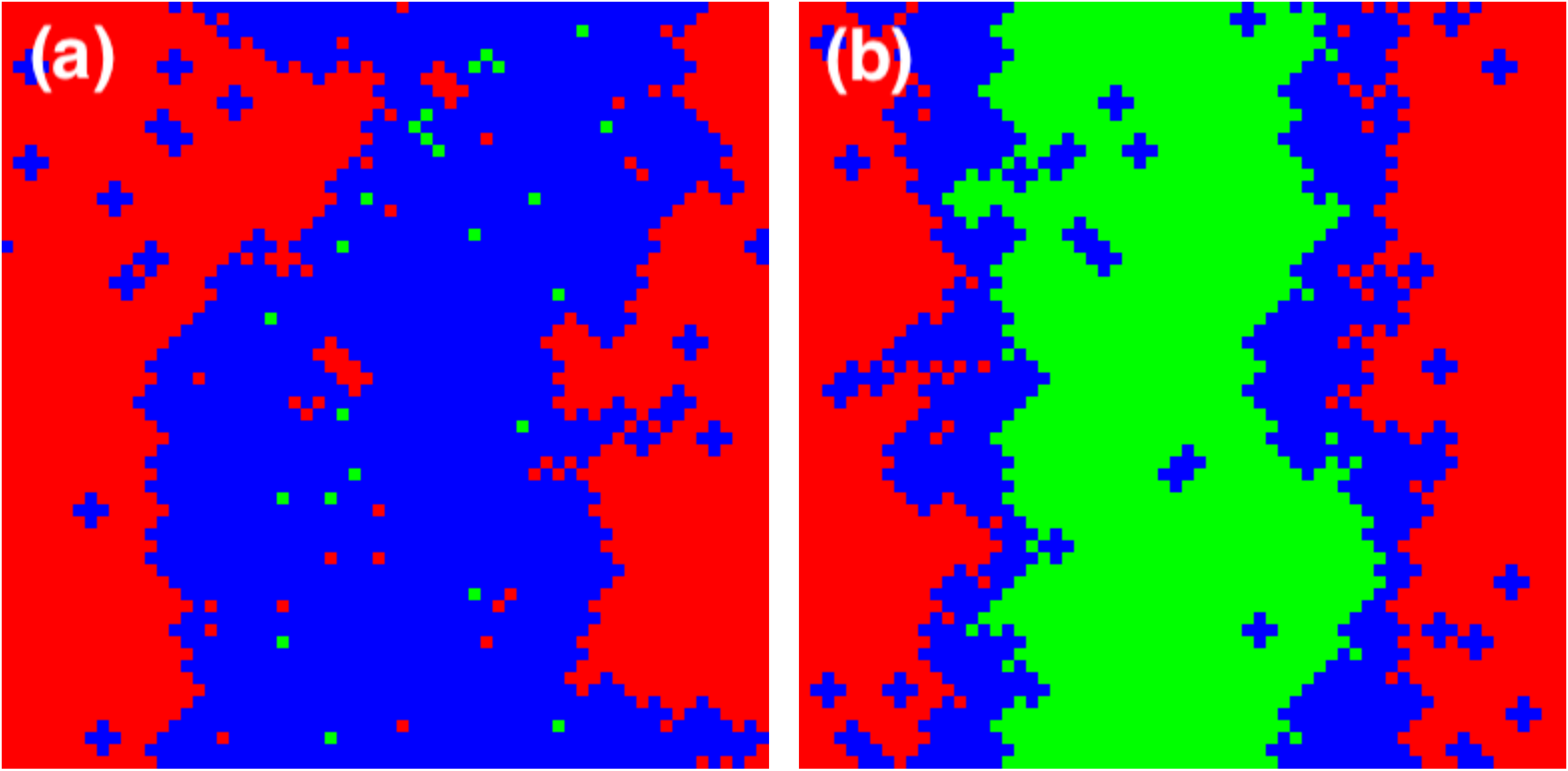}
\caption{
(a) Snapshot of $\cb \cc$ coexistence at $H_s=0$ and $H=3.987$ in a system with $m_s\simeq 0.5$.
(b) Snapshot of $\ca\cc$ coexistence at $H_s=0$ and $H=3.96$ in a system with $m_s\simeq 0$.  $L=64$ for both (a) and (b).}
\label{pics1}
\end{figure}

The surface tension (or interfacial free energy) $\sigma$ between two coexisting phases can be estimated from the height of the free energy barrier that separates the two successive minima corresponding to the coexisting phases in a plot of $G(m_s)$ or $G(m)$~\cite{Binder:2008p5313,Binder:2011du,Binder:2012ey}.
As shown in Figs.~\ref{sigmaCC} and \ref{sigmaUC}, the top of this barrier is flat for a system in which two phases coexist, indicating the range of $m_s$ or $m$ values where two flat interfaces separate the two phases in our periodic system.  Example snapshots of coexisting phases in our $L=64$ simulations are shown in Fig.~\ref{pics1}.

We define the interfacial free energy $\sigma$ such that the height of the barrier in $\beta G(m_s)$ or $\beta G(m)$ 
is $2L\beta \sigma$; the factor of $2L$ accounts for the two interfaces that occur in a system with periodic boundaries.
Fig.~\ref{sigmaUC} 
shows that for the coexistence of $\cb$ and $\cc$ in a $L=64$ system at the triple point conditions, 
$2L\beta\sigma_{\cb \cc}=25\pm 1$, 
and exhibits little observable variation along the $\cb\cc$ coexistence curve in the range of $H$ and $H_s$ studied here.

\begin{figure}
\includegraphics[scale=0.4]{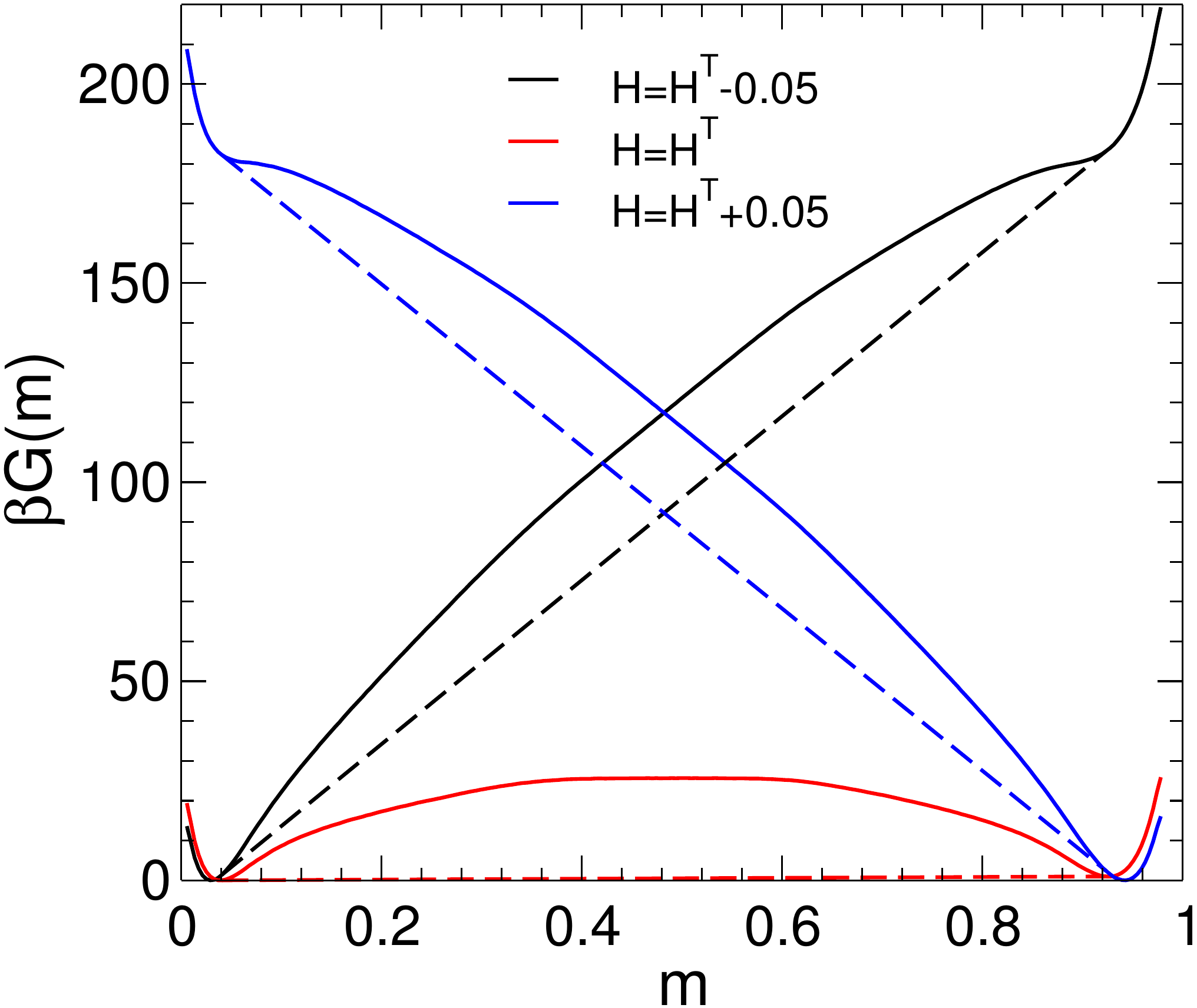}
\caption{
$G(m)$ at $H_s=0$ at various $H$ for system size $L=64$.  Dashed lines are common tangent constructions for each $G(m)$ curve.}
\label{Gsigma}
\end{figure}

In Fig.~\ref{Gsigma} we show $G(m)$ for various $H$ at $H_s=0$.  For any $H$, 
$\sigma_{\cb \cc}$ 
may be estimated from the difference between the $G(m)$ curve and a common-tangent line bounding the $G(m)$ curve from below, at a value of $m$ corresponding to a coexisting system of $\cb$ and $\cc$, e.g. $m=0.5$.  
Further, at fixed $H_s$, if $G(m)$ is computed at one value of $H=H_1$, it can be found (up to an irrelevant constant $C$) 
at a new value $H=H_2$ using,
\begin{equation}
G(m;H_2)=G(m;H_1)-N(H_2-H_1)m+C.
\label{g12}
\end{equation}
As a consequence of the form of Eq.~\ref{g12}, 
$\sigma_{\cb \cc}$ 
is independent of $H$ at fixed $H_s$.  Also, since we have observed that $\sigma_{\cb \cc}$ varies little on the $\cb\cc$ coexistence curve (Fig.~\ref{sigmaUC}), along which $H_s$ is changing, we conclude that $\sigma_{\cb \cc}$
is approximately constant for all $H$ and $H_s$ studied here.  
Furthermore, the symmetry of the system Hamiltonian ensures that $\sigma_{\ca \cb}=\sigma_{\cb \cc}$. 
We thus use the value $2L\beta\sigma_{\ca \cb}=25$, or $\sigma_{\ca \cb}/J=0.195$ (per unit lattice site of interface) in all of our analysis.  

\begin{figure}
\includegraphics[scale=0.4]{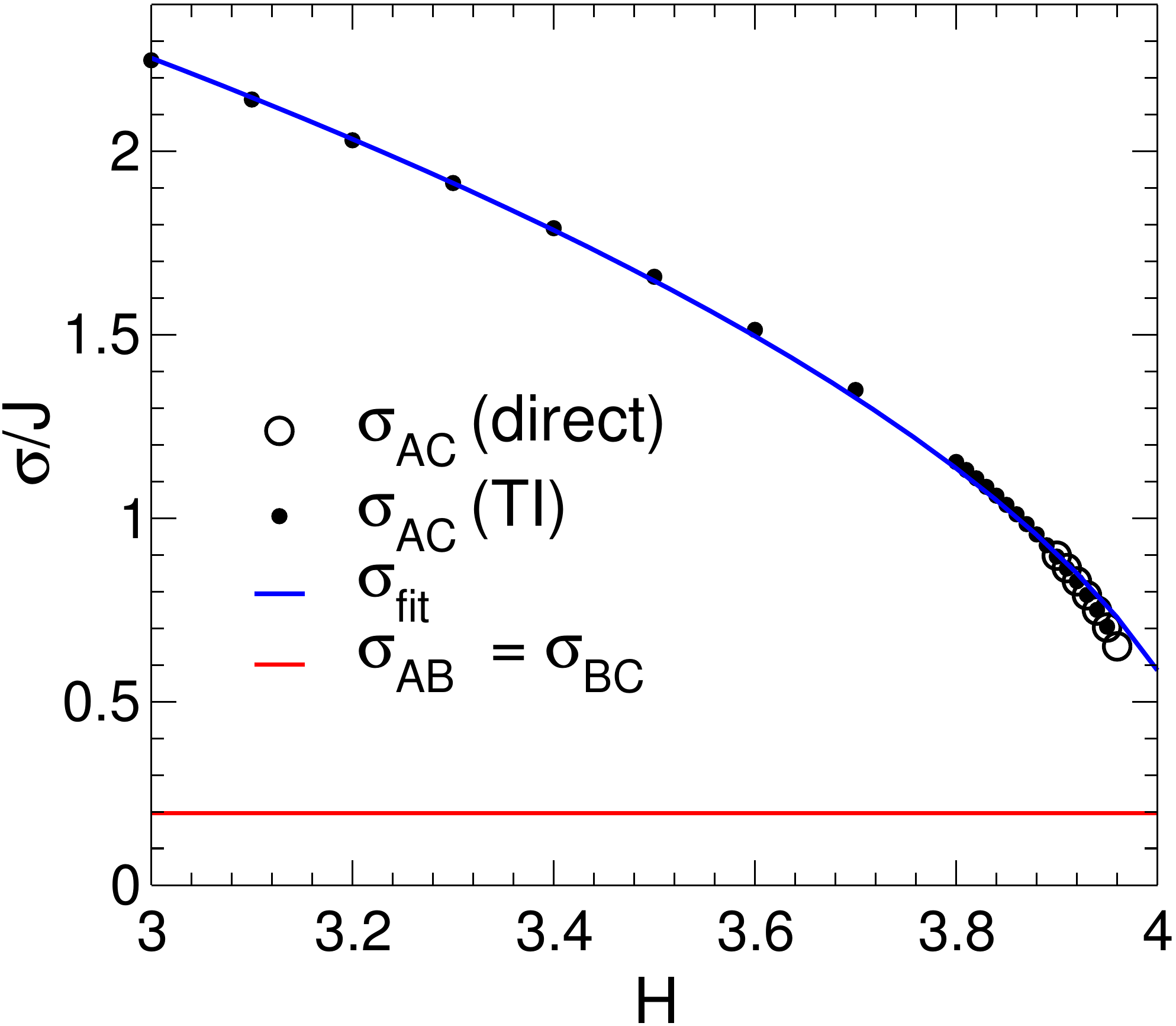}
\caption{Interfacial tensions versus $H$ at $H_s=0$ for $L=64$.
Open circles are values of $\sigma_{\ca \cc}$ found directly from the plots of $G(m_s)$ presented in Fig.~\ref{sigmaCC}(a).
Filled circles are values of $\sigma_{\ca \cc}$ found by thermodynamic integration (TI) using Eq.~\ref{td}.
The blue curve is $\sigma_{\rm fit}/J=(19.299-4.739H)^{1/2}$, fitted to data obtained via TI for $H=2$ to $H=4$.
The red curve gives the value of $\sigma_{\ca \cb}=\sigma_{\cb \cc}$.}
\label{sigma}
\end{figure}

We next estimate $\sigma_{\ca \cc}$ 
as a function of $H$ along the $H_s=0$ axis for $H<H_T$.
In Fig.~\ref{sigmaCC} 
we plot  $G(m_s)$ on the $\ca \cc$ coexistence curve at $H=3.94$, a point at which the $\cb$ phase is unstable for a system of size $L=64$.  We find $2L\beta\sigma_{\cal C \bar C}=96\pm 1$.  We also find $\sigma_{\ca \cc}$ for other values of $H$ from the $G(m_s)$ curves in Fig.~\ref{sigmaCC} that are flat near $m_s=0$, and plot the results in Fig.~\ref{sigma}.  

To estimate $\sigma_{\ca \cc}$ 
as a function of $H$ along the entire ${\ca \cc}$ coexistence curve we use,
\begin{equation}
2L\sigma_{\ca \cc}(H) = 2L\sigma_{\ca \cc}(H_0) + \Delta G_{\rm coex}(H) - \Delta G_\ca(H),
\label{td}
\end{equation}
where,
\begin{equation}
\Delta G_{\rm coex}(H) = -N\int_{H_0}^H m_{\rm coex}(H')\, dH',
\label{TI1}
\end{equation}
and,
\begin{equation}
\Delta G_\ca(H) = -N\int_{H_0}^H m_\ca(H')\, dH'.
\label{TI}
\end{equation}
That is, we choose as a reference value $H_0=3.94$ where we already know $\sigma_{\ca \cc}$.  
We then use thermodynamic integration to estimate the change in the interfacial free energy as we move the system to a different value of $H$.
$\Delta G_{\rm coex}(H)$ estimates the free energy of a system containing coexisting $\ca$ and $\cc$ phases, relative to its value at $H_0$.  To estimate $m_{\rm coex}(H)$, we evaluate $m$ as a function of $H$ for a coexisting system of $\ca$ and $\cc$ phases along the $H_s=0$ coexistence curve.  We constrain this coexisting system to remain within the range of $m_s$ consistent with the occurrence of a pair of ${\ca\cc}$ interfaces by applying a simple square-well biasing potential that prevents the system from sampling microstates with $|m_s|>0.1$.
$\Delta G_{\ca}(H)$ estimates the free energy of the homogeneous $\ca$ phase, relative to its value at $H_0$.  To estimate $m_{\ca}(H)$, we evaluate $m$ as a function of $H$ for the homogeneous $\ca$ phase along the $H_s=0$ coexistence curve.  Note that this calculation is carried out on the $\ca\cc$ coexistence curve, where the free energies of the homogeneous $\ca$ and $\cc$ phases are equal.  In computing the free energy change from the homogeneous $\ca$ system to the coexisting $\ca\cc$ system, the conversion of part of the system from $\ca$ to $\cc$ therefore makes no bulk contribution to the free energy change.

\begin{figure}
\includegraphics[scale=0.28]{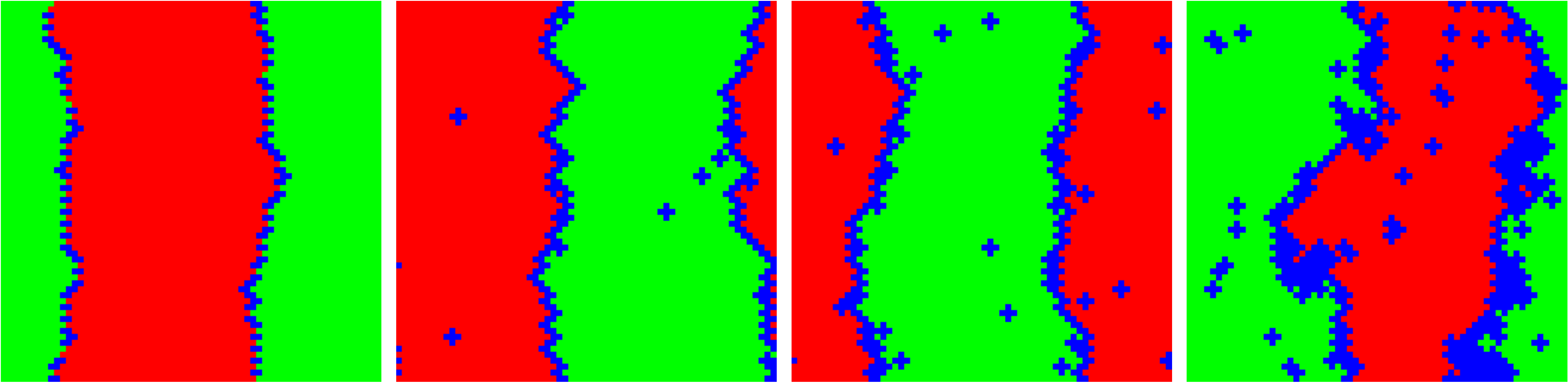}
\caption{Snapshots of $\ca\cc$ coexistence when $L=64$ and $H_s=0$ for various $H=\{1.0, 3.0, 3.5, 3.9\}$, from left to right.}
\label{sigma1}
\end{figure}

The result for $\sigma_{\ca \cc}$ 
is shown in Fig.~\ref{sigma}.  We show snapshots of the coexisting $\ca$ and $\cc$ phases at different values of $H$ in Fig.~\ref{sigma1}.
We note the complexity of the $\ca\cc$ interface.
Depending on $H$ the interface may contain a significant wetting layer of $\cb$ between the $\ca$ and $\cc$ regions.
Approaching the triple point
$\sigma_{\ca \cc}$ decreases but remains more than twice the value of $\sigma_{\ca \cb}$
 at the triple point.  Given the emergence of the wetting layer of $\cb$ as $H\to H^T$, the behavior of $\sigma_{\ca \cc}$ 
 makes sense:  In this regime the ${\ca \cc}$ interface can be approximated as the superposition two interfaces, one $\ca\cb$ and the other ${\cb\cc}$.  Since 
$\sigma_{\ca \cb}= \sigma_{\cb \cc}$, it therefore seems likely that the condition 
$\sigma_{\ca \cc} \ge 2\sigma_{\ca \cb}$ 
holds under all conditions studied here.

In order to compare the behavior of our lattice model to the predictions of CNT, it is useful to have an analytic model of the dependence of $\sigma_{\ca \cc}$ 
on $H_s$ and $H$ throughout the phase diagram.
By an argument analogous to that used above to establish that $\sigma_{\cb \cc}$ is independent of $H$ at constant $H_s$ (see Eq.~\ref{g12}), it can be shown that $\sigma_{\ca \cc}$ 
is independent of $H_s$ at fixed $H$.  To model the dependence of $\sigma_{\ca \cc}$ on $H$, we notice empirically that $\sigma_{\ca \cc}^2$ is approximately linear in $H$ between $H=2$ and $4$.   We fit a straight line to our data for $\sigma_{\ca \cc}^2$ in this range and obtain 
$\sigma_{\rm fit}/J=(19.299-4.739H)^{1/2}$, shown in Fig.~\ref{sigma}.  We use $\sigma_{\rm fit}$ to compute the CNT estimate of $n^*_{\ca\cc}$ plotted in Fig.~\ref{nonCNT}.



We note that our quantitative estimates for $\sigma$ should be considered preliminary.  All of our estimates for $\sigma$ are based on square systems with $L=64$, and assume an interface that is, on average, flat and oriented parallel to a lattice axis. 
A more detailed and accurate analysis is possible by monitoring system-size effects, the influence of the system shape and boundary conditions, as well as considering the influence of the orientation of the interface to the lattice axes~\cite{Binder:2011du}.  
That said, for the purposes of this work, it is sufficient that we have shown that $\sigma_{\ca\cb}<\sigma_{\ca\cc}$ throughout the range of the phase diagram studied here.

\section{Identifying local fluctuations occurring within the $\ca$ phase}
\label{viz}

Here we focus on the $\ca$ phase, and develop a definition for identifying local regions that deviate in structure from that expected in the $\ca$ phase.  

\begin{figure}
\includegraphics[scale=0.28]{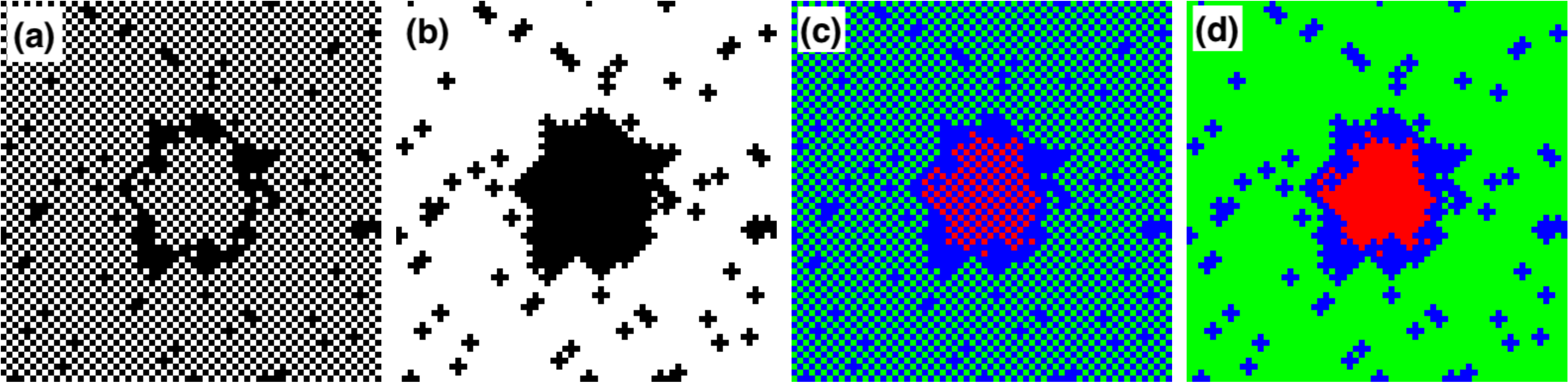}
\caption{Example $L=64$ system configuration.  
(a) White sites satisfy $s_i=-1$.
Black sites satisfy $s_i=1$.
(b) Black sites satisfy the definition for belonging to a fluctuation.  White sites do not.
(c) Green sites satisfy $s_i=-1=-\sigma_i$.  
Blue sites satisfy $s_i=1$.  
Red sites satisfy $s_i=-1=\sigma_i$.
(d) Same as (c), except that all blue sites in (c) totally surrounded by green sites are rendered as green in (d);
and all blue sites in (c) totally surrounded by red sites are rendered as red in (d).}
\label{op}
\end{figure}

In the perfect $\ca$ phase, all sites satisfy $s_i=-\sigma_i$.  
In the perfect $\cc$ phase, all sites satisfy $s_i=\sigma_i$.  
In the perfect $\cb$ phase, all sites satisfy $s_i=1$.  
We therefore define a local fluctuation occurring within the $\ca$ phase as any contiguous cluster of sites for which 
$s_i=\sigma_i$ or $s_i=1$. 
The one exception to this definition is
a single site at which $s_i=1$ and for which all 4 nn's have $s_i=-1$.  Half of the sites in the perfect $\ca$ phase have this property, and we exclude them from our definition of a fluctuation.
An example system configuration is shown in Fig.~\ref{op}, and illustrates our cluster definition.

The number of sites in a cluster is denoted by $n$.  The number of sites in the largest cluster in the system is $n_{\rm max}$.  The composition of a cluster is defined as $f={\bar n}/n$, where $\bar n$ is the number of sites in the cluster that correspond to the $\cc$ phase.  We define 
$\bar n=2n_{\rm down}$, where $n_{\rm down}$ is the number of cluster sites for which $s_i=-1$.  The reason for this definition of $f$ is that cluster sites satisfying $s_i=1$
may also satisfy $s_i=\sigma_i$, and so it is ambiguous if these sites belong to the fraction of sites inside the cluster that belong to the $\cc$ phase or to the $\cb$ phase.  Since cluster sites with $s_i=-1$ unambiguously belong to the $\cc$ 
phase, and since the fraction of $s_i=-1$ sites in the perfect $\cc$ phase is $1/2$, we estimate the total number of 
$\cc$ sites within a cluster to be $2n_{\rm down}$.  The quantity $f$ therefore characterizes the cluster composition in terms of how much of the cluster is taken up by the $\cc$ phase:  $f=0$ is a pure $\cb$ cluster, while $f=1$ is a pure $\cc$ cluster.

We note that cluster sites with $s_i=1$ on the cluster perimeter are always considered part of the cluster, even though half of them (on average) might reasonably be associated with the surrounding $\ca$ phase.  
For computational efficiency, we do not apply this correction, which if implemented would decrease the values of 
$n$ and $n_{\rm max}$ from those used here.

For visualization purposes, we render system configurations as shown in Fig.~\ref{op}(d).  The rules we use to assign a color to each site are stated in the figure caption.  In the resulting color scheme, the $\ca$ phase is green, the $\cb$ phase is blue, and the $\cc$ phase is red. 

\section{2D umbrella sampling simulations to find $G(n_{\rm max},f)$}
\label{2d}

$G(\nmax,f)$ is estimated from 2D umbrella sampling simulations using the biasing potential given in Eq.~\ref{us}.
We choose $\kappa_n=0.0005$ and 
$\kappa_f=500$.  
For each choice of $(L,H_s,H,T)$
we conduct 900 simulations for 
$\nmax^*=100i$ where the integer $i\in [0,99]$, and for 
$f^*=j/10$ where the integer $j\in [0,8]$.  Each run is initiated from a perfect $\ca$ configuration, into which a seed cluster is inserted.  The seed cluster is a square of sites with $s_i=1$ of a size chosen to be closest to $\nmax^*$.  At the centre of the seed cluster there is a square region with $s_i=\sigma_i$ of a size chosen so that the seed cluster has a value of $f$ closest to $f^*$.  This system is equilibrated for $5\times 10^4$~MCS, and then the time series of $\nmax$ and $f$ 
is recorded every 100~MCS for $10^6$~MCS.  
Trial configurations are accepted or rejected using the umbrella potential every 1~MCS.
One MCS corresponds to $L^2$ attempts to flip the spin of a randomly chosen lattice site.
Our time series for $\nmax$ and $f$ 
are analyzed using WHAM to evaluate $P(n_{\rm max},f)$ and $G(n_{\rm max},f)$.  
We estimate that the error in $G(n_{\rm max},f)$ is not more than $1kT$.
We exclude from the WHAM analysis any run for which the acceptance rate for the umbrella sampling is less that $0.1$, which occurs in a few cases when the local variation of 
$G(n_{\rm max},f)$ is very steep.
Our system size for these 2D umbrella sampling runs is $L=128$ or $200$, as indicated in the legends or captions of the figures.

\begin{figure}
\includegraphics[scale=0.4]{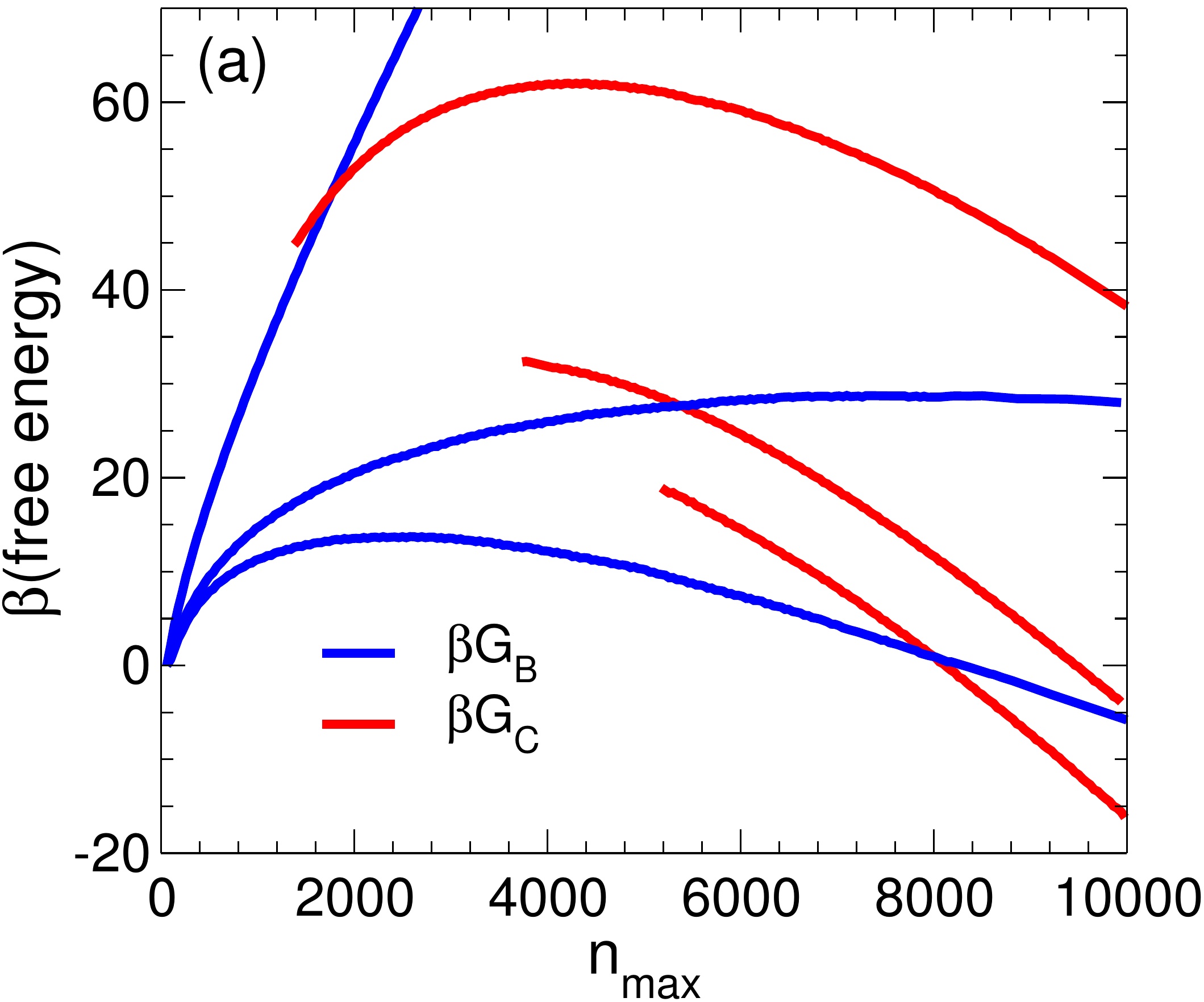}\\
\vspace{0.8cm}
\includegraphics[scale=0.4]{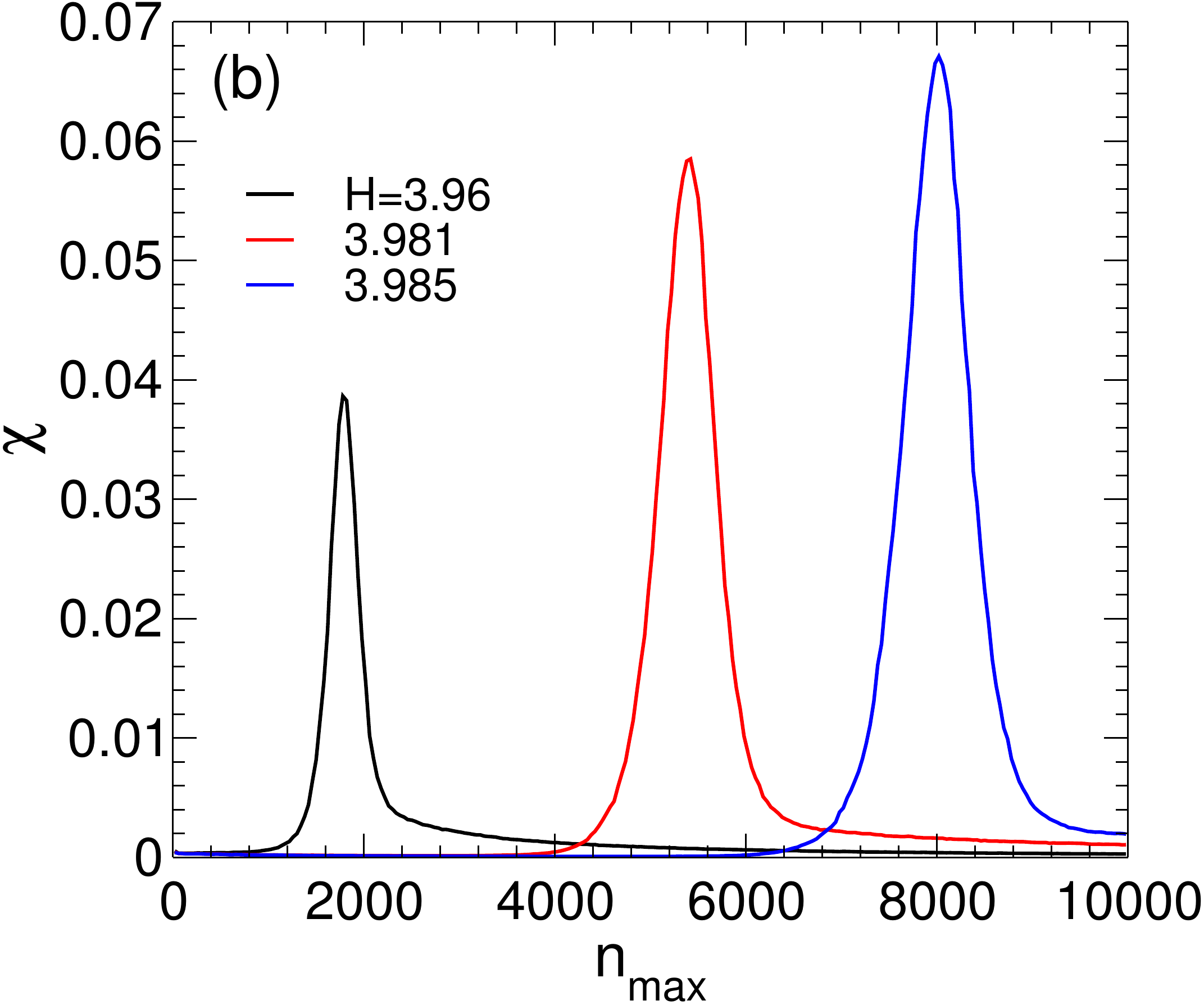}
\caption{(a) $G_\cb$ and $G_\cc$ for $H_s=0.01$ and $L=200$.  
From top to bottom $H=\{3.96,3.981,3.985\}$
(b) $\chi$ for $H_s=0.01$ and $L=200$.  
From left to right $H=\{3.96,3.981,3.985\}$.
}
\label{ncchi}
\end{figure}

\section{Finding $n_c$ from $\chi$}
\label{ncfromchi}

Fig.~\ref{ncchi}(a) reproduces the data for $G_\cb$ and $G_\cc$ from Fig.~\ref{G1D}(a), in which the crossing of these two curves identifies $n_c$.  Fig.~\ref{ncchi}(b) shows $\chi$ as a function of $\nmax$ for the same three cases shown in Fig.~\ref{ncchi}(a).  We find that the maximum of $\chi$ corresponds within error to the value of $n_c$ obtained by finding the intersection of $G_\cb$ and $G_\cc$.  
Based on this correspondence, all values of $n_c$ reported in this work are computed by finding the maximum of $\chi$.  This definition allows us to estimate $n_c$ from both 1D and 2D umbrella sampling simulations.

\begin{figure*}
\newcommand\x{0.25}
\includegraphics[scale=\x]{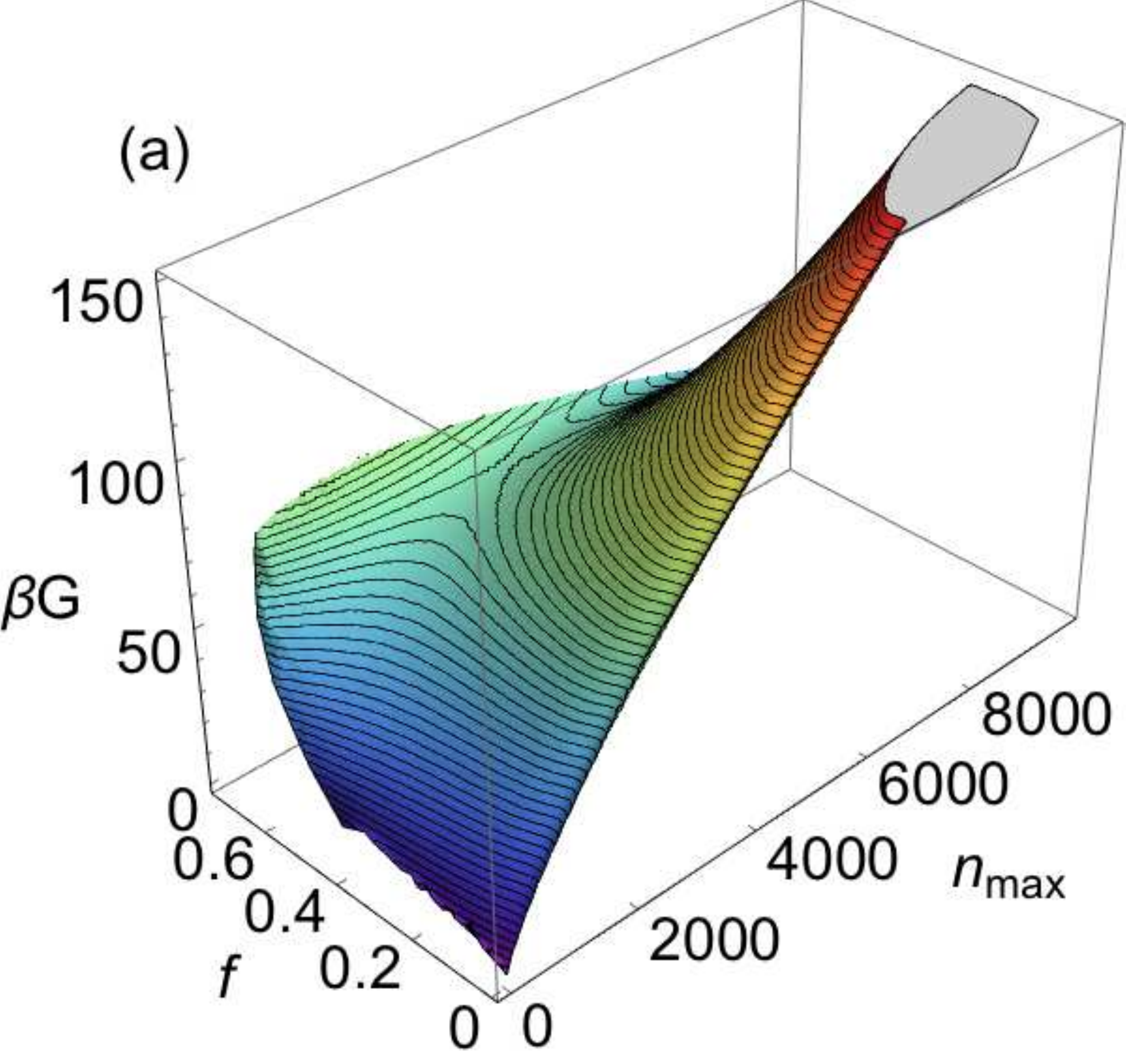}
\includegraphics[scale=\x]{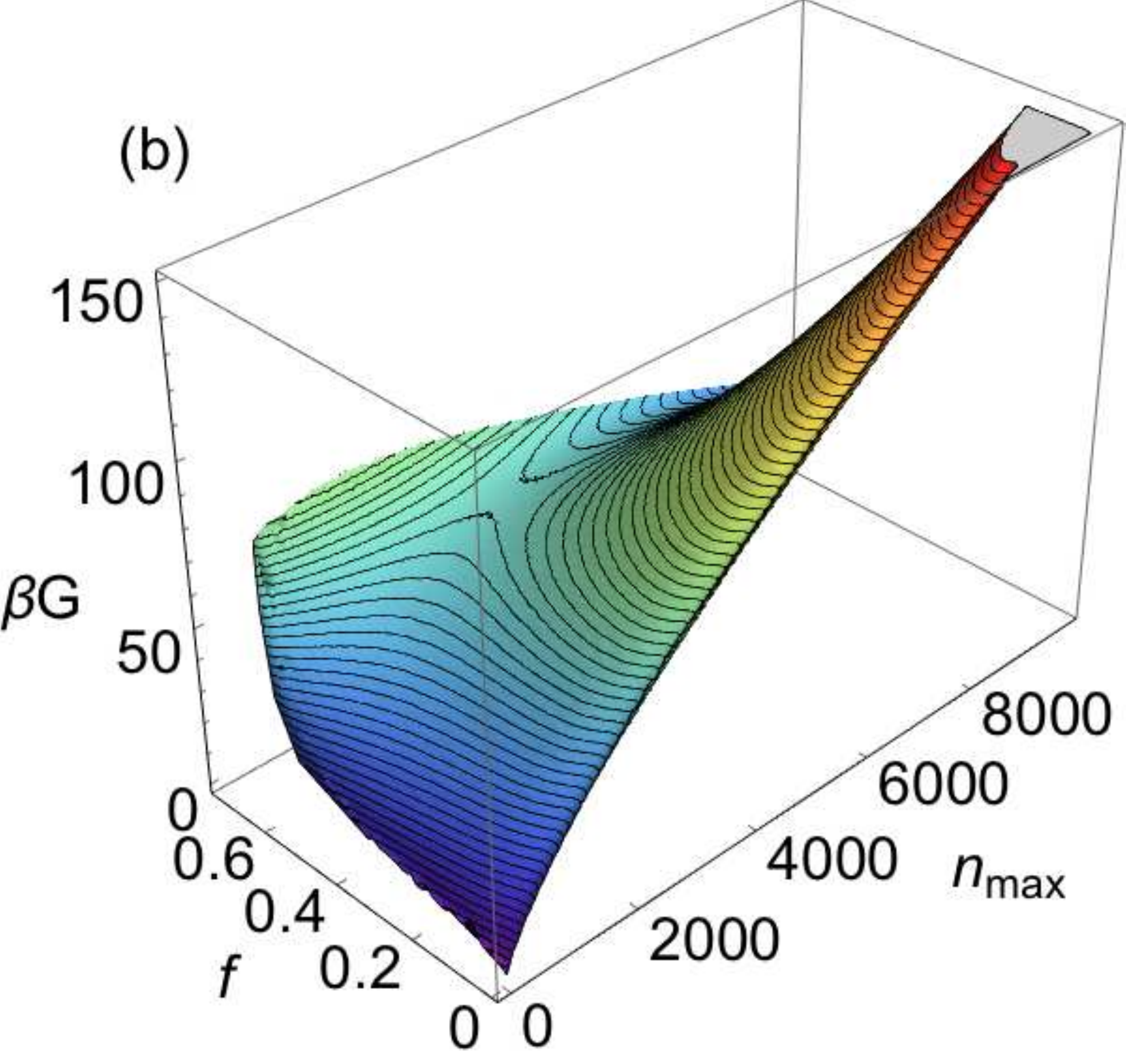}
\includegraphics[scale=\x]{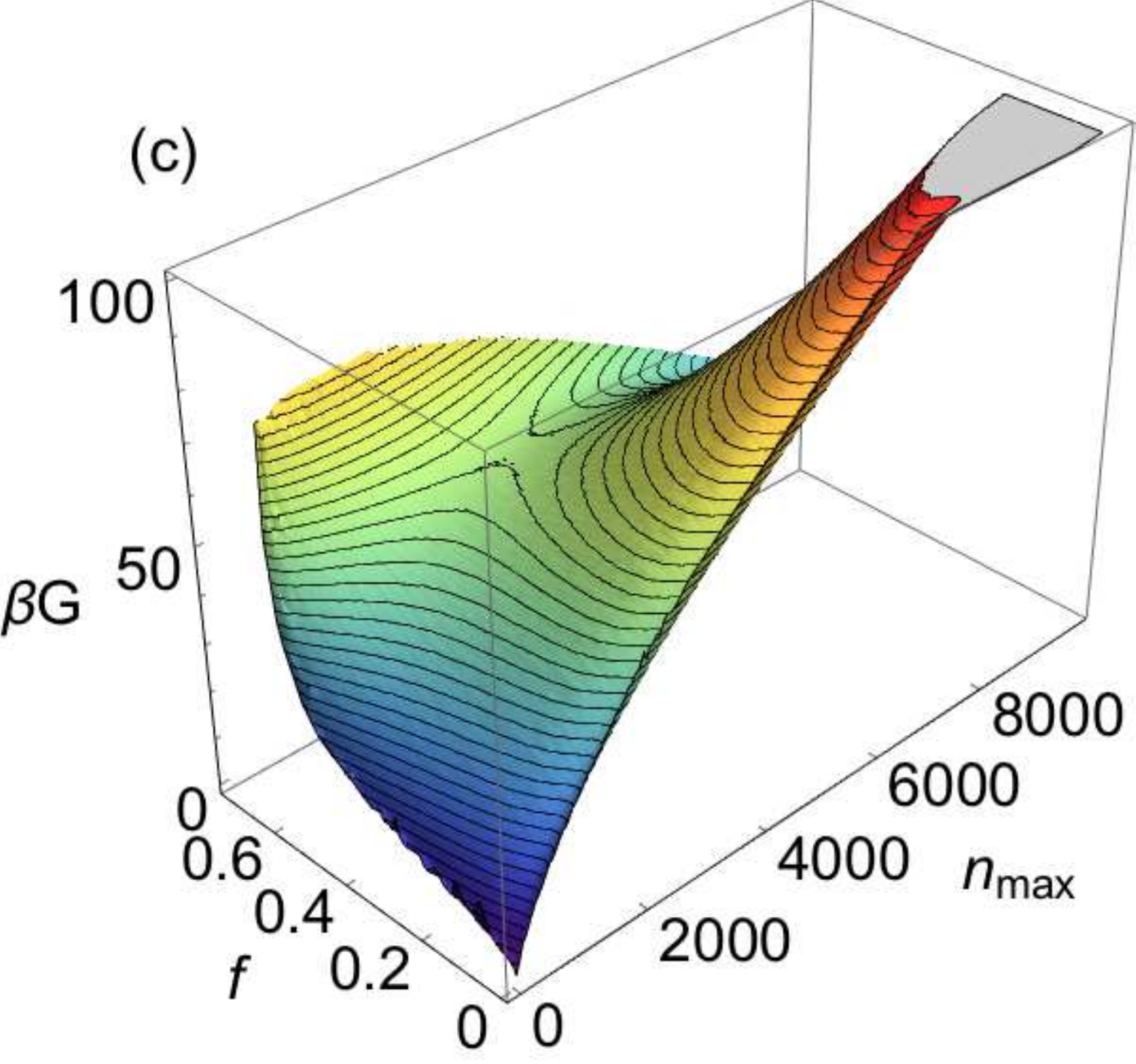}
\includegraphics[scale=\x]{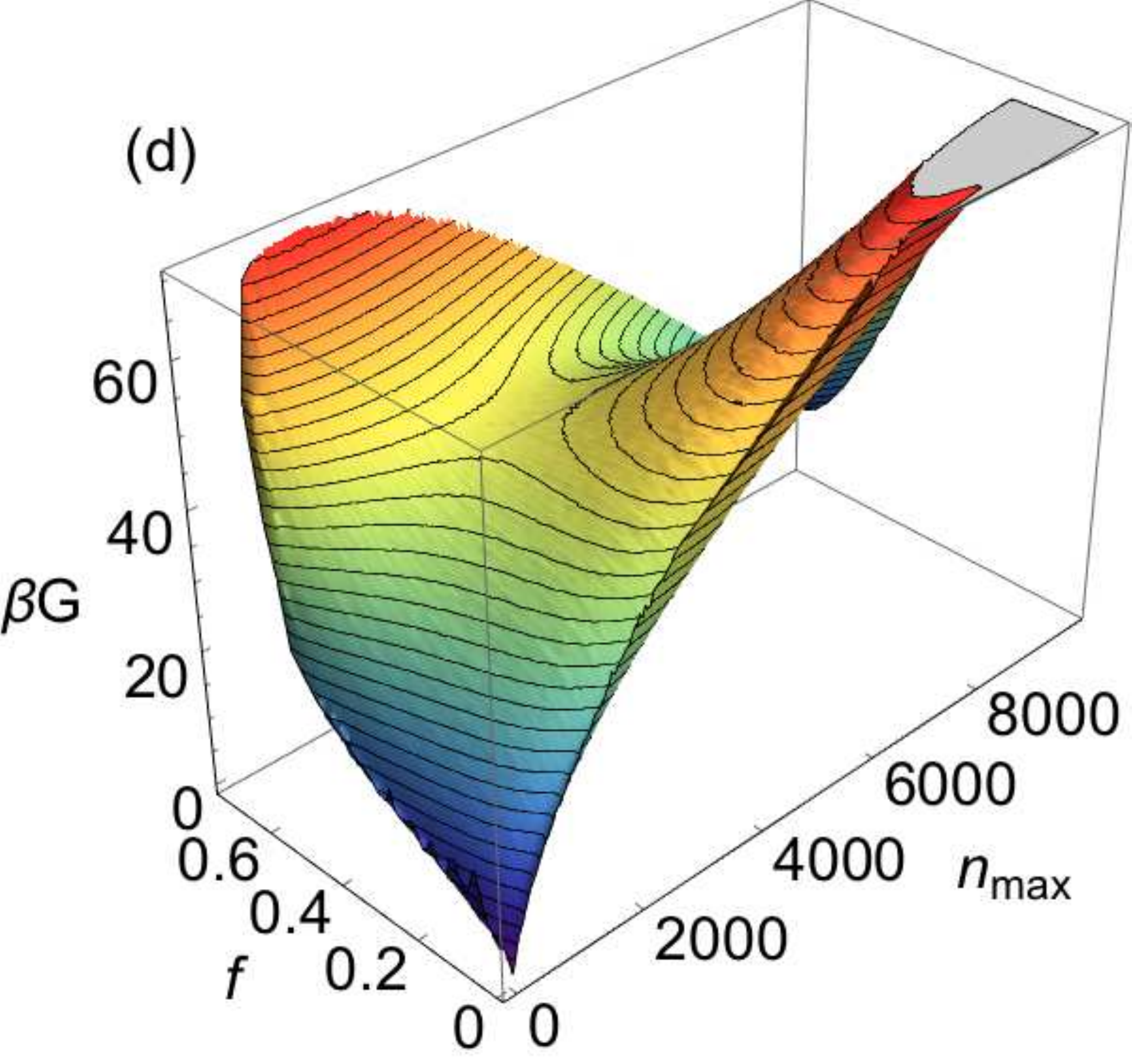}
\includegraphics[scale=\x]{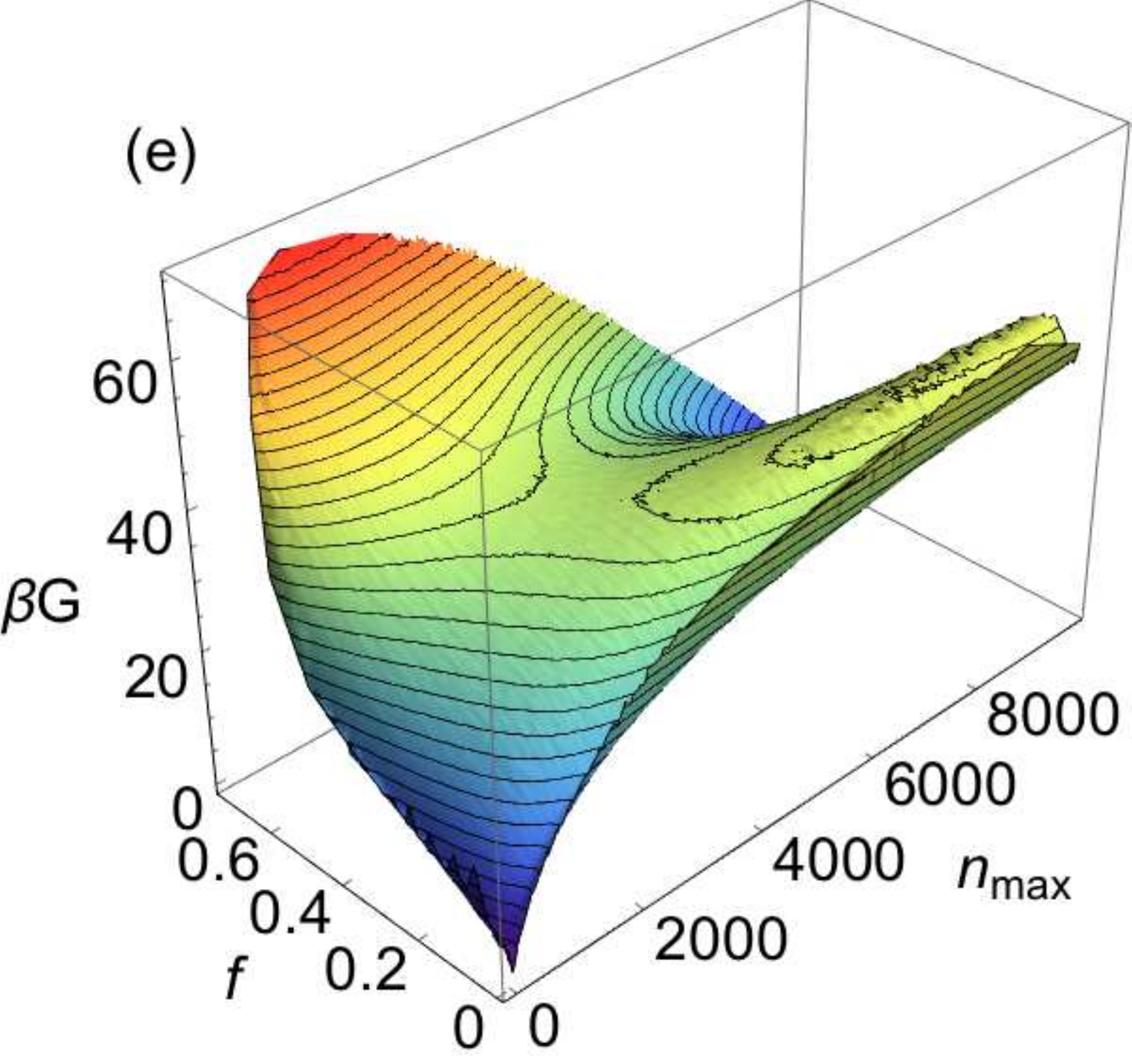}
\includegraphics[scale=\x]{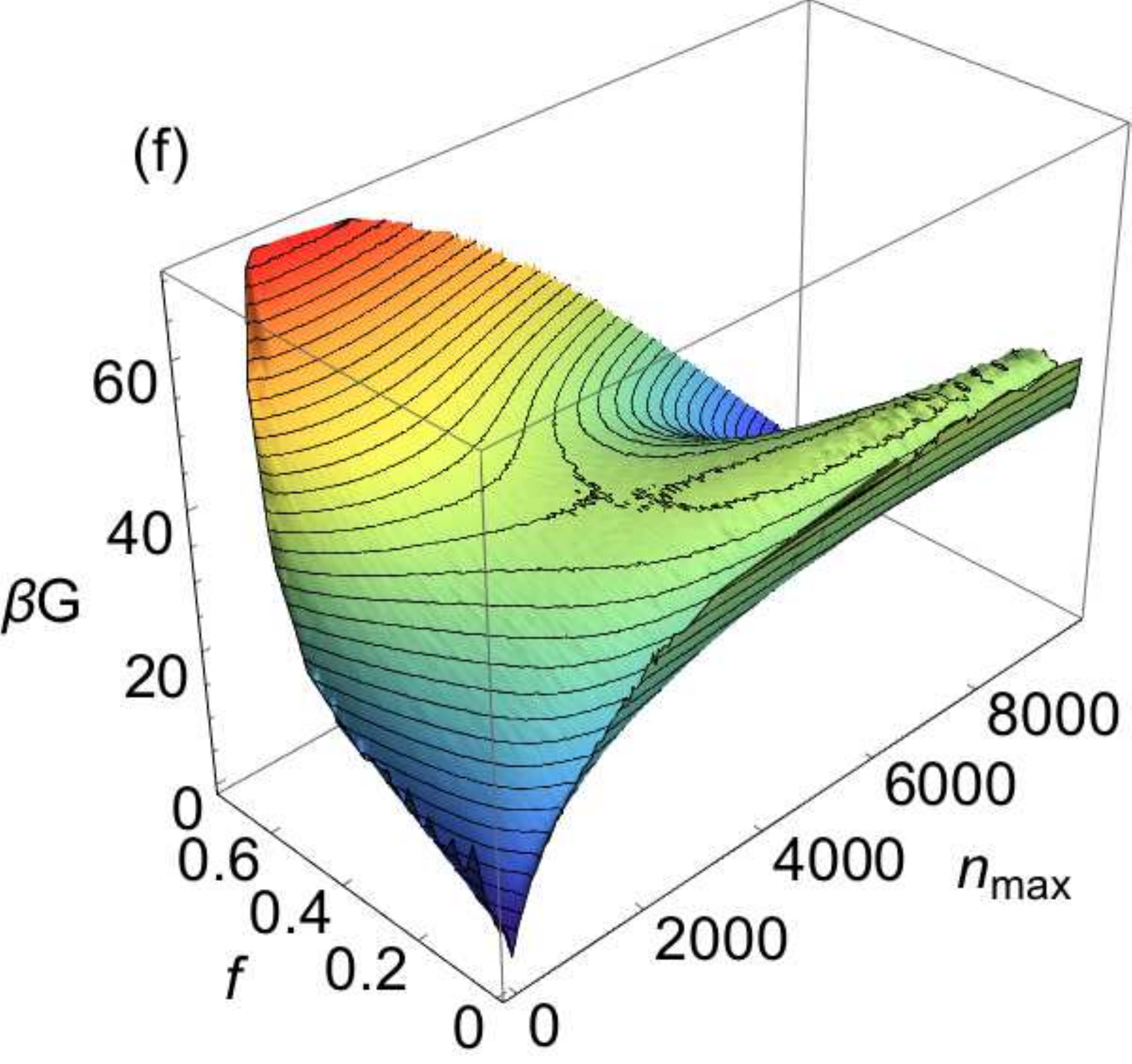}
\includegraphics[scale=\x]{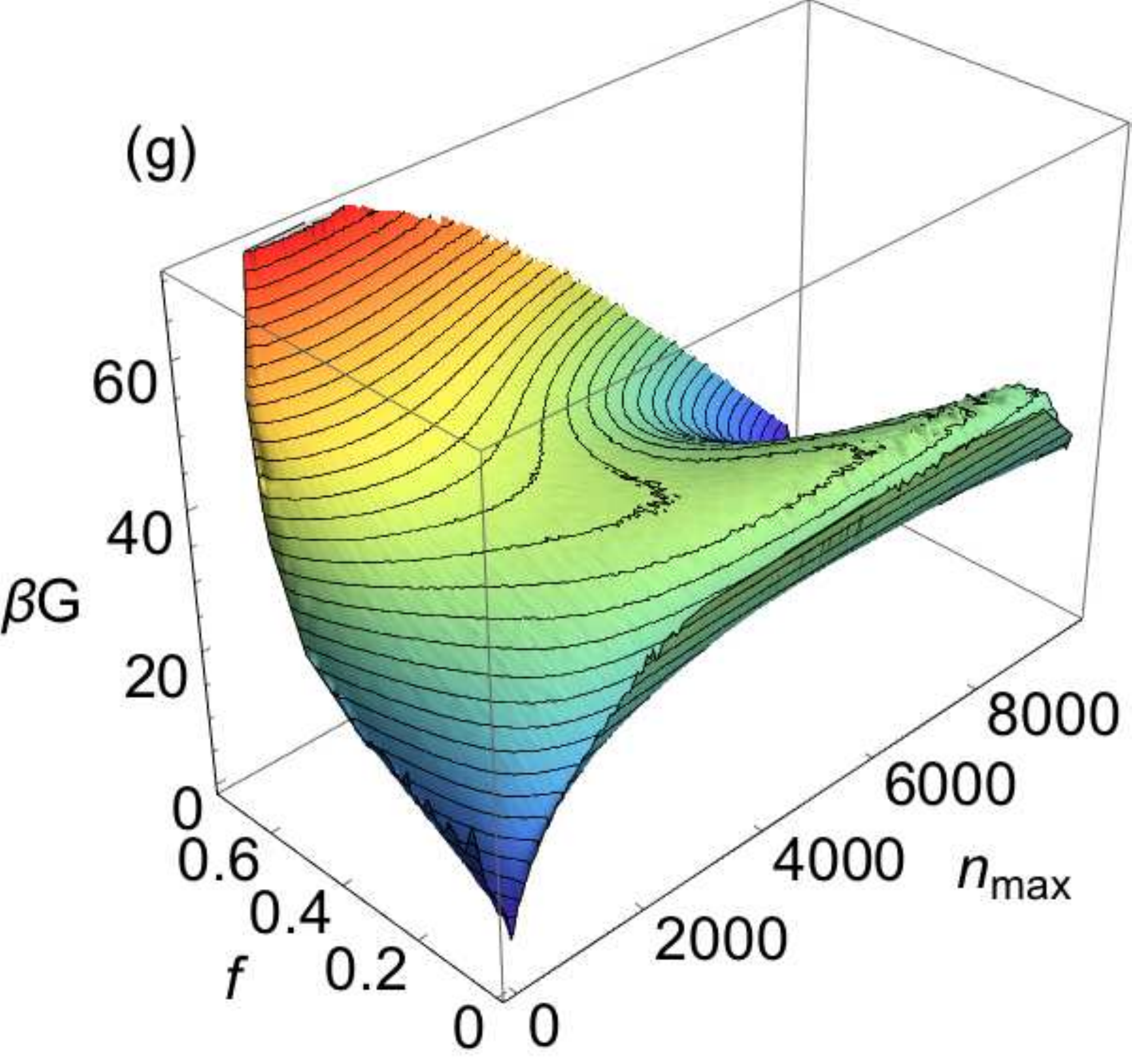}
\includegraphics[scale=\x]{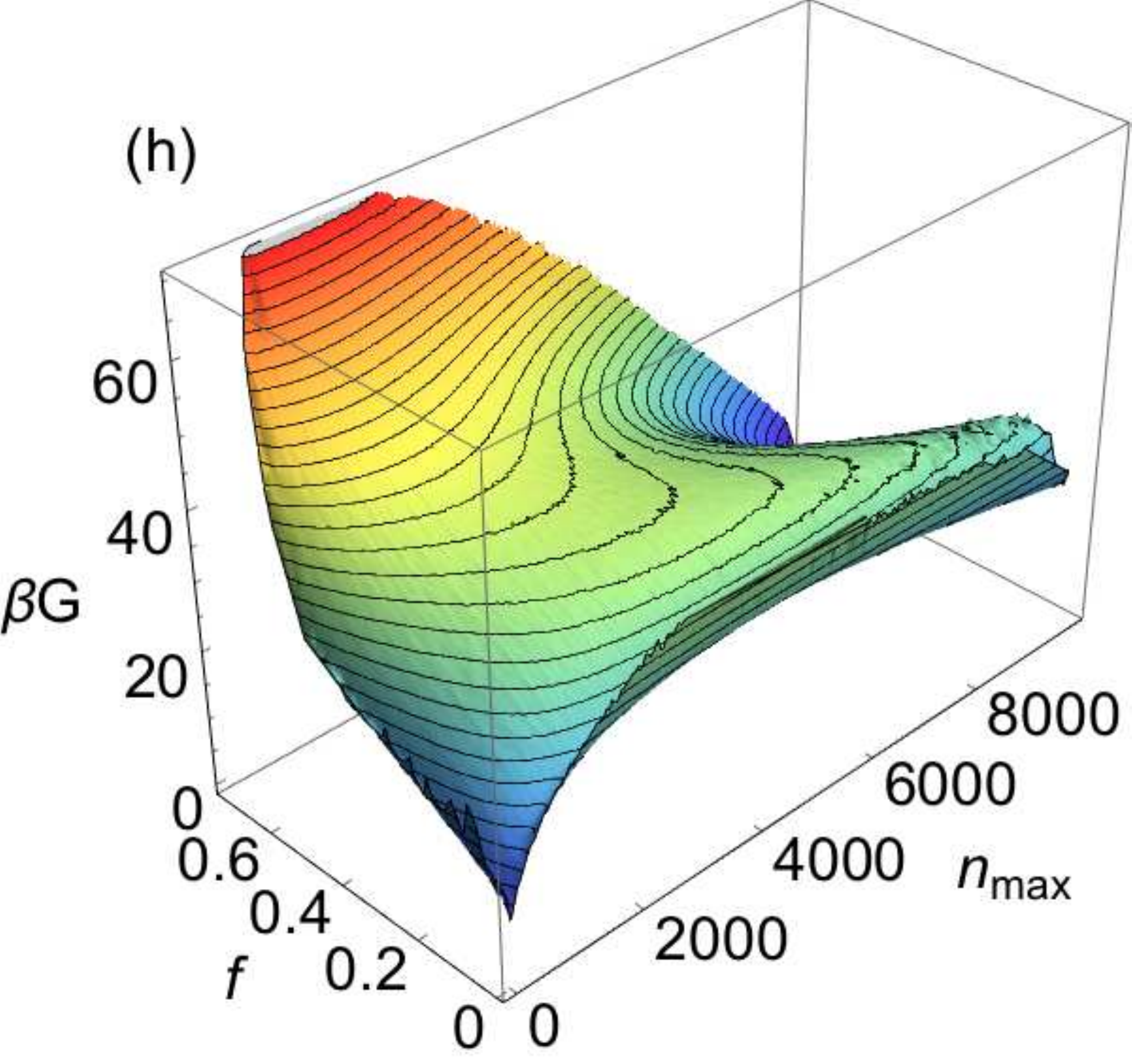}
\includegraphics[scale=\x]{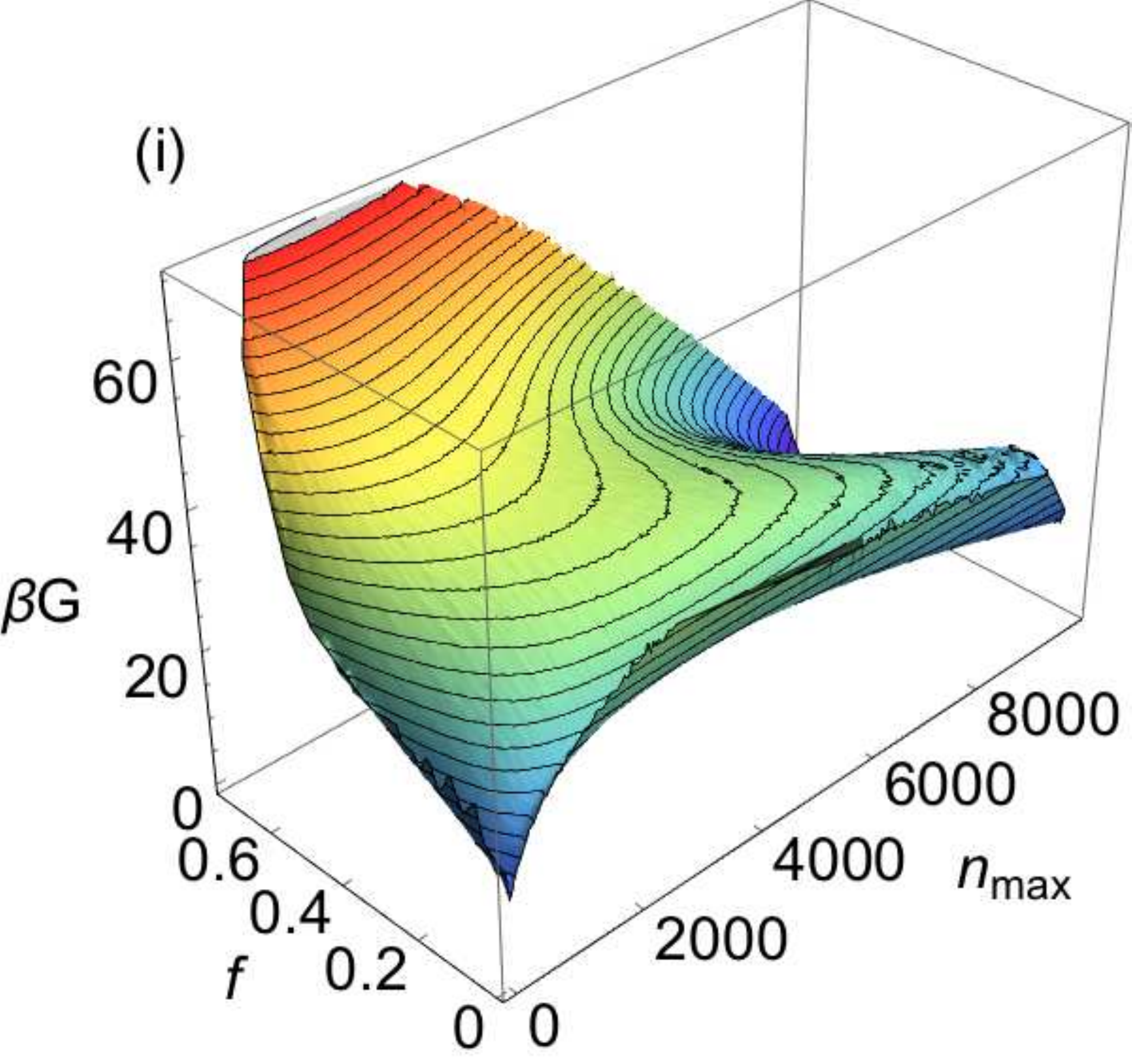}
\includegraphics[scale=\x]{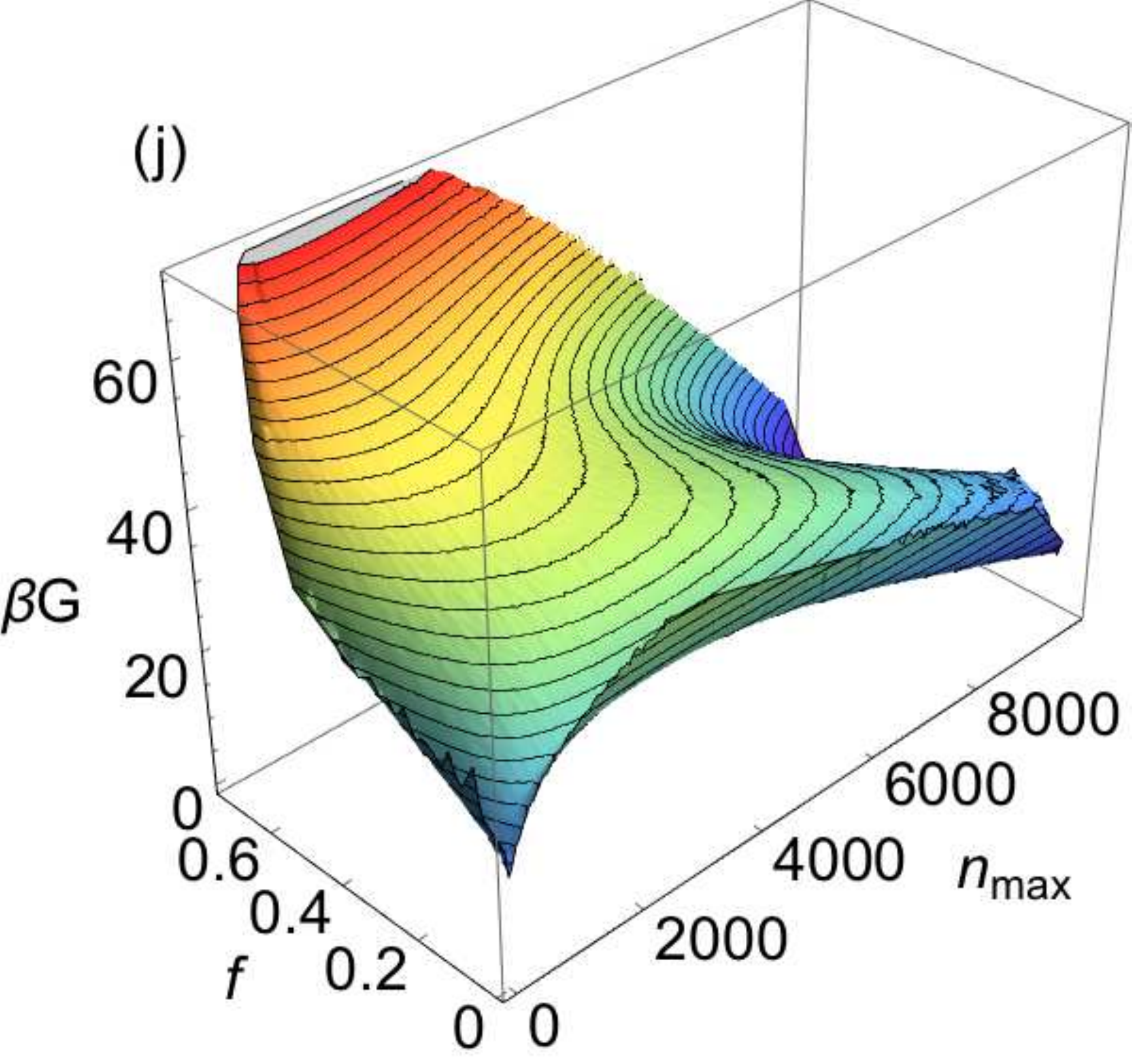}
\caption{$G(n_{\rm max},f)$ for $H_s=0.01$ and $L=200$.  
Panels (a) through (j) correspond respectively to $H=\{3.96$, 3.965, 3.97, 3.975, 3.98, 3.981, 3.982, 3.983, 3.984, 3.985$\}$. 
Contours are $2kT$ apart.
}
\label{GCbar2}
\end{figure*}

\begin{figure*}
\newcommand\x{0.27}
\includegraphics[scale=\x]{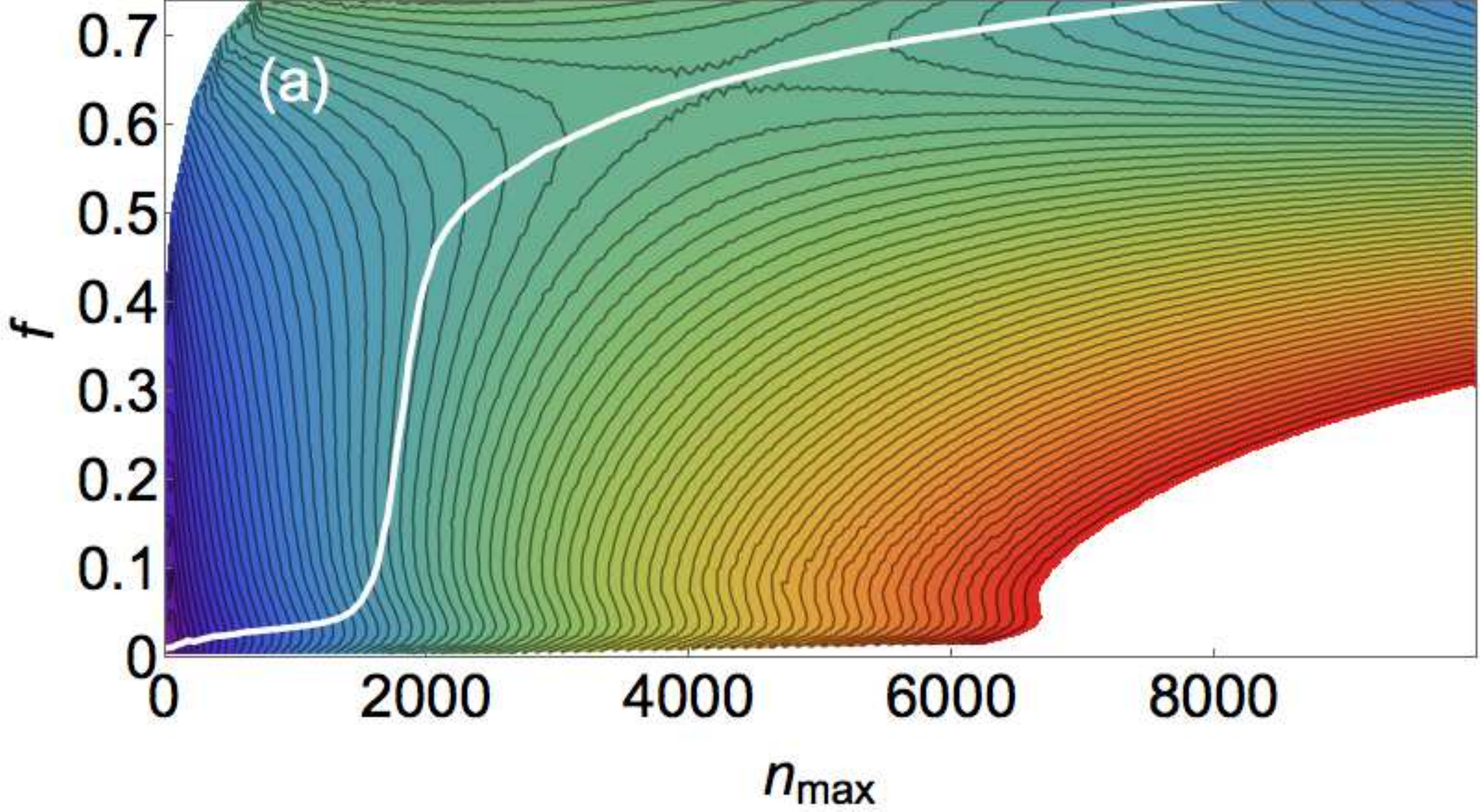}
\includegraphics[scale=\x]{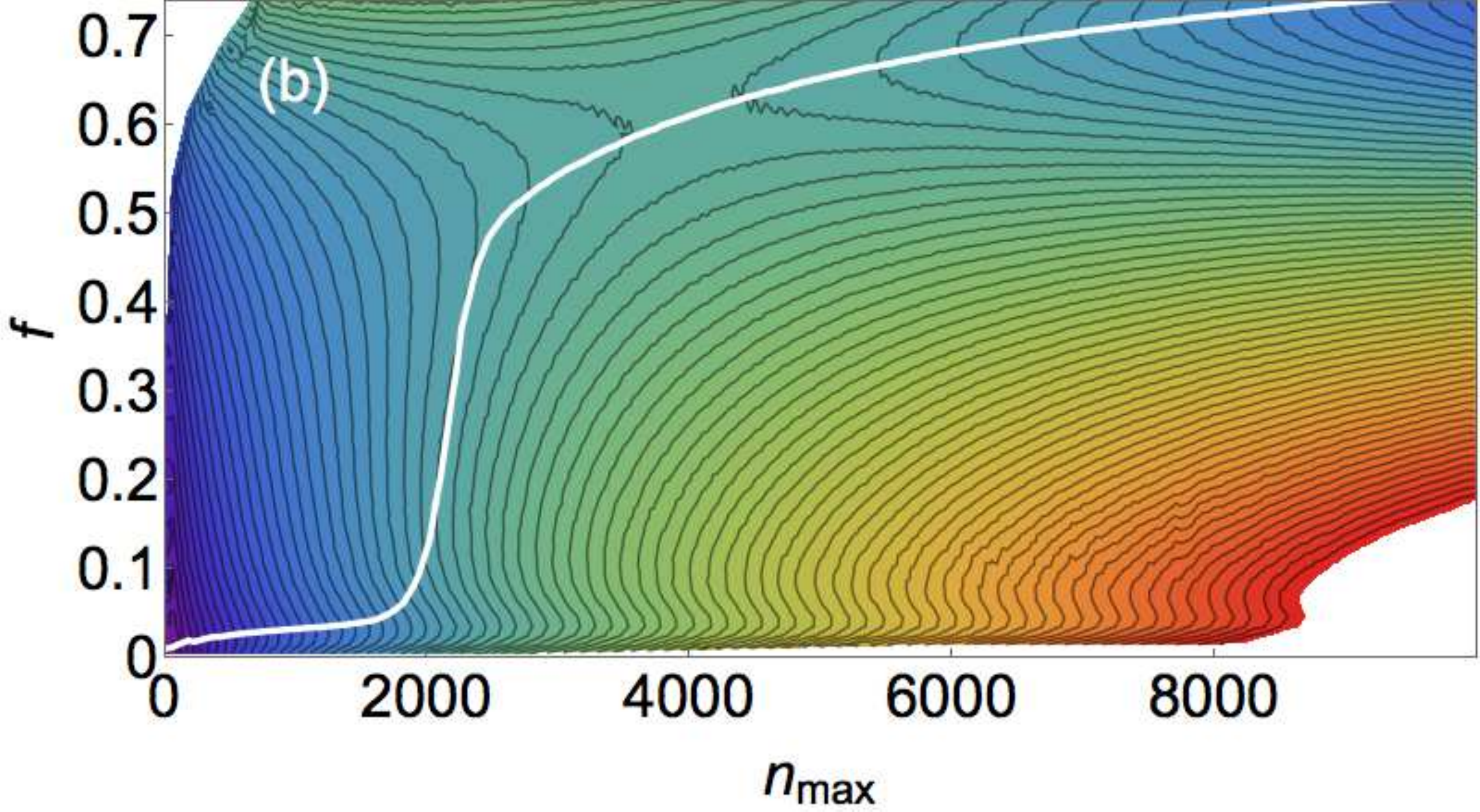}
\includegraphics[scale=\x]{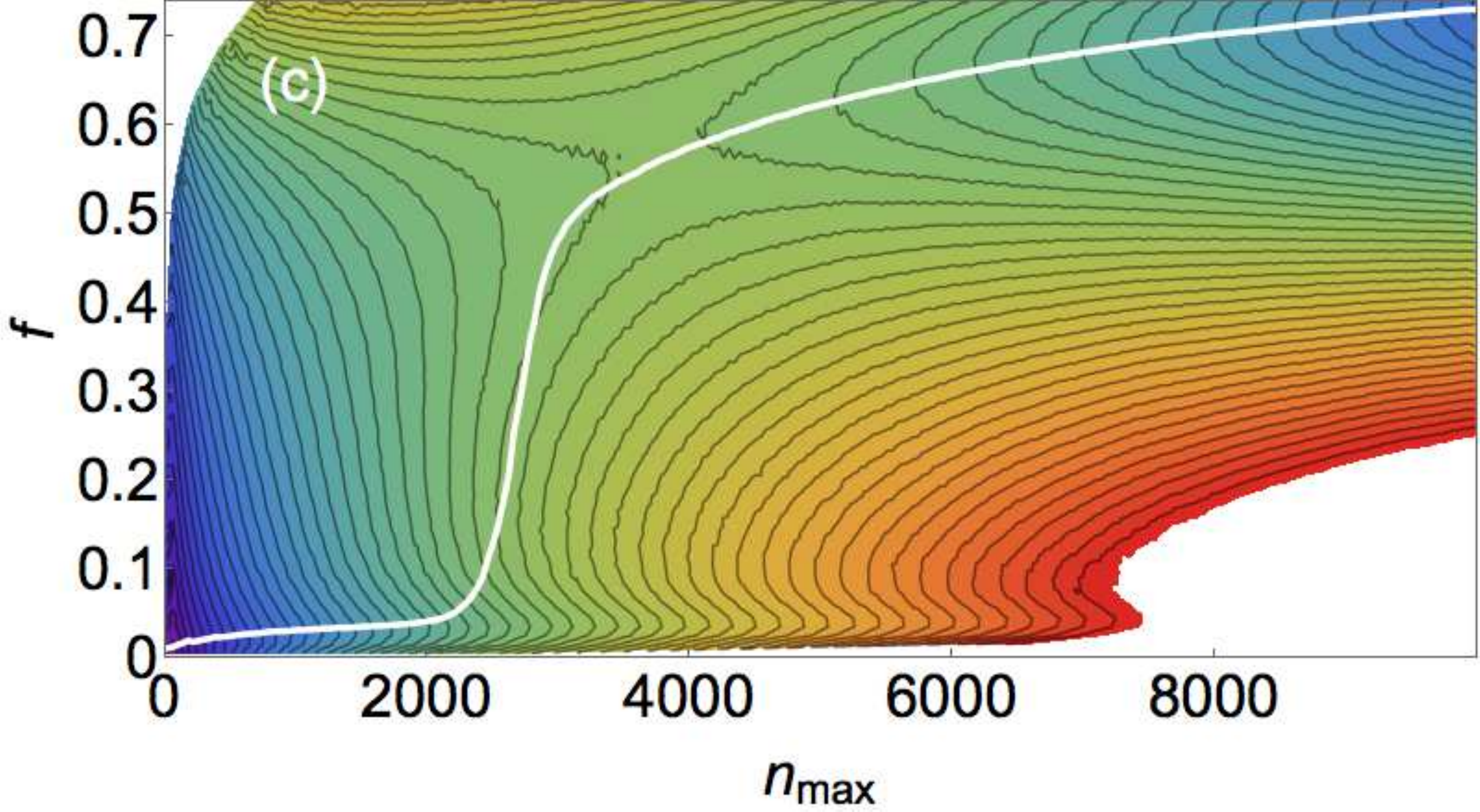}
\includegraphics[scale=\x]{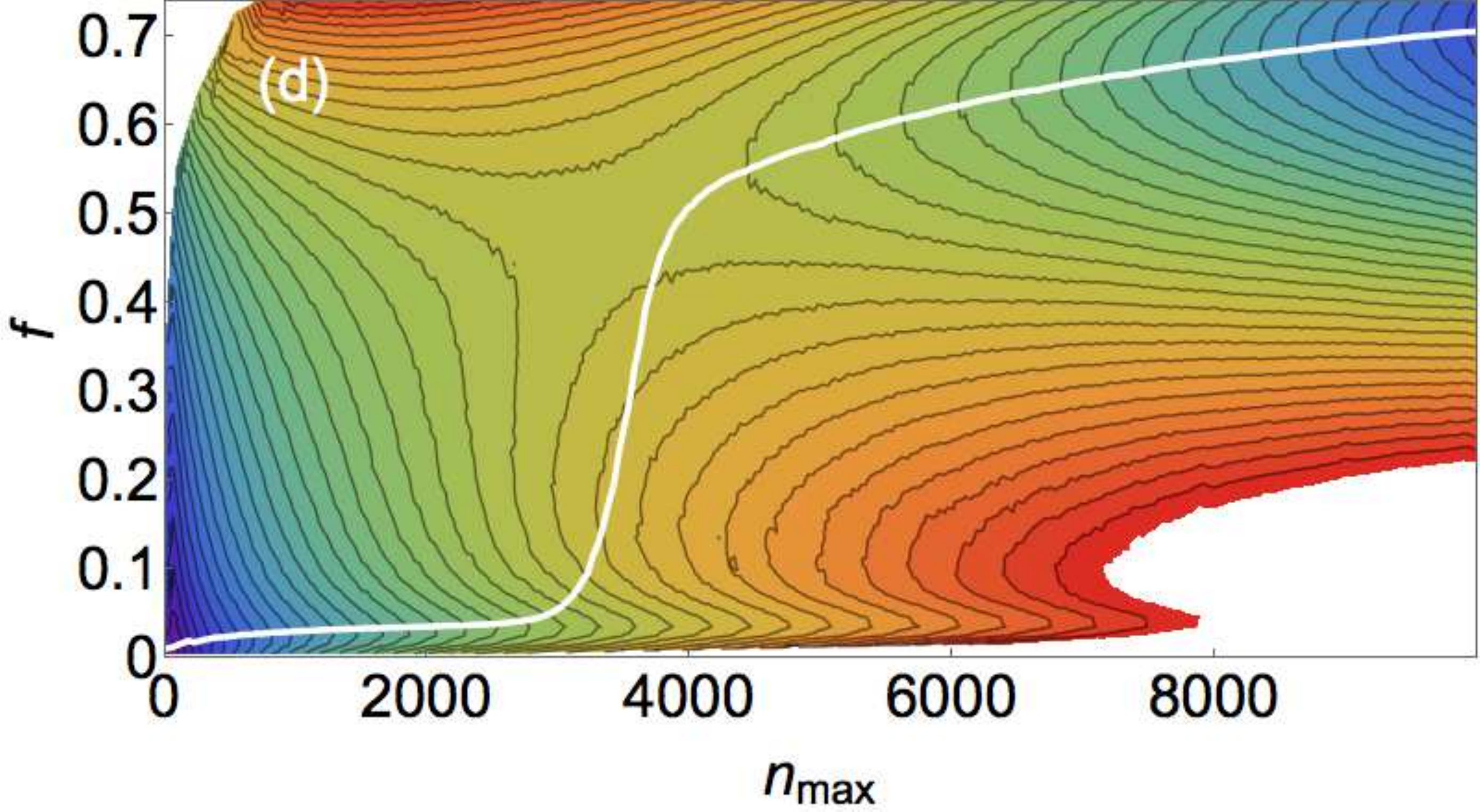}
\includegraphics[scale=\x]{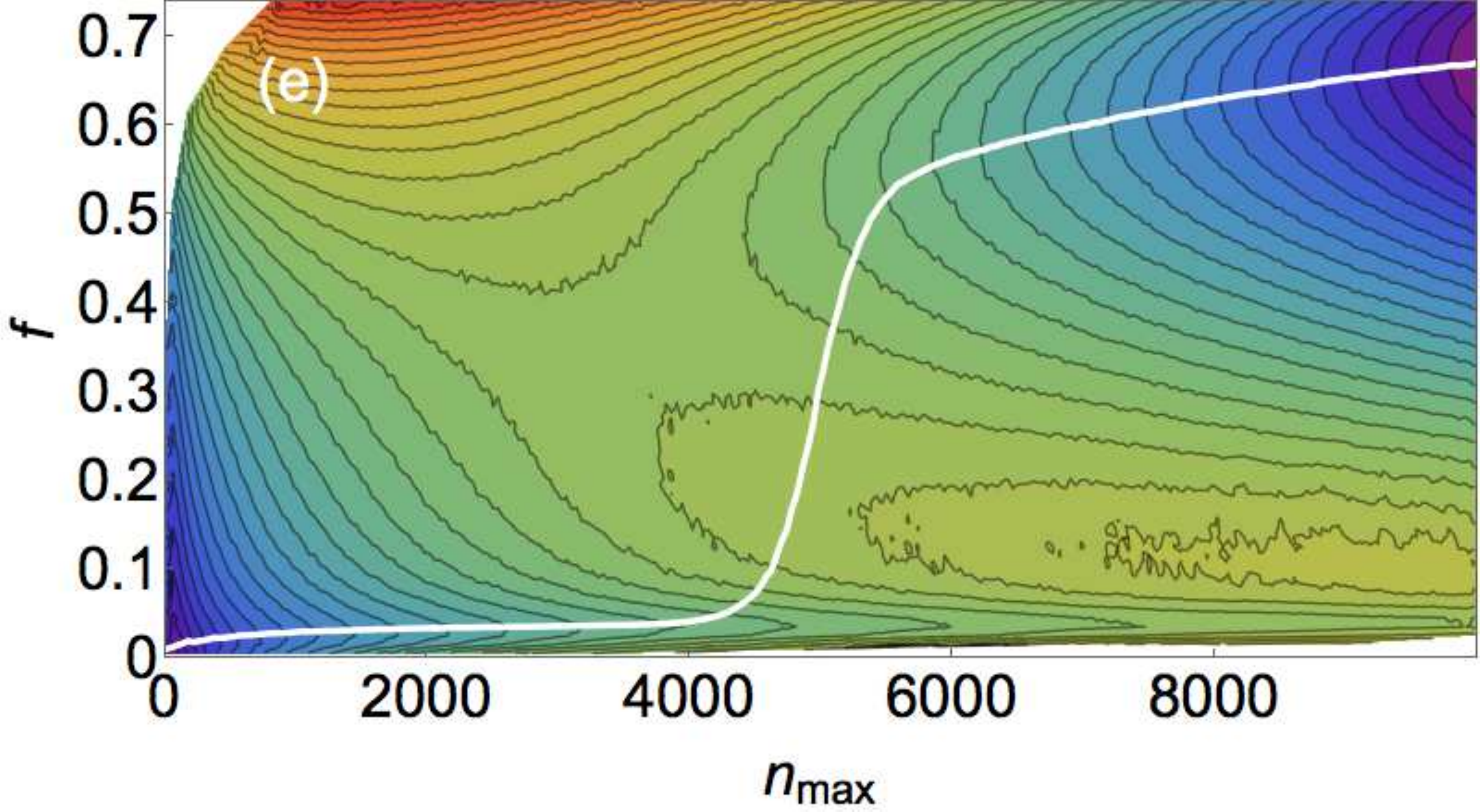}
\includegraphics[scale=\x]{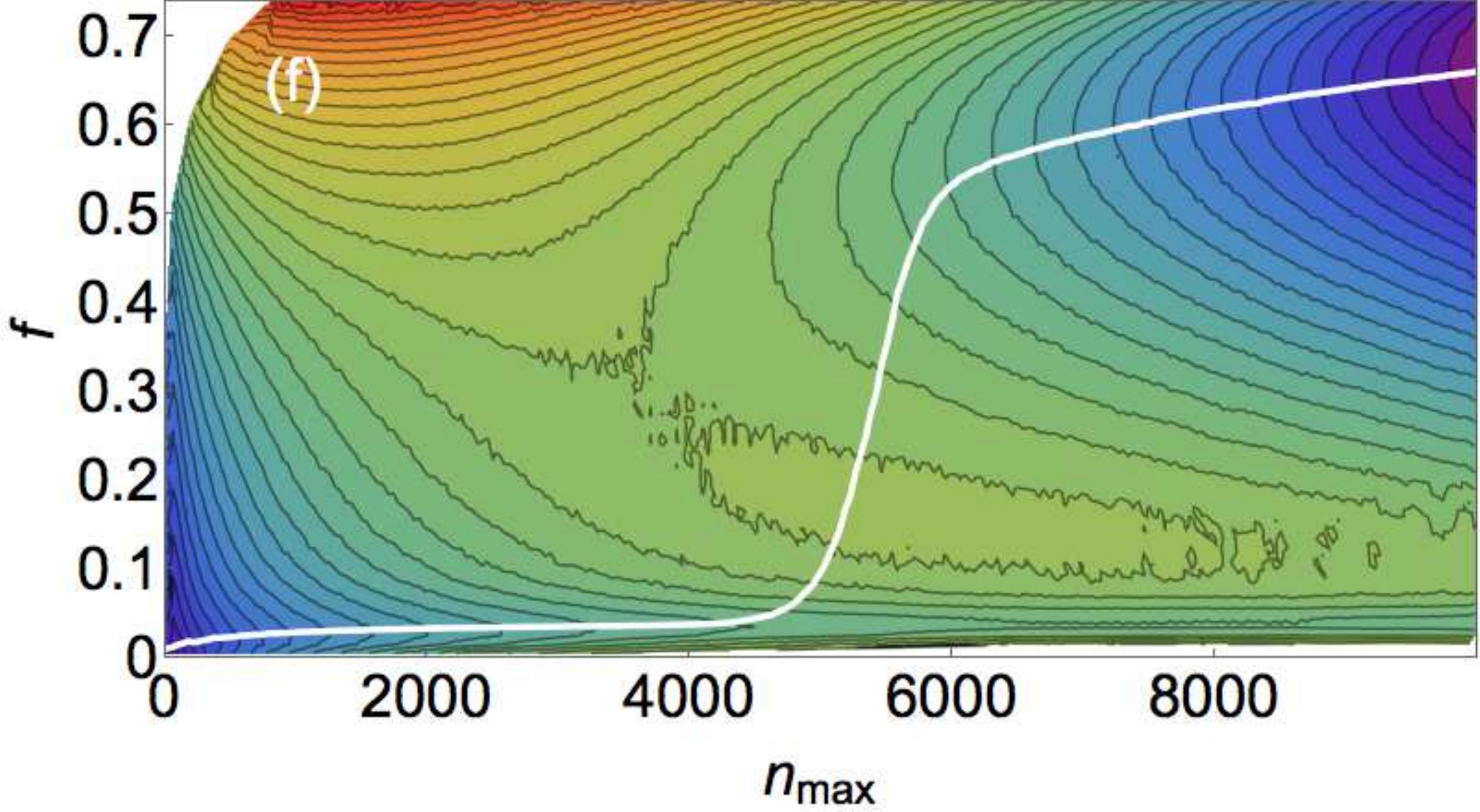}
\includegraphics[scale=\x]{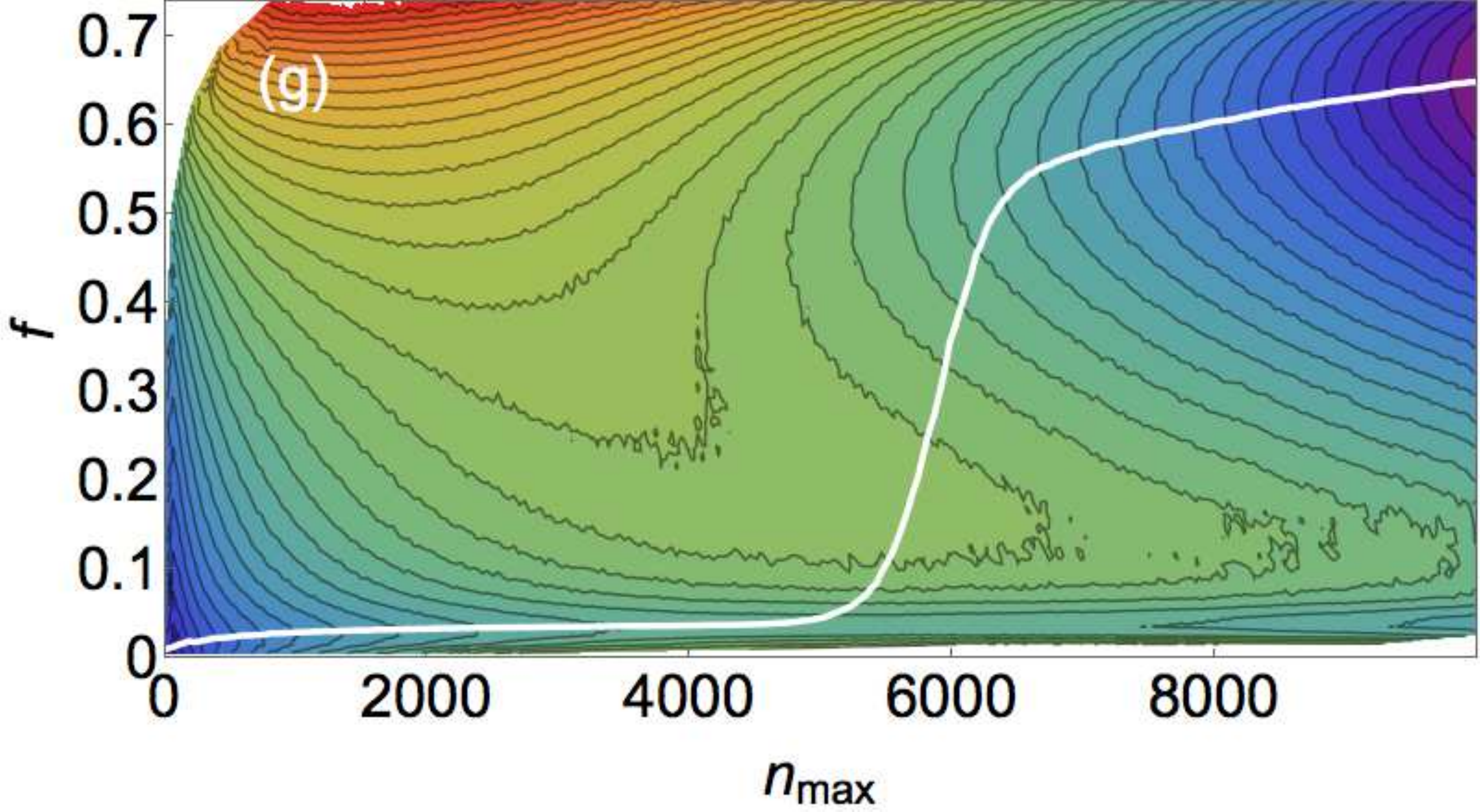}
\includegraphics[scale=\x]{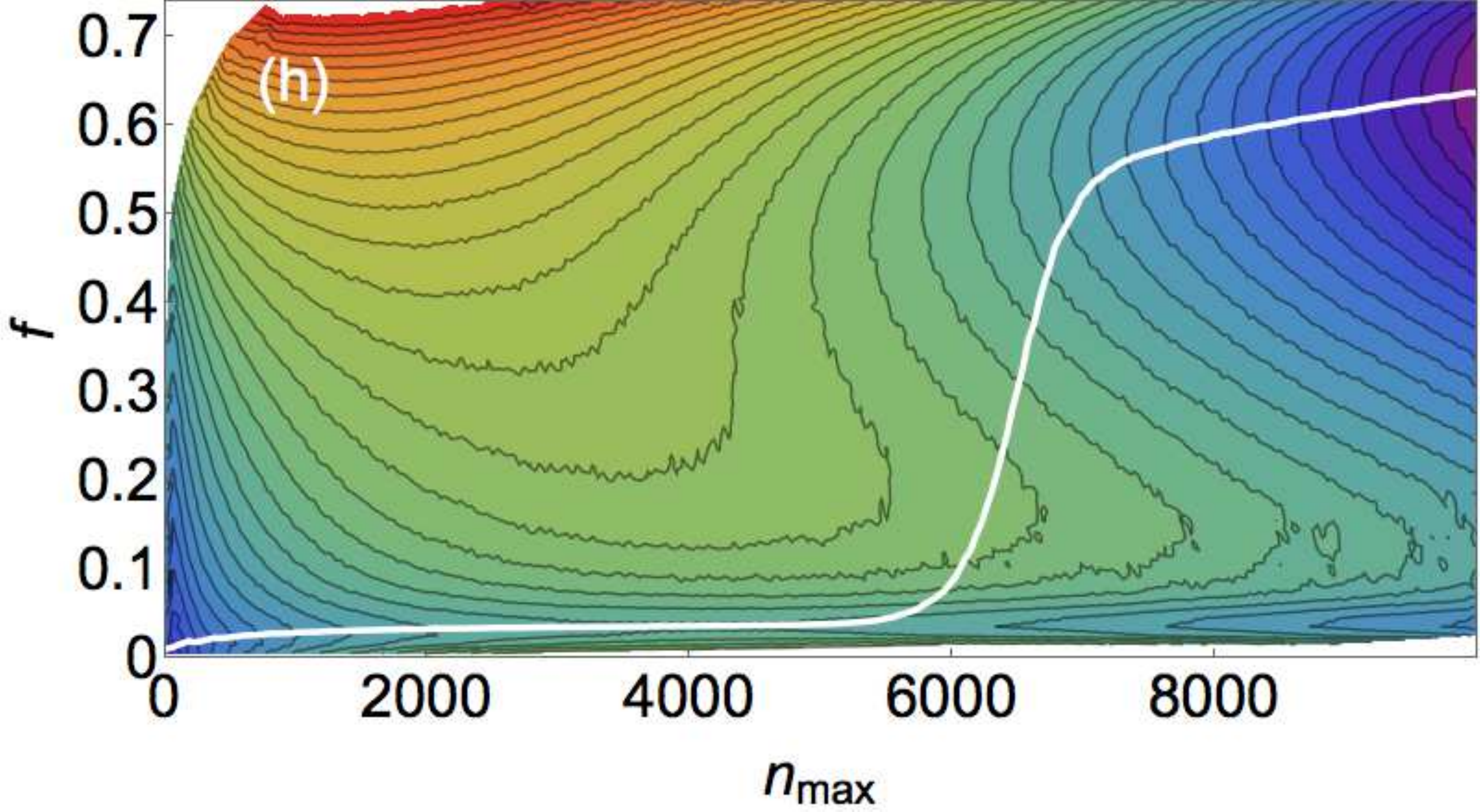}
\includegraphics[scale=\x]{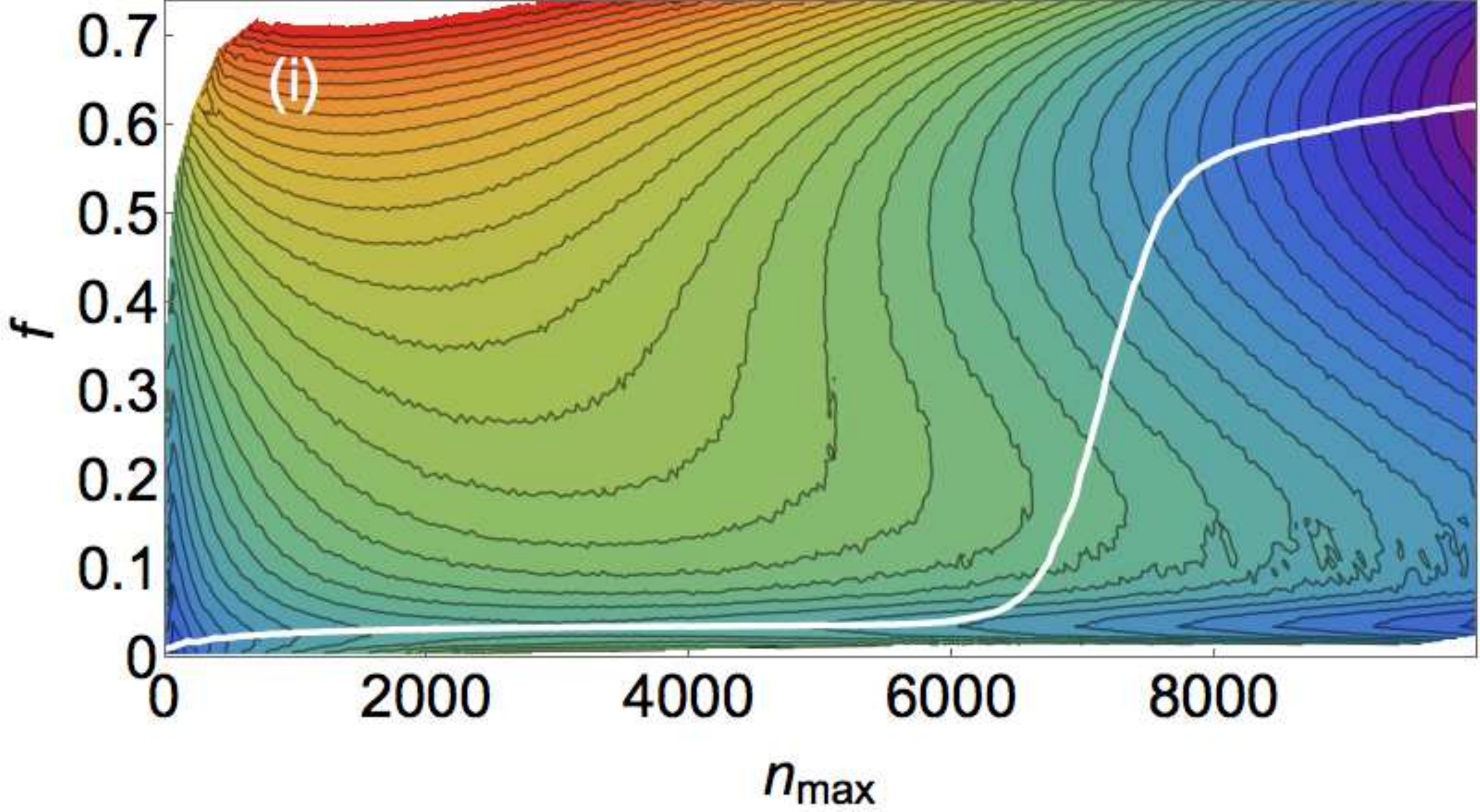}
\includegraphics[scale=\x]{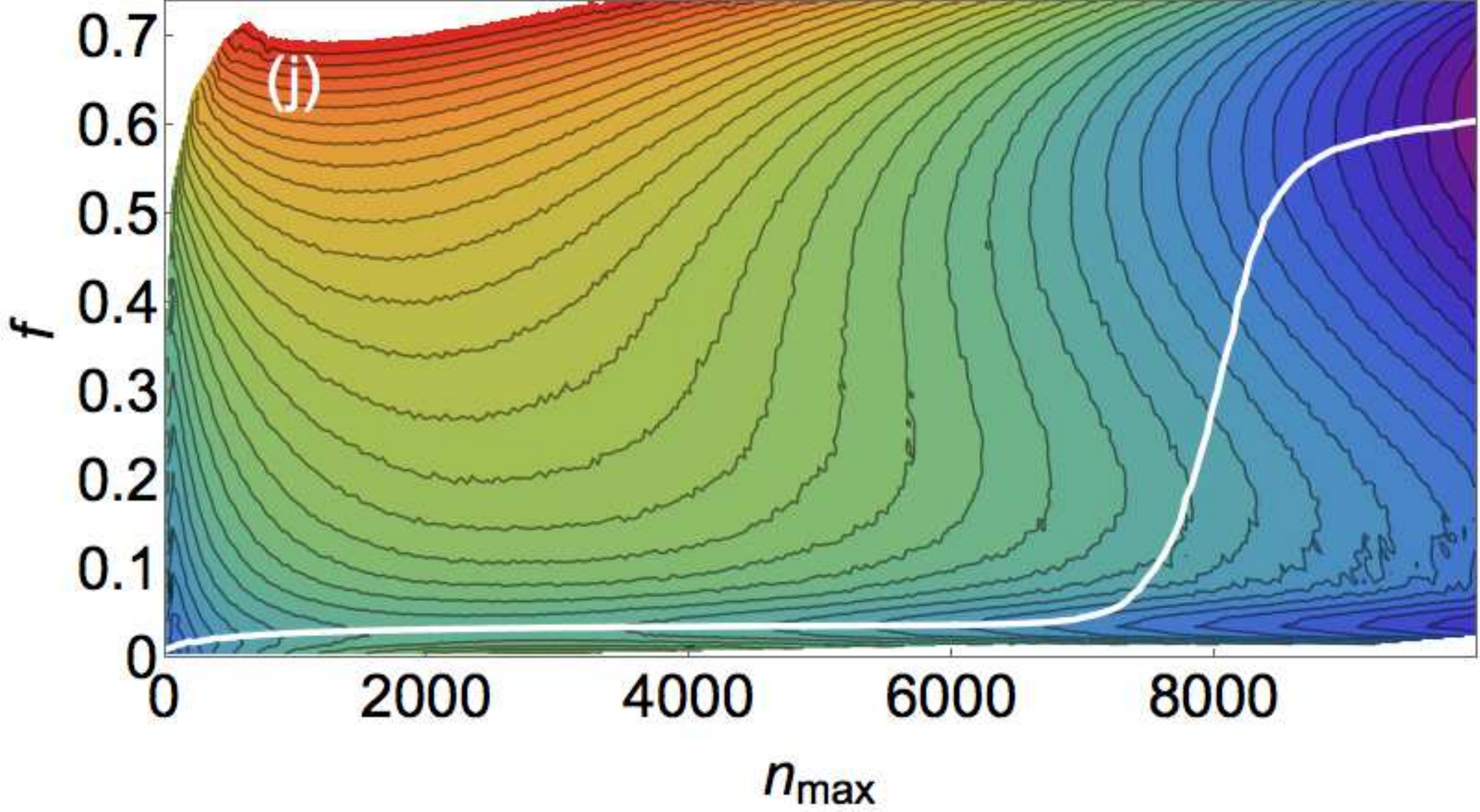}
\caption{$G(n_{\rm max},f)$ for $H_s=0.01$ and $L=200$.  The white line is $\langle f\rangle$. 
Panels (a) through (j) correspond respectively to
$H=\{3.96$, 3.965, 3.97, 3.975, 3.98, 3.981, 3.982, 3.983, 3.984, 3.985$\}$. 
Contours are $2kT$ apart.
}
\label{GCbar}
\end{figure*}

\section{Additional plots of $G(\nmax,f)$}
\label{morefes}

We present in Figs.~\ref{GCbar2} and \ref{GCbar} surface and contour plots of $G(\nmax,f)$ over additional values of $H$ between 3.960 and 3.985.

\section{Finding $n_c$ and $n^*$ over a wide range of $H_s$ and $H$}
\label{1d}

To estimate $n_c$ and $n^*$ over a wide range of $H_s$ and $H$, we conduct 1D umbrella sampling simulations in which $\nmax$ alone is constrained.  In this approach, we directly compute $G_1(\nmax)$ using,
\begin{equation}
\beta G_1(\nmax)=-\log[P_1(\nmax)] +C,
\end{equation}
where $P_1(\nmax)$ is proportional to the probability to observe a microstate with a given value of $\nmax$.  We find $n^*$ from the maximum in $G_1(\nmax)$.  We also monitor $f$ during these runs, which allows us to compute $\chi$ and thus to find $n_c$ from the maximum in $\chi$. 

To proceed, we use a biasing potential,
\begin{equation}
U_B''=\kappa_n(\nmax-\nmax^*)^2,
\label{us1}
\end{equation}
where $\kappa_n=0.002$.
Trial configurations are accepted or rejected using the umbrella potential every 1~MCS.
Each run is initiated from a perfect $\ca$ configuration, into which a seed cluster is inserted.  The seed cluster is a square of sites with $s_i=1$ of a size chosen to be closest to $\nmax^*$.  
We carry out runs using $L=64$ or $128$, as indicated in the legends or captions of the figures.
For each choice of $(L,H_s,H,T)$,
we conduct simulations for each value of 
$\nmax^*=50i$ where the integer $i\in [0,24]$ when $L=64$,
and $i\in [0,80]$ when $L=128$.
Each simulation is equilibrated for $5\times 10^4$~MCS, and then the time series of $\nmax$ is recorded every 100~MCS for $10^6$~MCS.  
These time series are analyzed using WHAM to evaluate $P_1(n_{\rm max})$ and $G_1(n_{\rm max})$.  
We exclude from the WHAM analysis any runs for which the acceptance rate for the umbrella sampling is less than $0.1$.
As shown in Fig.~\ref{nn}, we use these 1D umbrella sampling runs to evaluate $n_c$ and $n^*$ for $H_s$ in the range $[0.01,0.10]$ and $H$ in the range $[3.7,4.0]$.  

\section{height of the nucleation barrier}
\label{gstar}

\begin{figure}
\includegraphics[scale=0.40]{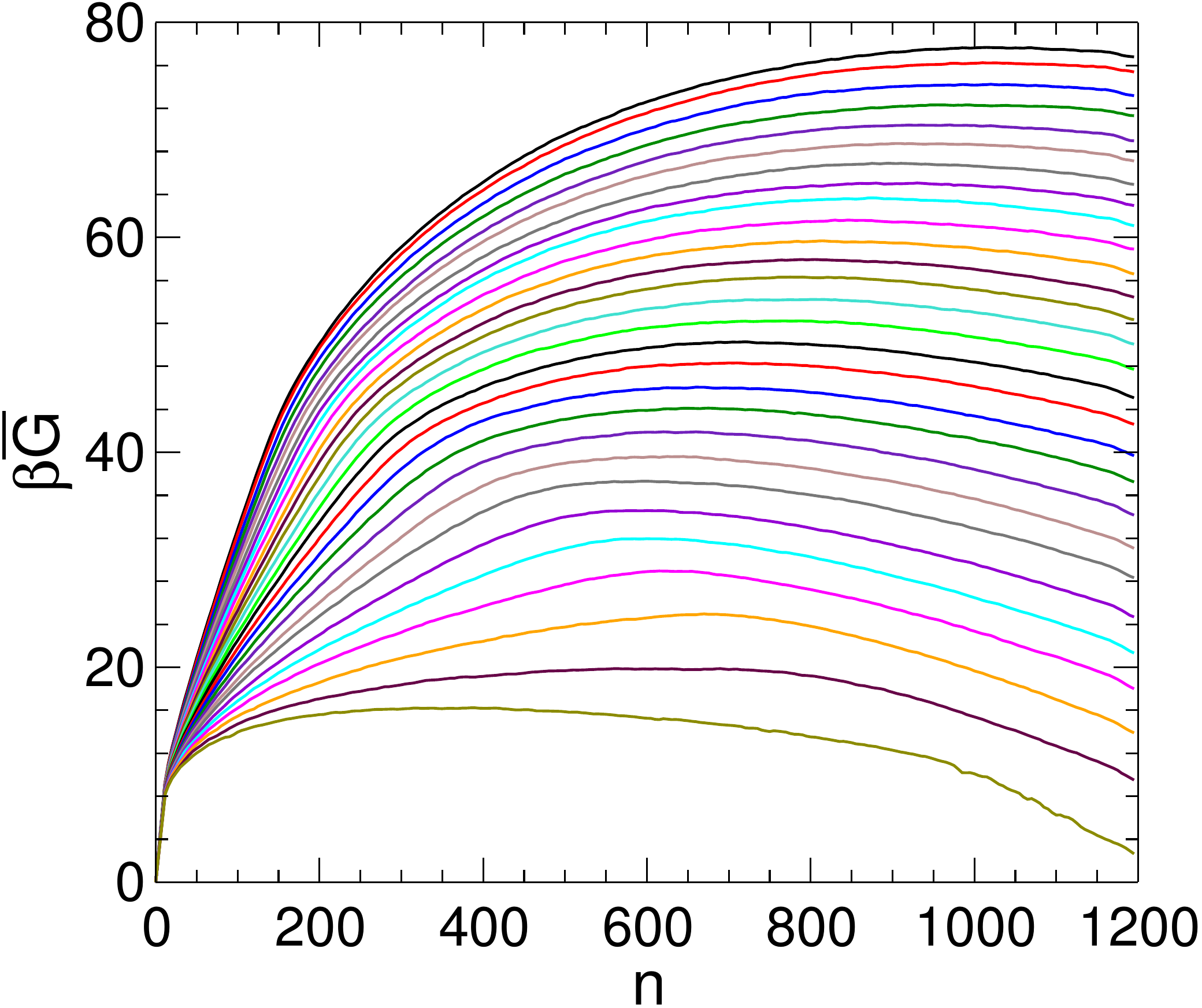}
\caption{${\bar G}(n)$ for $L=64$ and $H_s=0.04$.  $H=3.70$ to 3.97 in steps of 0.01, from top to bottom.}
\label{non}
\end{figure}

Although the maximum in $G_1(\nmax)$ properly identifies $n^*$, it is important to note that $G_1(n^*)$ does not give the height of the nucleation barrier.  
The nucleation barrier is more accurately estimated by computing,
\begin{equation}
\beta \bar G(n)=-\log\frac{{\cal N}(n)}{N},
\end{equation}
where ${\cal N}(n)$ is the average number of clusters of size $n$ in a system of size 
$N$~\cite{Wolde:1996p3069,Auer:2004db,Lundrigan:2009p5256}.
The height of the nucleation barrier is then defined as $G^*=\bar G(n^*)$.  We evaluate ${\cal N}(n)$ from our 1D umbrella sampling simulations, and so we are able to estimate $\bar G(n)$ for these cases.  An example is shown in Fig.~\ref{non}, where we show $\bar G(n)$ for several value of $H$ at $H_s=0.04$.  We obtain $G^*$ from the maxima of these curves, and by interpolation identify the value of $H$ at which $\beta G^*=20$.  The locus of points in the $(H_s,H)$ plane at which $\beta G^*=20$ is shown in Fig.~\ref{pd1}.

\end{document}